\documentclass[sigconf]{acmart}
\usepackage{balance}
\usepackage{expl3}
\usepackage{amsmath,amssymb,amsfonts}
\usepackage{hyperref}
\usepackage{soul}
\usepackage{xcolor}
\usepackage[capitalize]{cleveref}
\usepackage{float}
\usepackage{tabularx}
\usepackage{pseudo}
\pseudodefinestyle{fullwidth}{
  begin-tabular = \tabularx{\linewidth}{@{}
    r % Labels
    >{\pseudosetup} % Indent, font, ...
    X % Code (flexible)
    >{\leavevmode\small\color{black!60}} % Comment styling
    p{0.45\linewidth} % Comments (fixed)
    @{}},
  end-tabular = \endtabularx,
  setup-append = \pseudoeq
}
\pseudoset{
  ctfont=\color{black!55}\textit,
  ct-left=//\ ,
  ct-right=
}

\floatstyle{plaintop}
\newfloat{algorithm}{tbp}{alg}[section]
\floatname{algorithm}{Algorithm}

\usepackage{multirow}
\usepackage{multicol}
\usepackage[labelformat=simple]{subcaption}

\def\BibTeX{{\rm B\kern-.05em{\sc i\kern-.025em b}\kern-.08emT\kern-.1667em\lower.7ex\hbox{E}\kern-.125emX}}

\copyrightyear{2020}
\acmYear{2020}
\setcopyright{acmlicensed}
\acmConference[ICPE '20]{Proceedings of the 2020 ACM/SPEC International Conference on Performance Engineering}{April 20--24, 2020}{Edmonton, AB, Canada}
\acmBooktitle{Proceedings of the 2020 ACM/SPEC International Conference on Performance Engineering (ICPE '20), April 20--24, 2020, Edmonton, AB, Canada}
\acmPrice{15.00}
\acmDOI{10.1145/3358960.3379141}
\acmISBN{978-1-4503-6991-6/20/04}

\begin{document}

\fancyhead{}
% do not delete this code.

% The "title" command has an optional parameter, allowing the author to define a "short title" to be used in page headers.
\title{Throughput Prediction of Asynchronous SGD in TensorFlow}

\settopmatter{authorsperrow=4}

\author{Zhuojin Li}\authornote{Authors with equal contribution.}
\affiliation{%
  \institution{University of Southern California}
  \city{Los Angeles}
  \state{California}
}
\email{zhuojinl@usc.edu}

\author{Wumo Yan}\authornotemark[1]
\affiliation{%
  \institution{University of Southern California}
  \city{Los Angeles}
  \state{California}
}
\email{wumoyan@usc.edu}

\author{Marco Paolieri}
\affiliation{%
  \institution{University of Southern California}
  \city{Los Angeles}
  \state{California}
}
\email{paolieri@usc.edu}

\author{Leana Golubchik}
\affiliation{%
  \institution{University of Southern California}
  \city{Los Angeles}
  \state{California}
}
\email{leana@usc.edu}

\renewcommand{\shortauthors}{Li et al.}

\begin{abstract}
Modern machine learning frameworks can train neural networks using multiple nodes in parallel, each computing parameter updates with stochastic gradient descent (SGD) and sharing them asynchronously through a central parameter server. Due to communication overhead and bottlenecks, the total throughput of SGD updates in a cluster scales sublinearly, saturating as the number of nodes increases.
In this paper, we present a solution to predicting training throughput from profiling traces collected from a single-node configuration. Our approach is able to model the interaction of multiple nodes and the scheduling of concurrent transmissions between the parameter server and each node. By accounting for the dependencies between received parts and pending computations, we predict overlaps between computation and communication and generate synthetic execution traces for configurations with multiple nodes.
We validate our approach on TensorFlow training jobs for popular image classification neural networks, on AWS and on our in-house cluster, using nodes equipped with GPUs or only with CPUs. We also investigate the effects of data transmission policies used in TensorFlow and the accuracy of our approach when combined with optimizations of the transmission schedule.
\end{abstract}


%\ccsdesc[500]{Computing methodologies~Distributed algorithms}
%\ccsdesc[500]{Computing methodologies~Neural networks}
%\ccsdesc[500]{Networks~Network performance analysis}

% Separate the keywords with commas.


\maketitle
% -*- ispell-local-dictionary: "american"; TeX-master: "../tensorpredict.tex"; -*-

\section{Introduction} \label{introduction}

Deep learning \cite{lecun2015deep} has achieved breakthrough results
in several application domains, including computer vision, speech
recognition, natural language processing.
In contrast with traditional machine learning, Deep Neural
Networks~(DNNs) discover internal representations suitable for
classification from training data, without the need for manual feature
engineering.
This approach requires very large amounts of training data and
computation: for example, popular DNNs for image classification
include millions of model parameters (\emph{DNN weights}) trained
using datasets of millions of labeled images.
Training examples are grouped in small batches and used for
optimization steps with Stochastic Gradient Descent (SGD), which is
computationally expensive (gradients are computed by propagating
output errors back to each model parameter, through a sequence of
matrix multiplications \cite{lecun2015deep}).

\begin{figure}[bt!]
  \centering
  \vspace{1em}
  \begin{subfigure}[b]{.50\columnwidth}
    \centering
    \includegraphics[width=\linewidth]{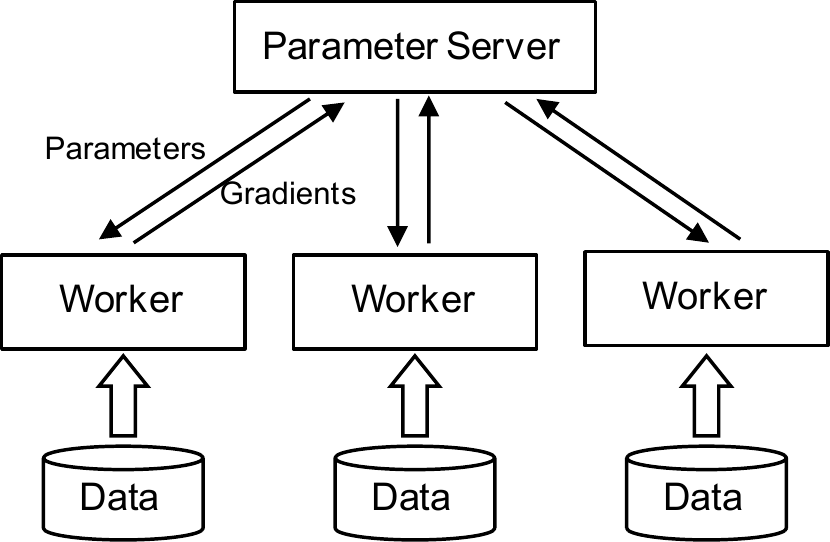}
  \end{subfigure}%
  \caption{Parameter Server Architecture}\label{fig:ps_architecture}
  \vspace{-1.5em}
\end{figure}

Training performance can be improved by using more powerful hardware,
such as GPUs and FPGAs. To improve performance even further, machine
learning frameworks such as TensorFlow~\cite{abadi2016tensorflow} can
use multiple \emph{worker nodes}, each performing SGD steps on a shard
of the training data.
A popular architecture to share model updates between worker nodes is
the \emph{parameter server}~\cite{li2014scaling}, illustrated in
\cref{fig:ps_architecture}.
The parameter server holds a global version of model parameters (the
DNN weights): each worker receives these parameters (\emph{downlink
  phase}), computes an update from a batch of labeled examples
(\emph{computation phase}), and transmits its update to the parameter
server (\emph{uplink phase}), where it is applied to the global model
(\emph{update phase}).
In \emph{asynchronous SGD} (the focus of our work), workers proceed
independently; in contrast, in synchronous SGD the parameter server
waits for updates from all the workers before sending an updated
model, introducing blocking at the workers.

\begin{figure}[b!]
  \vspace{-1em}
  \centering
  \begin{subfigure}[b]{.6\columnwidth}
    \centering
    \includegraphics[width=\linewidth]{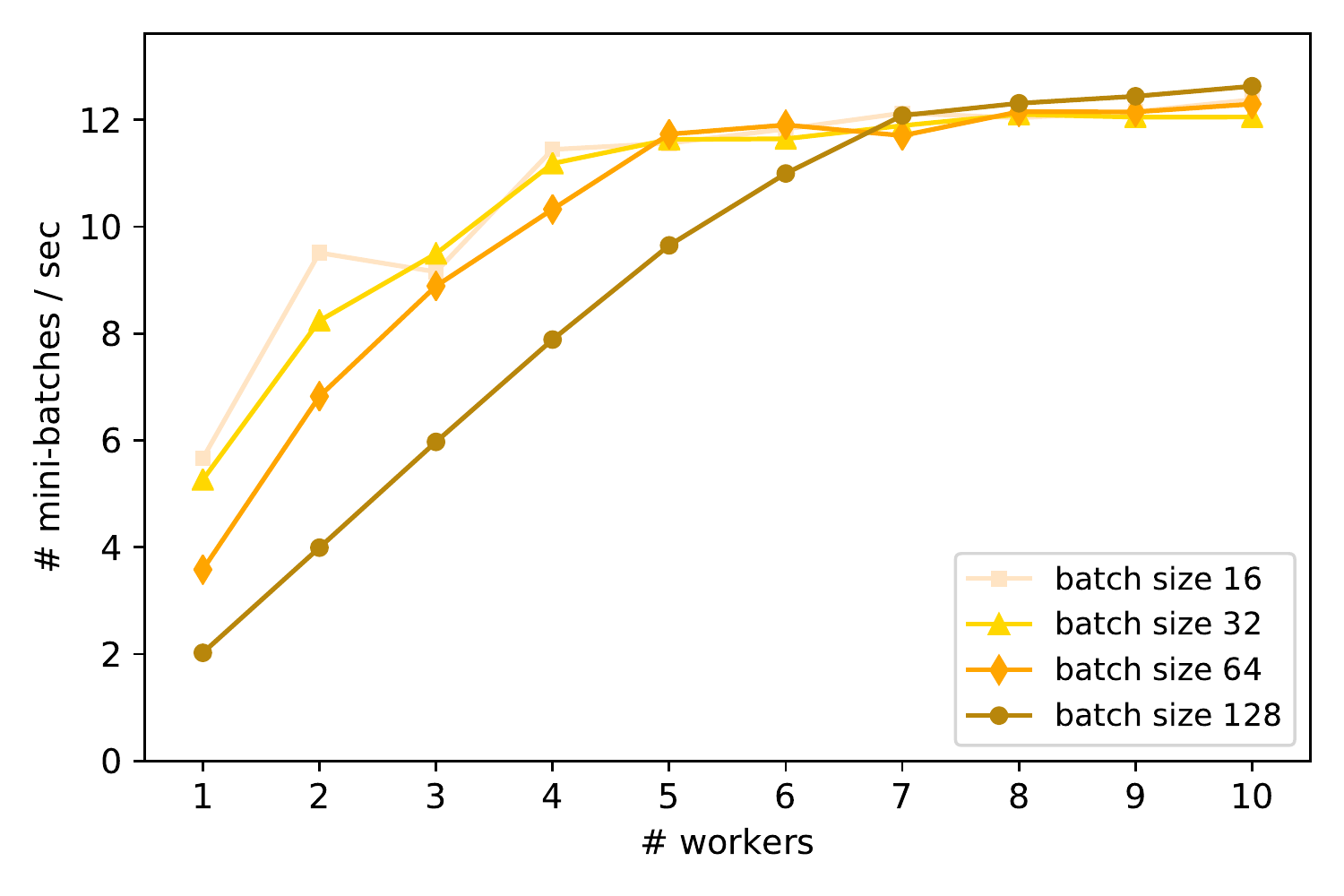}
  \end{subfigure}%
  \vspace{-1em}
  \caption{Training throughput of Inception-v3 on AWS
    \texttt{p3.2xlarge} GPU instances for different batch
    sizes}\label{fig:training_different_batch_sizes}\vspace{-1em}
\end{figure}

As the number of worker nodes increases, network traffic at the
parameter server also increases, resulting in sublinear scaling of
\emph{training throughput} (examples/s processed by all the workers).
For example, \cref{fig:training_different_batch_sizes} illustrates
the training throughput measured for the Inception-v3
model~\cite{googlenet} when training on AWS \texttt{p3.2xlarge}
instances (each equipped with NVIDIA V100 GPU) with TensorFlow and
asynchronous SGD, for batch sizes of 16, 32, 64, 128.
Throughput saturates at 4 workers for batch sizes 16 and 32; adding
more workers yields only marginal improvements. In contrast, for batch
sizes 64 and 128 throughput saturates at 5 and 7 workers,
respectively: in this case, workers access the network less frequently
(it takes longer to compute a model update), reducing network load.

The goal of our work is to provide an approach to predicting training
throughput of asynchronous SGD for any number of workers~$W$, from
quick job profiling performed in TensorFlow using a single worker
node.
This would allow users to avoid testing multiple configurations and
manually checking throughput; cost savings with respect to manual
benchmarks are particularly important in large cloud environments with
GPU nodes, where users submit multiple jobs and schedulers need to
decide how many nodes to assign to each job based on its size and
ability to scale.

\begin{figure}[tb!]
  \centering
  \begin{subfigure}[tb]{\columnwidth}
    \centering
    \includegraphics[width=\linewidth]{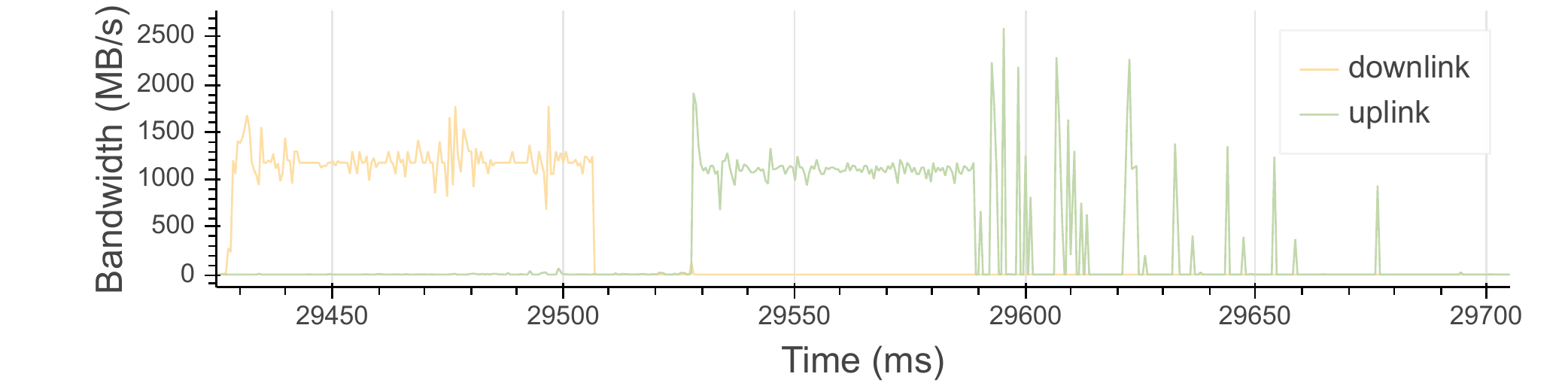}
    \caption{Downlink/uplink transmission bandwidth (from \texttt{tcpdump})\label{fig:inceptionv3_bandwidth_trace}}\vspace{.5em}
  \end{subfigure}
  \vspace{.5em}

  \begin{subfigure}[tb]{\columnwidth}
    \centering
    \includegraphics[width=\linewidth]{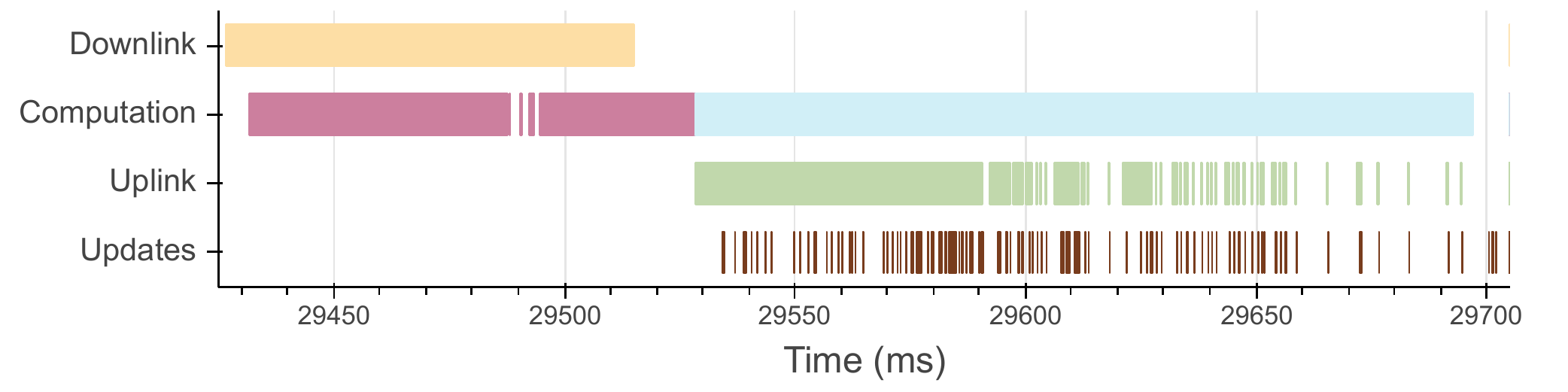}
    \caption{Summary of profiling data captured in TensorFlow\label{fig:inceptionv3_profiling_trace}}
  \end{subfigure}
  \vspace{.5em}

  \begin{subfigure}[tb]{\columnwidth}
    \centering
    \includegraphics[width=.7\linewidth]{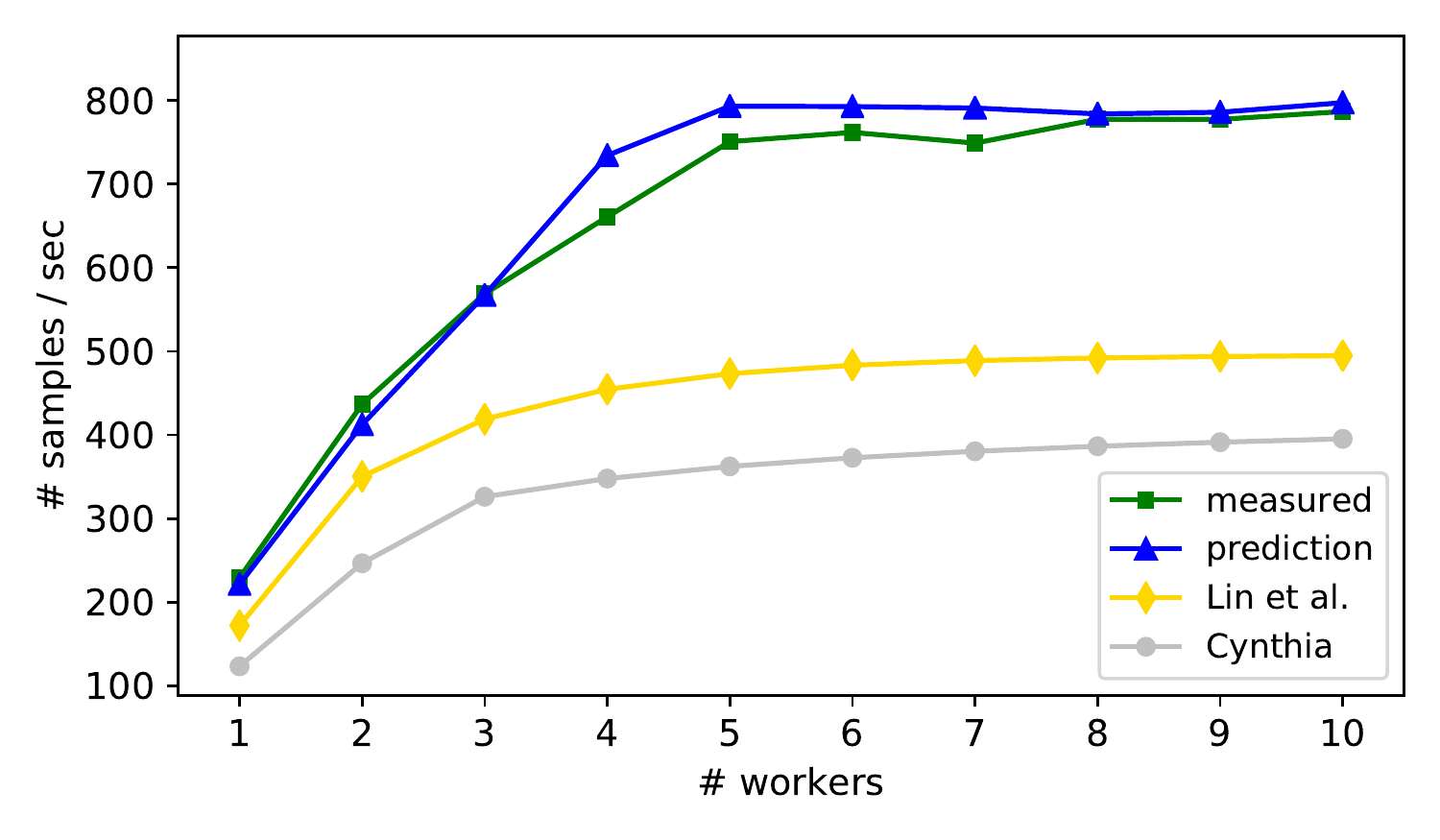}
    \vspace{-.2em}
    \caption{Throughput prediction with different methods\label{fig:prediction_comparison}}\vspace{-.5em}
  \end{subfigure}%
  \caption{Analysis of an SGD step (batch of 64 examples) of
    Inception-v3 training using 1 worker and 1 parameter server in
    TensorFlow (on AWS \texttt{p3.2xlarge}) and prediction results}
    \vspace{-2em}
\end{figure}

Existing approaches for performance prediction of asynchronous SGD are
based on very coarse models of computation and communication, not
accounting for dependencies and overlaps of fine-grained operations.
For example, previous work~\cite{lin2018model} infers the duration of
each SGD phase from profiling information collected using network
analysis tools such as \texttt{tcpdump}~\cite{tcpdump}: the time
interval between the end of the downlink transmission and the start of
the uplink transmission is interpreted as the single worker's
computation, and the durations of these phases are used as parameters
of a queueing model to predict throughput.
An even coarser model, proposed in \cite{zheng2019cynthia},
estimates throughput with $W$ workers and batch size $K$ from the
network utilization~$U_1$ measured for a single worker as
$WK/(T_P \max(1, W U_1) + 2T_C)$, where $T_P$ is the time required to
process a batch and $T_C$ is the model/updates transmission time.

\begin{figure}[tb]
  \centering
  \begin{subfigure}[b]{\columnwidth}
    \centering
    \includegraphics[width=\linewidth]{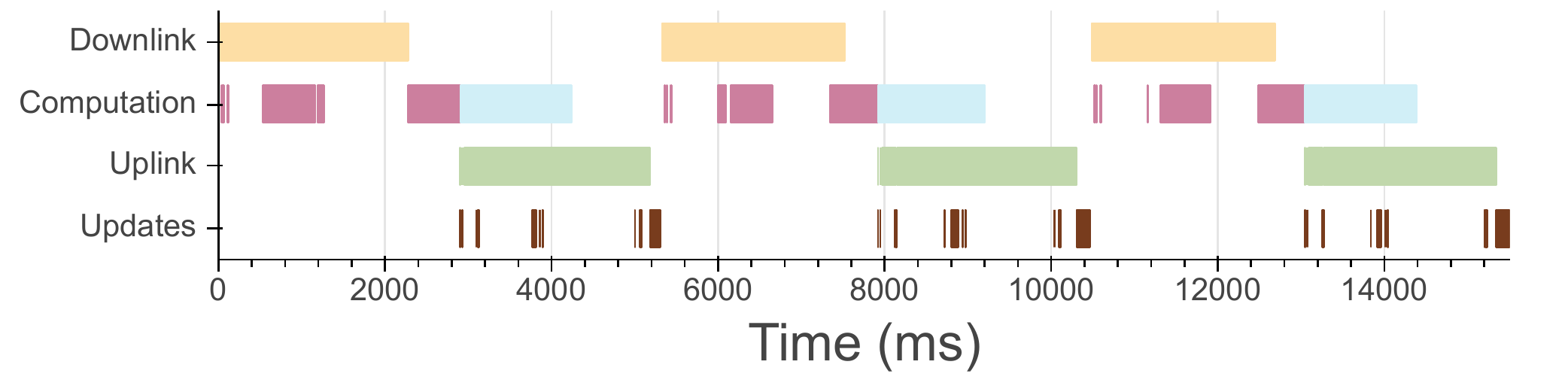}
    \caption{Batch size = 4}
  \end{subfigure}%%
  \hfill
  \begin{subfigure}[b]{\columnwidth}
    \centering
    \includegraphics[width=\linewidth]{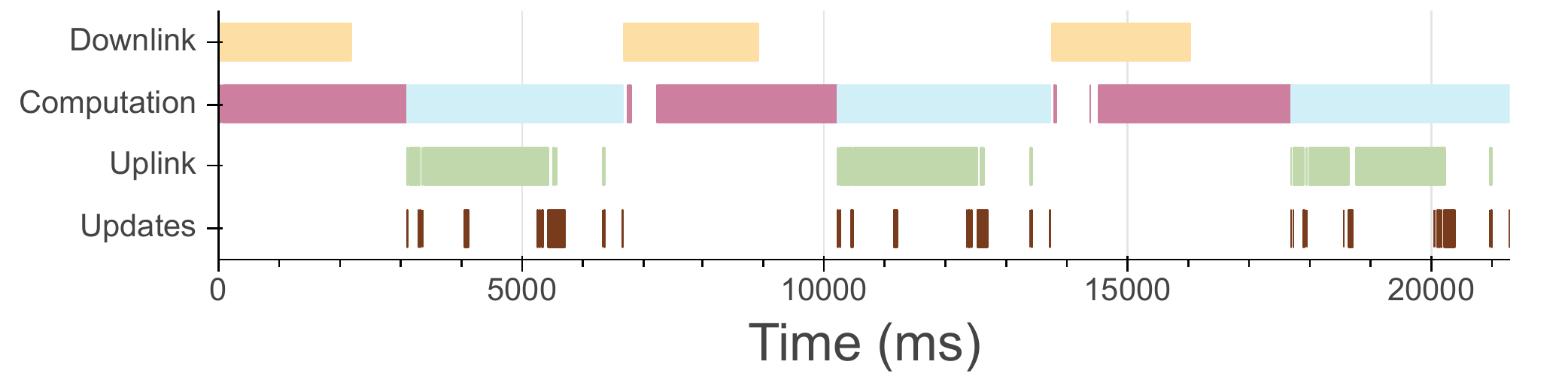}
    \caption{Batch size = 16}
  \end{subfigure}%
  \vspace{-1em}
  \caption{Summary of TensorFlow profiling data for 3~training steps
    (1~worker/1~server) on a private CPU-only
    cluster}\label{fig:alexnet_profiling_traces}
  \vspace{-1.5em}
\end{figure}

\vspace{2mm}\noindent
\textsf{\textit{Overview}}
We argue that these models overlook the complexity of computation and
communication in asynchronous SGD. As illustrated in
\cref{fig:inceptionv3_bandwidth_trace} for Inception-v3, the uplink
phase (green) is spread out over a long time interval.
By looking at the trace information collected with TensorFlow
(\cref{fig:inceptionv3_profiling_trace}), we observe that computation
overlaps with both uplink and downlink communication: the first part of the
computation (forward propagation, in red, computing the output error)
overlaps with the downlink, while the second part (backward
propagation, in cyan, propagating the error back to model parameters)
overlaps with the uplink.
As illustrated in \cref{fig:alexnet_profiling_traces}, these
overlaps are not limited to a specific DNN model, batch size, or
platform. In fact, TensorFlow starts each operation in an SGD step as
soon as its dependencies are satisfied: forward propagation at the
worker can start as soon as the initial layers are received;
similarly, as soon as backward propagation completes for one of the
final layers, its uplink transmission can start.

\begin{figure*}[htb]
  \begin{minipage}[b]{0.31\linewidth}
    \centering
    \includegraphics[width=\linewidth]{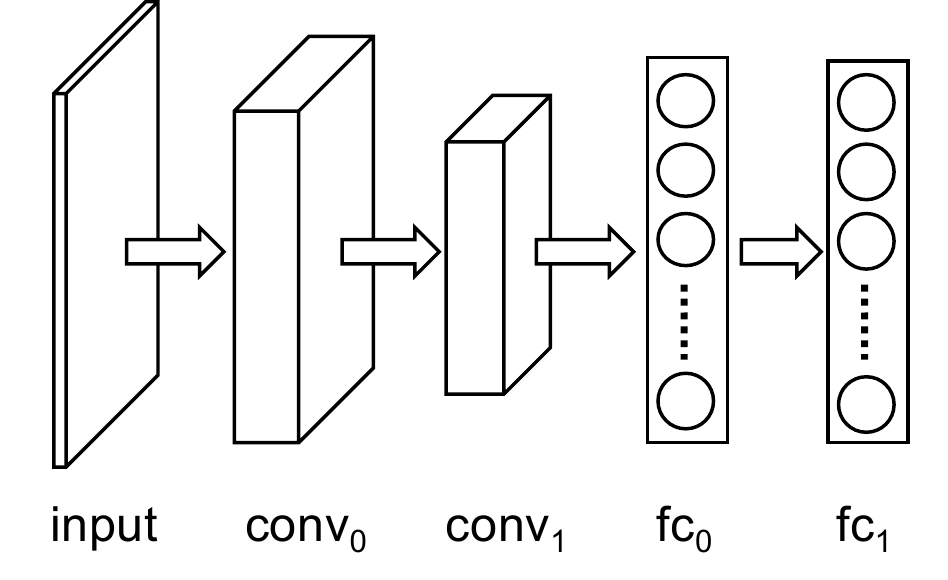}\vspace{-1.5em}
    \hfill
    \caption{A simple DNN model}\label{fig:simple_cnn}
  \end{minipage}
\hfill
\begin{minipage}[b]{0.65\linewidth}
  \includegraphics[width=\linewidth]{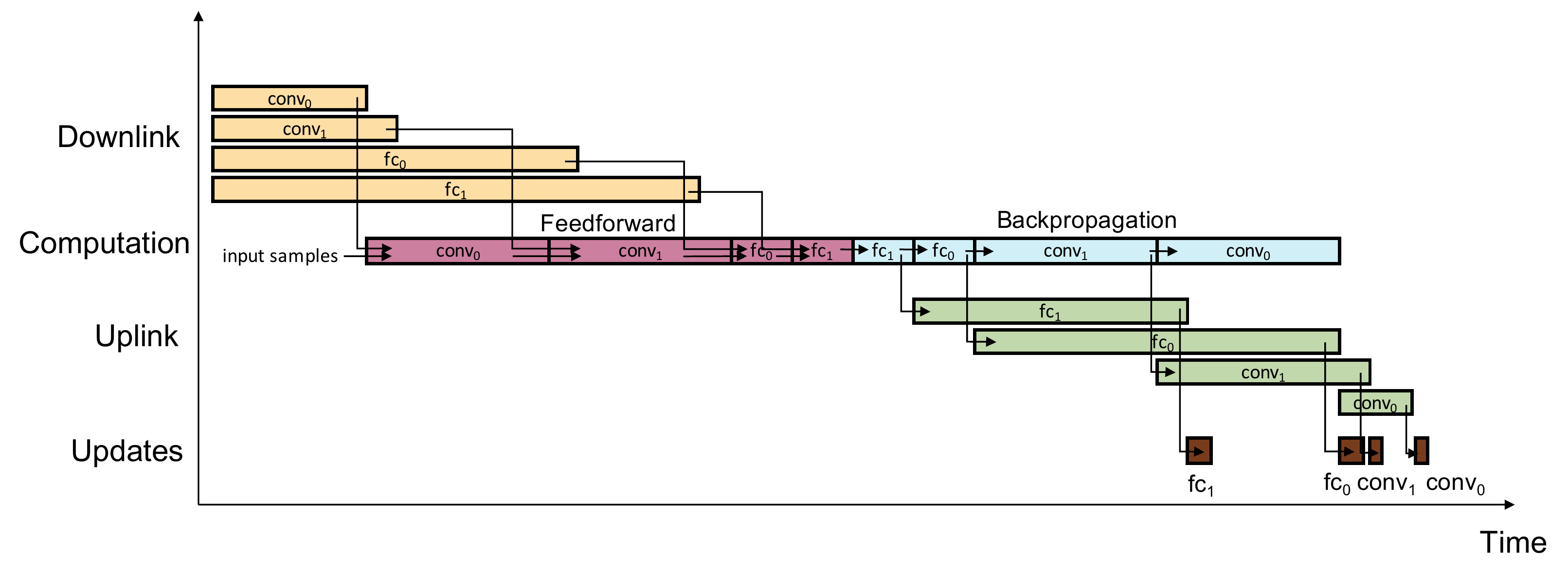}\vspace{-1.5em}
  \caption{Training steps profile: four phases of a training step}\label{fig:four_phases_overview}
\end{minipage}%
\end{figure*}

\begin{figure}[htb]
  \centering
  \begin{subfigure}[b]{.65\columnwidth}
    \centering
    \includegraphics[width=\linewidth]{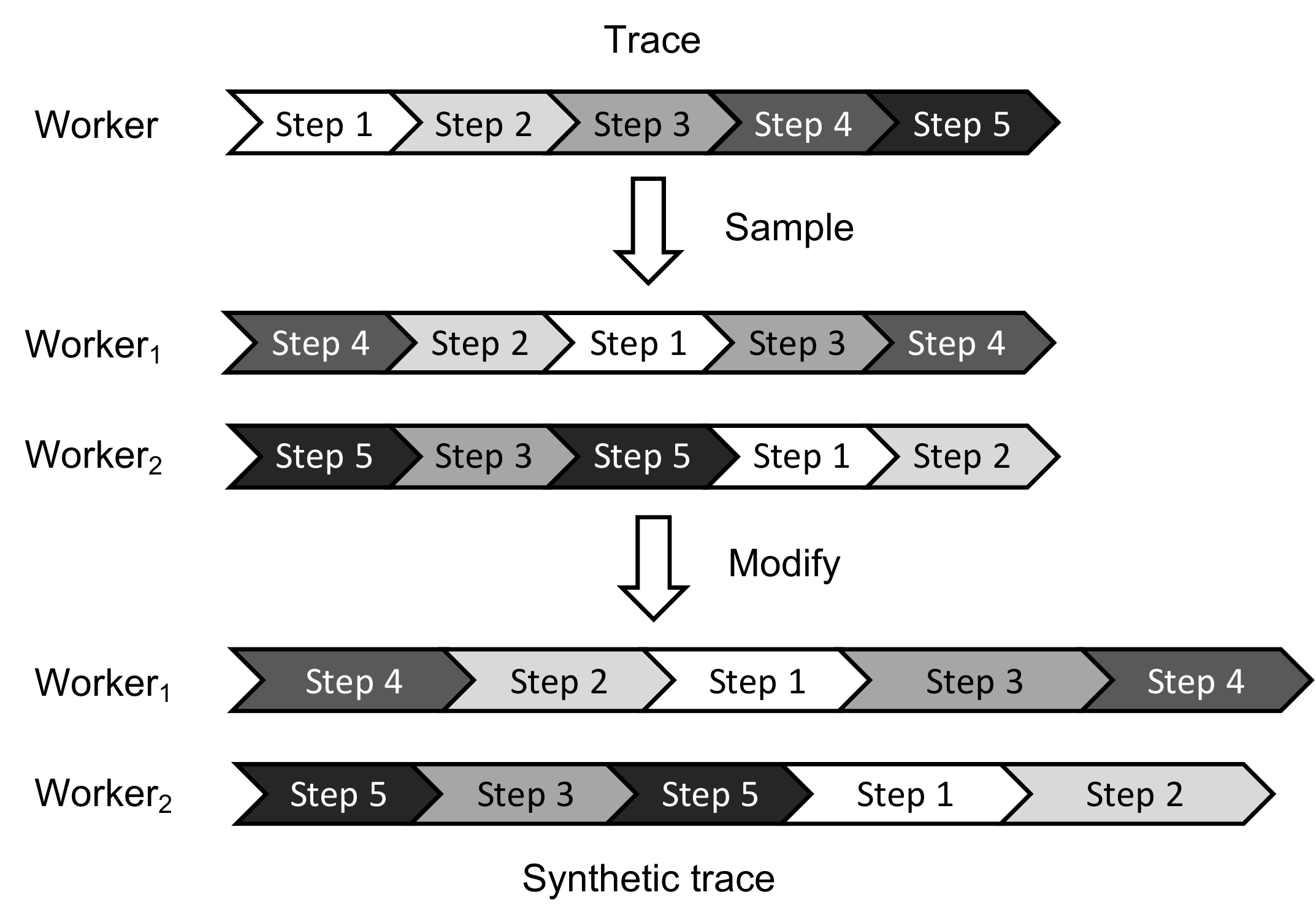}
  \end{subfigure}%
  \vspace{-.5em}
  \caption{Synthetic trace generated for 2 workers sampling and modifying steps from 1-worker profiling}\label{fig:synthetic_trace}
  \vspace{-1em}
\end{figure}

Our proposed approach collects fine-grained profiling information
using TensorFlow traces, recording dependencies of each operation in a
training step. For example, for the simple 4-layer DNN model of
\cref{fig:simple_cnn}, we collect 1-worker profiling such as in
\cref{fig:four_phases_overview} (but real-world DNNs include thousands of operations): each DNN layer results in multiple transmissions
(uplink/downlink) and computations (forward/backward propagation at
the worker, updates at the server).
Specifically,
we collect 1-worker profiling steps (e.g., 100 steps) and then sample
from them with replacement to generate synthetic traces for multiple
workers~(\cref{fig:synthetic_trace}).
To obtain accurate throughput estimates from synthetic traces, we
need to account for network sharing between workers and also for
limitations of recorded traces, which track only the start/end of
operations but not their active transmission intervals.
\cref{fig:prediction_comparison} compares our prediction results for
Inception-v3 with existing methods, clearly indicating the need for
a more fine-grained approach (like ours) rather than the coarse-grained approach in the current literature.

%\vspace{2mm}\noindent
%\textsf{\textit{Contributions}}
%
The \emph{contributions} of our work are as follows.
\begin{itemize}
\item We propose an approach for throughput prediction of asynchronous
  SGD based on fine-grained tracing information collected from
  1-worker profiling.
  We account for the type and dependencies of each operation in a
  discrete-event simulation with multiple workers, which allows us to
  predict delays caused by network congestion with great accuracy.
  To enable this approach, we overcome many limitations of
  TensorFlow trace profiling data, discussed in \cref{profile}.
  In particular, we provide a model for communication overhead due to
  message parsing (\cref{sec:overhead}) and for HTTP/2 multiplexing of
  multiple streams (\cref{simulating_communication}) in TensorFlow.

\item We include these models in a simulation algorithm
  (\cref{simulation}) for throughput prediction and we validate our
  approach using a large set of experiments for multiple DNN models, with
  many batch sizes, on private clusters and public cloud platforms,
  using CPU or GPU resources for each node, for different network
  speeds.
  The results highlight that our approach can accurately predict
  throughput and bottleneck points (\cref{results}).

\item We investigate the performance effects of state-of-the-art
  optimizations on the communication strategy of TensorFlow
  (\cref{communication_order_optimization}) and show that our approach
  can be modified to account for such changes, accurately predicting
  throughput in these settings (\cref{results}).

\item We investigate a model of bandwidth sharing in configurations
  with 2 parameter servers and evaluate the accuracy of our approach
  (\cref{sec:multiple_ps}); extension to a larger number of parameter
  servers is part of our ongoing efforts.

\end{itemize}

% -*- ispell-local-dictionary: "american"; TeX-master: "../tensorpredict.tex"; -*-

\begin{figure*}
\begin{minipage}[b]{0.69\linewidth}
    \begin{subfigure}[t]{.5\linewidth}
        \centering
        \includegraphics[width=\linewidth]{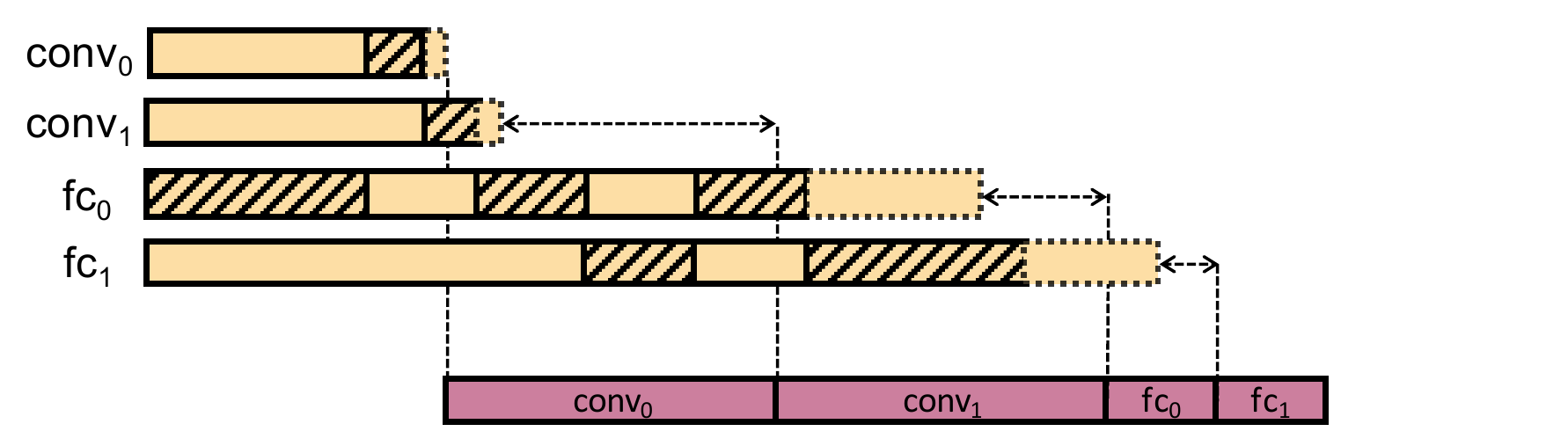}
        \caption{default}\label{fig:default_tensorflow}
    \end{subfigure}%
    \hfill
    \begin{subfigure}[t]{.5\linewidth}
        \centering
        \includegraphics[width=\linewidth]{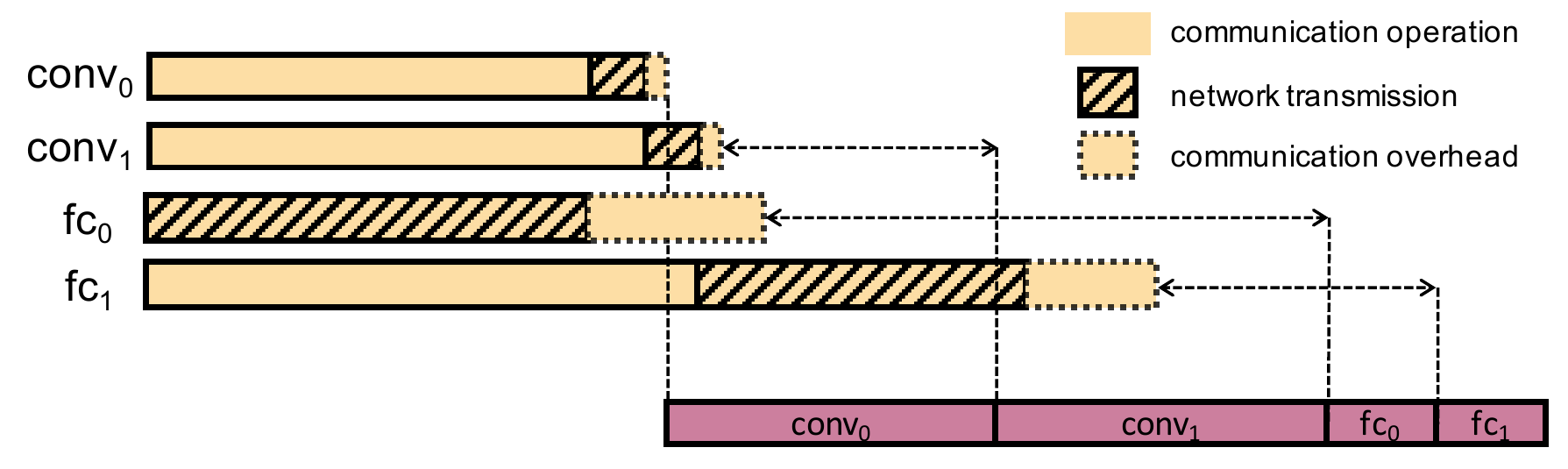}
        \caption{HTTP/2 flow control disabled}\label{fig:no_flow_control}
    \end{subfigure}%
    \hfill
    \begin{subfigure}[t]{.5\linewidth}
        \centering
        \includegraphics[width=\linewidth]{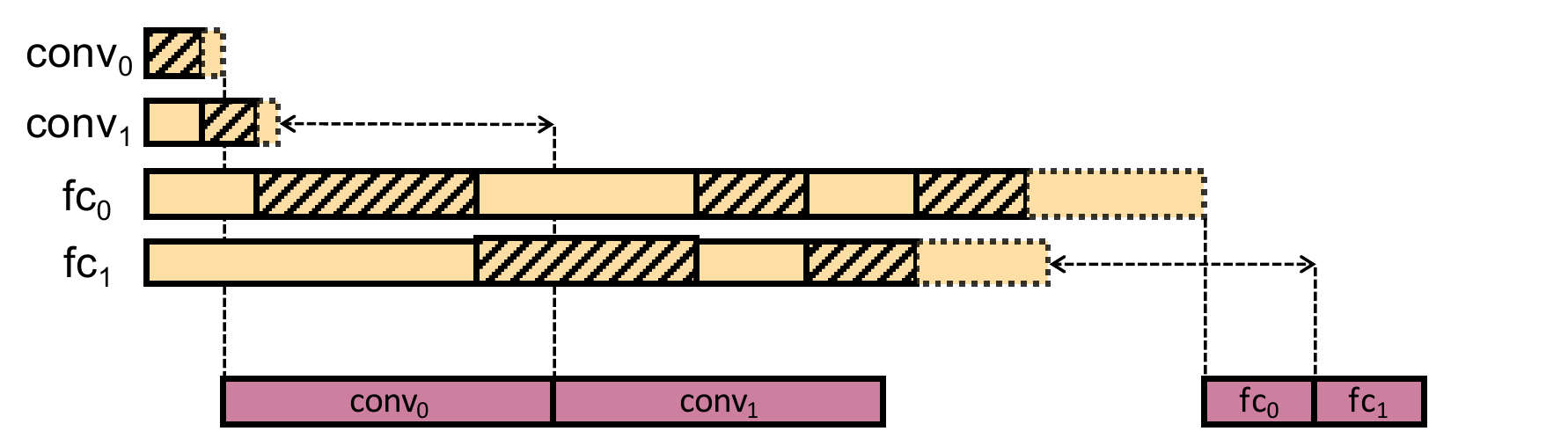}
        \caption{communication ordering enforcement}\label{fig:tictac}
    \end{subfigure}%
    \hfill
    \begin{subfigure}[t]{.5\linewidth}
        \centering
        \includegraphics[width=\linewidth]{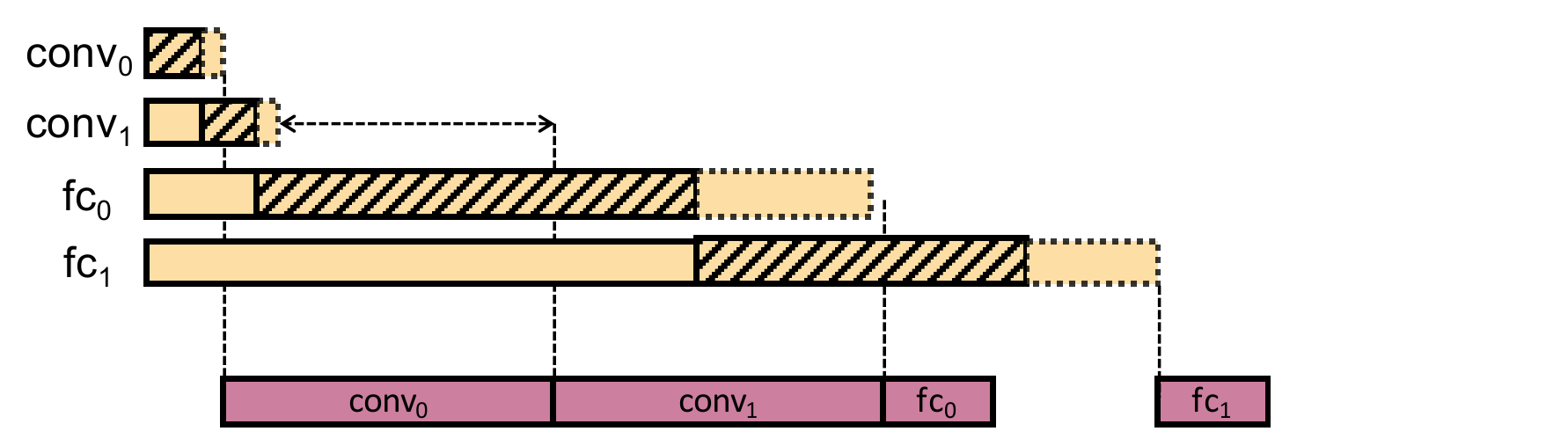}
        \caption{communication ordering enforcement,\\
        HTTP/2 flow control disabled}\label{fig:tictac_no_flow_control}
    \end{subfigure}%
    \caption{Different communication mechanisms}
    \label{fig:timeline}
\end{minipage}%
\hfill
\begin{minipage}[b]{0.28\linewidth}
  \centering
  \includegraphics[width=\linewidth]{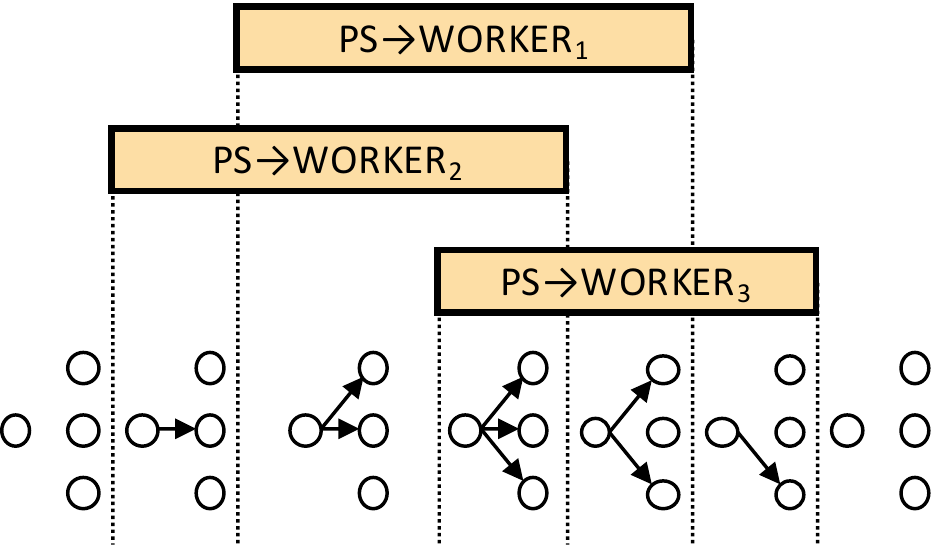}
  \caption{Active connections and bandwidth allocation change over time as transmissions start/end}
  \label{fig:downlink_many_states}
\end{minipage}%
\end{figure*}

\section{Profiling} \label{profile}

Dependencies among operations of DNN training (adjusting weights to
fit a dataset of input/output pairs) or inference (computing output
classifications for new inputs) are represented in TensorFlow as a
\emph{computation graph}: different operations (e.g., matrix
multiplications and activation functions in a layer) are nodes of this
graph, while tensors (multidimensional arrays) flow along directed
edges between nodes (i.e., the output of one operation is the input of
the next).  Execution is triggered by feeding data (e.g., an
input image) into the input nodes of the graph.

\cref{fig:four_phases_overview} illustrates the computation graph of a
step of \emph{distributed} training for the simple 4-layer DNN model
of \cref{fig:simple_cnn}.  There are two types of operations:
(1)~\textit{computation operations} (e.g., addition, matrix
multiplication, convolution) producing an output tensor from one or
more input tensors (in red, cyan, and brown); and
(2)~\textit{communication operations}, transferring data between nodes
(in orange and green).
During \emph{forward propagation}, a worker can compute the output of
each layer in order, after receiving its weights from the parameter
server; during \emph{backward propagation}, updates to each layer are
computed in reverse order and immediately scheduled for transmission
to the parameter server.
Note that dependencies are much more complex than shown in
\cref{fig:four_phases_overview}: the computation of a single layer
breaks down into many basic operations that are scheduled for
execution on the CPU and GPU.

We generate profiling information for a training job (a specific DNN
processing batches of training examples of a given size) by running
distributed TensorFlow for a few SGD steps using one parameter server
and one worker.
Each step can be described as a set of operations with mutual
dependencies: in the following, for each operation $op$ we denote by
\textit{op.waiting\_for} and \textit{op.dependent\_ops} the set of
operations that $op$ depends on, and the set of operations that can
start only after the completion of $op$, respectively.  Note that the
dependencies are the same for all training steps since training steps
are generated from the same DNN model.

Each operation $op$ in the collected profiling traces uses exactly one
resource,
$op.res \in \{\cn{downlink}, \cn{worker}, \cn{uplink}, \cn{ps}\}$
where: \cn{downlink} models the transmission channel of the parameter
server (used by the workers to receive up-to-date model parameters),
\cn{worker} models the computation unit at the worker (CPU cores or
GPU used for SGD), \cn{uplink} models the receiving channel of the
parameter server (used by the workers to transmit model updates), and
\cn{ps} models the computation unit at the parameter server
(used to apply model updates).
For communication operations, the size $op.size$ of the transmitted
tensor is also recorded.

Unfortunately, communication operations recorded by TensorFlow
profiling tools do not accurately represent the exact timings for the
transmission of the corresponding parameters. In fact, TensorFlow uses
gRPC \cite{grpc}, a framework for remote procedure calls (RPC) over
HTTP/2, to transfer tensors and manage connections.
Tensor transfers are triggered at the
beginning of each training step: first, TensorFlow finds all
tensors that need to be transferred to another device; then, it starts
a communication operation for each such tensor through gRPC.

\cref{fig:default_tensorflow} shows that, when training the DNN model
from \cref{fig:simple_cnn}, each layer triggers a communication
operation. For each transfer, TensorFlow creates a corresponding RPC
request: the recorded start time of the communication operation
corresponds to when the tensor is ready on the sender side, while the
recorded end time tracks the time when the data is available to the
receiver.
In fact, these start/end times do not correspond to the underlying
network transmissions:
(1)~Transmission can start after the recorded start time, since gRPC
API calls are asynchronous and start times only indicate when
transmissions are \emph{requested} by the sender; for example,
\cref{fig:default_tensorflow} shows that, at the beginning of each
step, all tensors are available to be transmitted from parameter
server to the workers, so that the profiler records the beginning of
their transmission at the same time, although only one starts
transmitting data.
(2)~The duration of each recorded transmission does not necessarily
represent transmission time; not only can transmissions start well
after their recorded start time, but they can be performed in parallel
or suspended, since each gRPC transfer is assigned to a different
HTTP/2 stream subject to multiplexing.
(3)~In addition, the recorded end time is also increased due to the
latency introduced by parsing operations performed after the data has
been transferred to the receiver (including deserialization and memory
copies).

In order to perform accurate predictions, we need to infer the real
start/end times of network transmissions (i.e., which communication
operation is being served at each time) based on the limited
information provided by TensorFlow profiling.
To account for~(1) and~(2), we propose a model of HTTP/2 multiplexing
in gRPC.
To account for~(3), we propose a linear model of communication
overhead.
We present and evaluate these models in \cref{sec:reconstructing}.

% -*- ispell-local-dictionary: "american"; TeX-master: "../tensorpredict.tex"; -*-

\section{Prediction} \label{prediction}

From profiling information collected in a single-worker configuration,
we extract detailed information on the communication and computation
operations of each SGD step. In this section, we use profiling
information to construct synthetic traces for multiple SGD steps in a
configuration with an arbitrary number of workers~$W$.
To do so, we perform a discrete-event simulation of the operations at
each worker, accounting for the reduction in bandwidth due to the
presence of multiple workers transmitting or receiving data. In turn,
extended communication times at a worker can delay dependent
operations in an SGD step.
First, we address bandwidth sharing between multiple workers; then, we
analyze the effects of HTTP/2 multiplexing of concurrent transmissions
at each worker.

\subsection{Bandwidth Sharing among Workers} \label{share_of_bandwidth}

During our profiling phase, only a single worker communicates with the
parameter server: in this case, the uplink/downlink operations of the
worker can use the entire network bandwidth in each direction.
In contrast, in a distributed SGD configuration with multiple workers
networking resources are shared.

In order to adapt networking of single-worker profiling traces for
multiple-workers prediction, we need to track the number of workers
\emph{currently active} in the uplink/downlink direction.
In fact, as illustrated by \cref{fig:inceptionv3_profiling_trace},
communication with the parameter server is intermittent: the worker
sends updates for each layer/tensor of the model as soon as they are
ready.
With multiple workers, many network states (active/inactive links) are
possible (\cref{fig:downlink_many_states}).
We keep track of the number of workers~$n$ active in each direction
(uplink/downlink) and assume that each active worker receives a
fraction $\frac{1}{n}$ of the available bandwidth. Our model assumes
that network capacity is shared equally, without significant
background traffic and with similar round-trip times (RTTs).
Although in practice workers may split network bandwidth unevenly
because of background traffic and heterogeneous RTTs, we find this
model to be accurate for throughput prediction (\cref{results}).

\begin{figure}[tb!]
\begin{minipage}[b]{\linewidth}
  \centering
    \includegraphics[width=.5\linewidth]{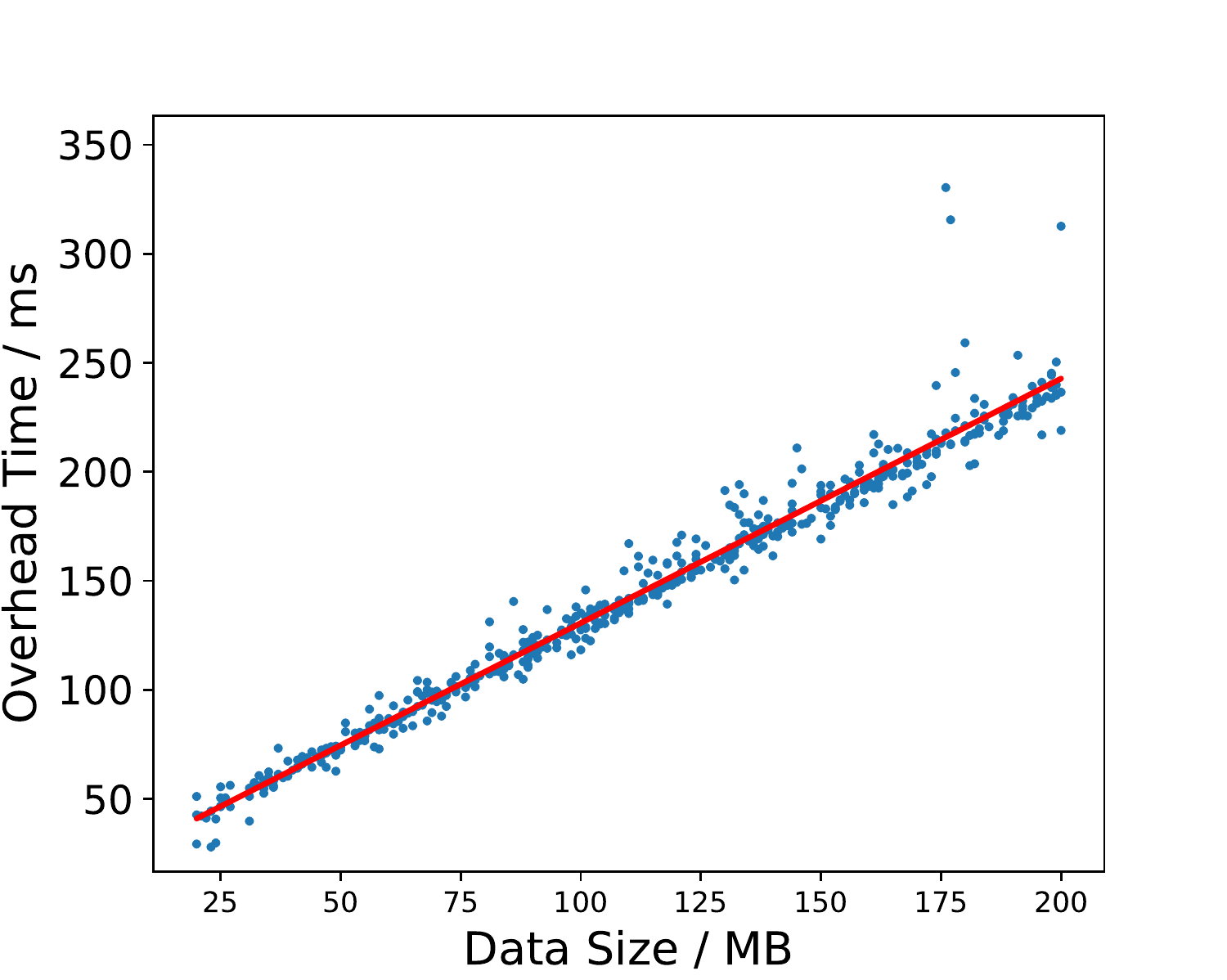}
    \vspace{-.6em}
    \caption{Overhead for each transmitted tensor size}\label{fig:overhead}
    \vspace{-1.5em}
\end{minipage}%
\end{figure}

\begin{figure*}
\begin{minipage}[b]{.38\linewidth}
        \centering
        \includegraphics[width=\linewidth]{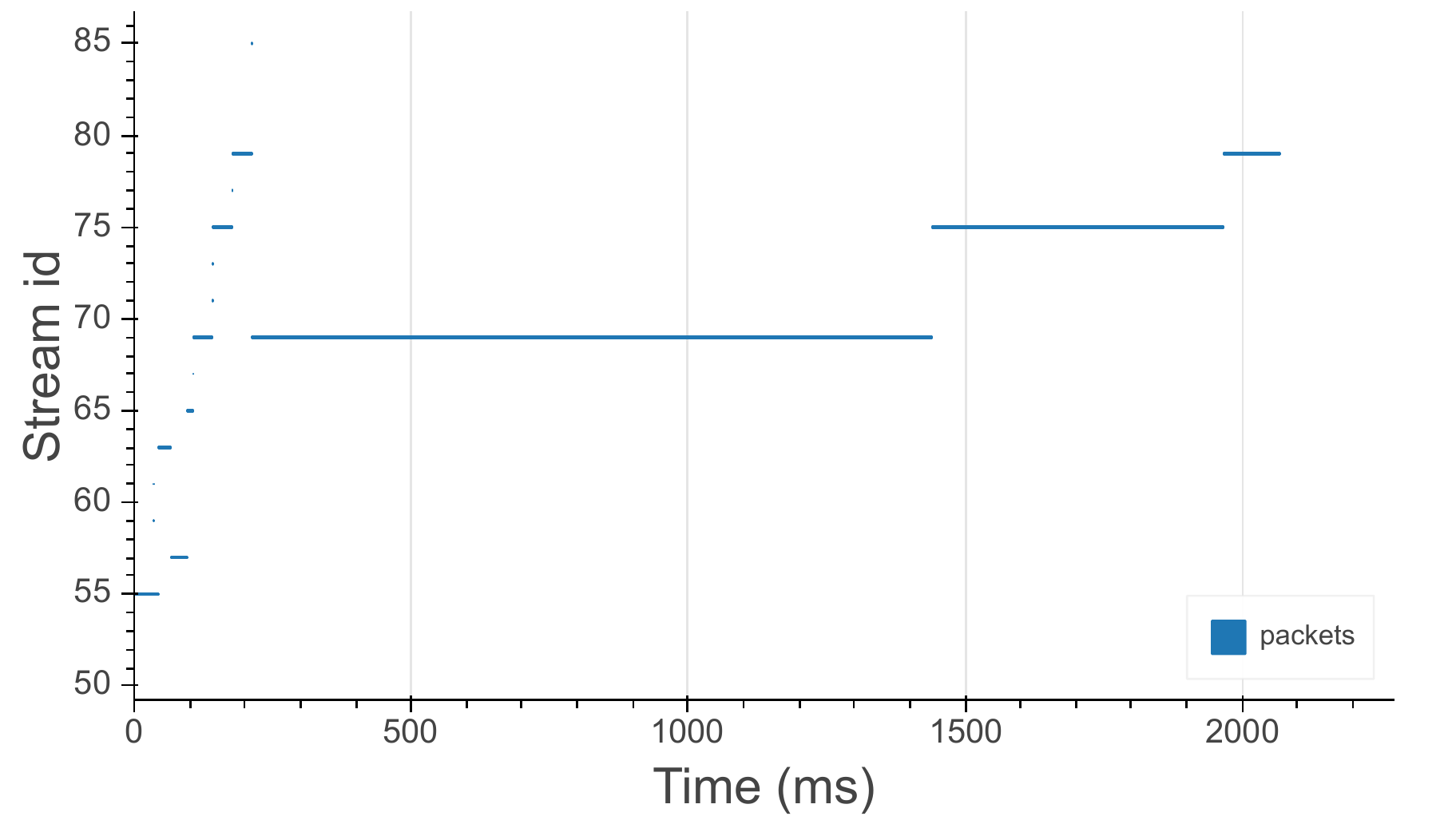}
        \caption{Multiplexing behavior of HTTP/2 streams during the
          downlink phase for AlexNet using a single
          worker}\label{fig:multiplex}
\end{minipage}%
\hfill
\begin{minipage}[b]{.59\linewidth}
        \centering
        \includegraphics[width=\linewidth]{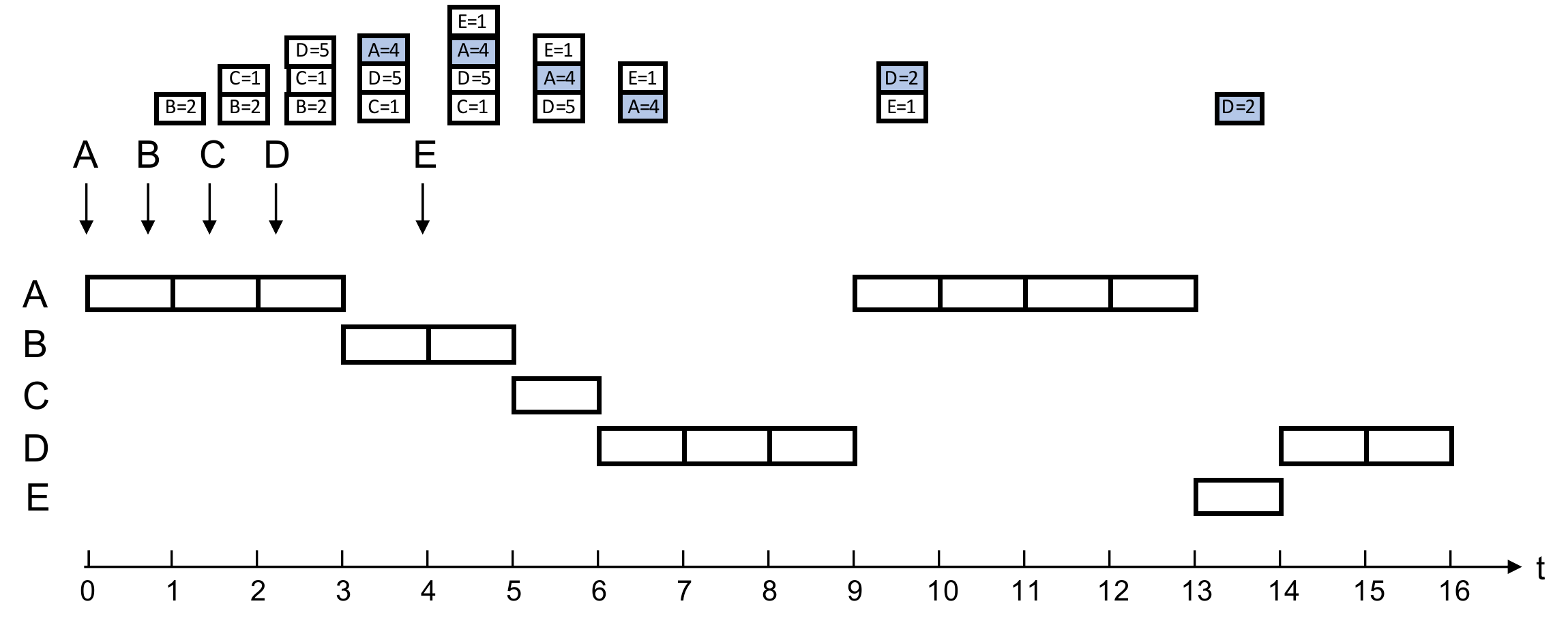}
        \caption{Multiplexing model of streams $A,B,C,D,E$ with
          $WIN = 3$. Streams waiting to be transmitted (and remaining
          sizes) are represented on top; each is allowed to transmit
          up to \textit{WIN} during the first service; streams waiting
          for their second service (to completion) are marked in
          blue.}\label{fig:stream_multiplexing_heuristic}
\end{minipage}%
\end{figure*}

\subsection{Analysis of Single-Worker Profiling Traces}
\label{sec:reconstructing}

As noted in \cref{profile}, communication operations recorded in
TensorFlow during the profiling phase do not accurately represent the
actual transmission times because HTTP/2 flow control alternates the
transmission of different streams.
This is a major obstacle to predicting communication times in
configurations with multiple workers, since extending recorded
communication operations according to the available bandwidth yields
inaccurate results. We illustrate these challenges and our
proposed solution below.

\subsubsection{Parsing Overhead of Received Tensors}
\label{sec:overhead}

We benchmark the parsing overhead of communication operations for data
transfers of different sizes. The results, illustrated in
\cref{fig:overhead}, suggest a linear model
$op.overhead = \alpha \times op.size + \beta$ with respect to the size
of the data transferred by the operation $op$.
The parameters $\alpha$ and $\beta$ are independent of the specific DNN
model, and they can be estimated once for the nodes used in the
cluster.
We remove this parsing overhead from the duration of a communication
operation and assign it to a dependent computation operation.

\subsubsection{Downlink and Uplink Multiplexing}
\label{simulating_communication}

HTTP/2 achieves better performance than HTTP/1.1 (especially for web
browsers) due to the introduction of multiplexing, so that multiple
streams (e.g., images of a web page) can be transmitted simultaneously
within a single connection between client and server without
``head-of-line blocking'' due to large files being requested before
smaller ones.

Stream multiplexing is the mechanism used in HTTP/2 for flow
control. The receiver side of every stream advertises a flow control
window \textit{WIN}, which is a credit-based value that specifies the
amount of data it is prepared to receive, in order to prevent the
stream from overwhelming the receiver and blocking other
streams. HTTP/2 defines only the format and semantics of the flow
control window, while implementations are free to decide how it should
adapt over time to current network and memory conditions (usually
based on the bandwidth-delay product and memory pressure), and how to
switch between multiple streams.

In gRPC, the remote procedure call (RPC) library used by TensorFlow
for distributed SGD training, there are two connections (one in each
direction) between each parameter server and worker. TensorFlow
creates a gRPC request, which initiates an HTTP/2 stream for each
tensor that needs to be transferred to a different node. To illustrate
the stream multiplexing behavior of gRPC, we perform an experiment
that transmits concurrent HTTP/2 streams in TensorFlow: we perform a
training step of the model AlexNet~\cite{krizhevsky2012imagenet} with
one parameter server and one worker, capture the packets of downlink
transmission using \texttt{tcpdump} \cite{tcpdump} and analyze HTTP/2
frames using Wireshark \cite{wireshark}.

The results, presented in \cref{fig:multiplex}, illustrate that (in a
single worker scenario) HTTP/2 switches between gRPC streams
intermittently.  We observe that streams smaller than the flow control
window \textit{WIN} finish without switching, while for streams larger
than \textit{WIN}, HTTP/2 transmits \textit{WIN} bytes and then
switches to another stream. Furthermore, stream preemption happens
only once for each stream: when a stream is selected again for
transmission, it will transmit to completion, even if its remaining
size is larger than \textit{WIN}.

We adopt the following model for the multiplexing mechanism of HTTP/2
in gRPC. We define a \emph{scheduler} for each link (i.e., for each
sender/receiver pair), multiplexing streams of multiple
transmissions. Each stream (which carries data of a communication
operation) is assigned to the scheduler as soon as its corresponding
operation starts.
While the scheduler is not empty, a \emph{chunk} (e.g., the initial
portion of a stream) is selected from one of the active streams for
transmission. The first time that a stream is selected, a chunk of
size up to \textit{WIN} is selected by our scheduler modeling HTTP/2
multiplexing. If the remaining size of the stream is less than the
current \textit{WIN}, or if the stream is selected for the second
time, the entire stream is consumed as a chunk scheduled for
transmission. Once transmission of the stream completes, another
stream is selected by the scheduler in our synthetic trace generation.
An example of this model is illustrated in
\cref{fig:stream_multiplexing_heuristic}.

This model allows us to predict HTTP/2 stream multiplexing, which is
crucial for our trace generation algorithm because of the dependencies
between different operations in an SGD step.
We extract the flow control window \textit{WIN} (on average, 28~MB)
from HTTP/2 headers captured by \texttt{tcpdump}.
When multiple streams are multiplexed within an HTTP/2 connection
between a worker and the parameter server, given the start time of
each stream and \textit{WIN}, the model infers which stream is
transmitted at each time, eventually predicting its end time. We
validate the estimation accuracy of our model on various platforms, by
comparing the end times of downlink streams resulting from our model
with those measured during single-worker profiling. Results are
presented in \cref{tab:stream_multiplexing_estimation} with error
statistics obtained from 100 training steps; most HTTP/2 multiplex
transmissions are modeled correctly and this model works for different
DNNs.

We observe three sources of modeling error:
(1) Our model is based on the assumption that \textit{WIN} does not
change over time; in fact, \textit{WIN} fluctuates as network
conditions change. If a stream is slightly larger than \textit{WIN},
its remaining portion may be transmitted at a much later time; thus,
errors in the estimation of \textit{WIN} can greatly affect end time
predictions.
(2) The estimation of parsing overhead can be inaccurate because our
simple linear model does not account for fluctuations of CPU usage on
the worker.
(3) Another source of error is network instability: transmission times
may be affected by background traffic, especially on cloud platforms.

\begin{table}[tb]
\caption{Error of endtime prediction of downlink streams}
\label{tab:stream_multiplexing_estimation}
\small
\begin{tabular}{clcc}
\toprule
DNN Model&&CPU Cluster&AWS Cloud\\
\midrule
\multirow{4}{*}{AlexNet}
&Average&1.82\%&2.89\%\\
&Median&1.18\%&0.76\%\\
&95th Percentile&3.35\%&9.71\%\\
&Maximum&47.48\%&45.44\%\\
\midrule
\multirow{4}{*}{GoogLeNet}
&Average&1.69\%&3.43\%\\
&Median&1.07\%&2.76\%\\
&95th Percentile&3.74\%&9.14\%\\
&Maximum&75.83\%&64.24\%\\
\midrule
\multirow{4}{*}{Inception-v3}
&Average&1.02\%&9.23\%\\
&Median&0.35\%&8.01\%\\
&95th Percentile&3.92\%&20.98\%\\
&Maximum&87.88\%&98.19\%\\
\midrule
\multirow{4}{*}{ResNet-50}
&Average&1.26\%&4.36\%\\
&Median&0.93\%&3.78\%\\
&95th Percentile&2.32\%&9.70\%\\
&Maximum&72.07\%&97.10\%\\
\bottomrule
\end{tabular}
\vspace{-1em}
\end{table}

\subsubsection{Worker Computation and Model Updates}

We assume that, for each additional worker, computation operations run
on independent resources not shared with other workers. In particular,
processing of a batch of training examples uses local resources at the
worker (CPU cores or a GPU); similarly, model updates run at the
parameter server independently (on separate cores).

\subsection{TensorFlow Networking Optimizations} \label{communication_order_optimization}

We observe that HTTP/2 stream multiplexing may introduce delays in
training steps, reducing DNN training throughput.

One feasible approach to improving training throughput is to maximize
the overlap between computation and communication. Ideally, once a
computation operation is completed (e.g., forward propagation of the
first DNN layer), the next computation should start immediately,
without being delayed by network transfers of required inputs.  For
simple DNN models in which layers are connected sequentially (i.e.,
without skip connections or branching~\cite{inceptionv4}), layers
should be transmitted in order during the downlink phase to reduce
blocking of computation operations; for example, the optimal
transmission order of layers in the model of \cref{fig:simple_cnn} is
$\textsc{conv}_0 \rightarrow \textsc{conv}_1 \rightarrow \textsc{fc}_0
\rightarrow \textsc{fc}_1$ (shown in
\cref{fig:tictac_no_flow_control}). For more complex DNN models, the
research prototype TicTac~\cite{hashemi2018tictac} was proposed as a
heuristic to derive efficient schedules for parameter transfers by
analyzing the \emph{critical path} of the computation; these schedules
achieve performance improvements by enforcing communication ordering
in TensorFlow.

However, this type of optimization cannot be fully implemented because
of the multiplexing features of HTTP/2 and of the inevitable switching
between pending communication operations. \cref{fig:tictac} shows an
example where HTTP/2 can suspend transmission of the current layer,
causing the computation to block.

We find that HTTP/2 stream switching can be eliminated from
distributed TensorFlow training by disabling HTTP/2 flow control in
gRPC, as in \cref{fig:no_flow_control}; in this case, there is no
multiplexing of downlink and uplink transmissions, as also illustrated
in \cref{fig:tictac_no_flow_control}, where the time required for a
training step is reduced with respect to \cref{fig:tictac}.
In \cref{sec:private-cluster}, we evaluate the accuracy of our
predictions when flow control is disabled, under multiple scheduling
policies (including TicTac).
To perform predictions in this setting, we modify the schedulers used
in our synthetic trace generation to process entire streams as a
single chunk (i.e., without interruptions), in the order in which they
are scheduled.

% Note that it can be seen as a special case where the flow control
% window \textit{WIN} is $\infty$.}

\subsection{Trace Generation for Multiple Workers} \label{simulation}

\begin{algorithm}

\begin{pseudo}[fullwidth,indent-length=1.3em]*
\hd{StartRandomStep}(S, Q, w)\\
step = \textsc{SampleWithReplacement}($S$)\\
\kw{for} op \kw{in} step.copy() \quad \ct{each step starts with downlinks}\\+
  \kw{if} op.res == \cn{downlink}\\+
    scheduler[$w$, \cn{downlink}].add(op)\\--
\ct{scheduler splits worker ops, runs $\leq 1$ chunk/worker/resource}\\
$Q$.add(\,scheduler[$w$, \cn{downlink}].remove\_chunk()\,)
\end{pseudo}

\vspace{1em}
\begin{pseudo}[fullwidth,indent-length=1.3em]*
\hd{Share}(r, active)\ct{fraction of r assigned to each worker}\\
    \kw{if} $r$ \kw{in} $\{ \cn{downlink}, \cn{uplink} \}$\\+
     \kw{return} 1/active[$r$] \quad\ct{uplink and downlink shared equally}\\-
    \kw{else} \kw{return} 1 \quad\ct{processing is independent for each worker}
\end{pseudo}

\vspace{1em}

\begin{pseudo}[fullwidth,indent-length=1.3em]*
\hd{GenerateTrace}(S, W)\\
$Q = \emptyset$ \quad\ct{set of scheduled operation chunks}\\
\kw{for} $w$ \kw{in} $W$ \quad\ct{setup for each worker}\\+
  completed\_steps[$w$] = 0\\
  \kw{for} $r$ \kw{in} $\{ \cn{downlink}, \cn{worker}, \cn{uplink}, \cn{ps} \}$\\+
  scheduler[$w$, $r$] = Scheduler($r$) \quad\ct{empty scheduler}\\-
  \textsc{StartRandomStep}($S, Q, w$) \;\ct{add first chunk of downlink}\\-

trace = \textsc{Trace}() \quad \ct{empty trace}\\
active = $\{ \cn{downlink} : |W|, \cn{uplink} : 0 \}$\\

\kw{while} $Q \neq \emptyset$\\+
  \kw{sort} chunks $x \in Q$ \kw{by} $x$.remaining / \textsc{Share}($x$.res, active)\label{line:sort}\\
  chunk = $Q$.remove\_min()\\
  $w$, $r$ = chunk.worker, chunk.res\\
  duration = chunk.remaining / \textsc{Share}($x$.res, active)\label{line:duration}\\
  \ct{$chunk.main\_op$ is the operation that the chunk is part of}\\
  trace.add($w$, $r$, chunk.main\_op, duration)\\
  \kw{for} chunk $x$ \kw{in} $Q$ \quad \ct{update remaining times}\\+
    $x$.remaining -= duration $\times$ \textsc{Share}($x$.res, active)\\-
  \kw{if} chunk.is\_last \quad \ct{this is the last chunk of the operation}\label{line:last_chunk}\\+
    \ct{dependent ops can be assigned to scheduler if ready}\\
    \kw{for} $d$ \kw{in} chunk.main\_op.dependent\_ops\\+
      $d$.waiting\_for.remove(chunk.main\_op)\\
      \kw{if} $d$.waiting\_for == $\emptyset$\quad\ct{no other dependency}\label{line:no_dependency}\\+
        \kw{if} scheduler[$w$, $d$.res] != $\emptyset$\;\,\ct{w already using $d$.res}\\+
          scheduler[$w$, $d$.res].add($d$)\;\,\ct{just queue d}\label{line:add_dependent}\\-
        \kw{else}\quad\ct{start running the first chunk of $d$}\\+
          \kw{if} $d$.res \kw{in} $\{\cn{downlink}, \cn{uplink}\}$\\+
            active[$d$.res] += 1\;\ct{w becomes active}\\-
          scheduler[$w$, $d$.res].add($d$)\\
          $Q$.add(\,scheduler[$w$, $d$.res].remove\_chunk()\,)\label{line:acquire_resource}\\----
  \kw{if} scheduler[$w$, $r$] != $\emptyset$ \quad\ct{w has more chunks to run on r}\\+
    $Q$.add(\,scheduler[$w$, $r$].remove\_chunk()\,)\label{line:take_next_chunk}\\-
  \kw{else}\quad\ct{no more chunks for $w$ to run on resource $r$}\\+
    \kw{if} $r$  \kw{in} $\{\cn{downlink}, \cn{uplink}\}$\\+
      active[$r$] -= 1\quad\ct{become inactive}\\-
    \kw{if} scheduler[$w$, $i$] = $\emptyset \;\forall i$\quad\ct{no more pending chunks}\label{line:no_more}\\+
      completed\_steps[$w$] += 1\quad\ct{step is over}\\
      \kw{if} completed\_steps[$w$] < \cn{n}\\+
        \textsc{StartRandomStep}($S, Q, w$)\label{line:new_step}\\----
\kw{return} trace
\end{pseudo}

\caption{Simulation for synthetic trace generation. {\normalfont Each
    operation $op$ from the profiling traces uses a resource
    \textit{op.res}, has some prerequisite operations
    \textit{op.waiting\_for}, and operations
    \textit{op.dependent\_ops} depend on it.} }
\label{alg:simulation}
\end{algorithm}

We generate a synthetic trace for each system configuration (network
bandwidth $B$, workers $W$, parameter servers $M$) through
discrete-event simulation.

A sequence of $N$ SGD steps is sampled with replacement for each
worker from the set of steps $S$ collected during job profiling (which
is performed only once, with a 1-server/1-worker configuration).
As illustrated in \cref{alg:simulation}, for each worker $w\in W$ and
resource $r\in\{ \cn{downlink}, \cn{worker}, \cn{uplink}, \cn{ps} \}$,
a separate scheduler $scheduler[w,r]$ keeps a queue of pending
operations.
Operations are split into smaller chunks by the scheduler; when a
chunk of worker $w$ is completed with resource $r$, another chunk
is selected by the $scheduler[w,r]$ and added to $Q$
(Line~\ref{line:take_next_chunk}).
This approach allows us to represent the scheduling policies observed
in gRPC when HTTP/2 multiplexing of multiple streams is enabled
(as in \cref{fig:multiplex}): first, a chunk of each stream is selected by the scheduler;
then the remaining data is transmitted until completion.

When the last chunk of an operation is completed
(Line~\ref{line:last_chunk}), dependent operations may become
available for execution (Line~\ref{line:no_dependency}) and are added
to schedulers for their required resources
(Line~\ref{line:add_dependent}). If worker $w$ is not using the
required resource, the first chunk of the operation can be processed
(Line~\ref{line:acquire_resource}).
At any time, $active[\cn{downlink}]$ and $active[\cn{uplink}]$ track
the number of workers using the downlink and uplink resources,
respectively: in our bandwidth sharing model (summarized by
$\textsc{Share}$($r, active$) in \cref{alg:simulation}), each worker
receives a fraction $1/active[r]$ of networking resource
$r \in \{\cn{downlink}, \cn{uplink}\}$ to transmit consecutive chunks
of tensors (in some order defined by the scheduler). In contrast,
computations (forward/backward propagation at the worker and model
update at the server) run on resources (worker CPU/GPU and parameter
server cores) reserved exclusively for each worker (i.e.,
$\textsc{Share}(\cn{worker}, active)=\textsc{Share}(\cn{ps}, active)=1$).
Execution times of running chunks are extended according to the
fraction of resource available to the worker (Lines~\ref{line:sort}
and \ref{line:duration}).
When no more chunks are pending for the operations of a step
(Line~\ref{line:no_more}), a new step is sampled and scheduled on
the worker (Line~\ref{line:new_step}).

The profiled SGD steps $S$ used by the simulation are pre-processed to
remove overhead from recorded transmission times and adjusted for the
network bandwidth $B$ available in the cluster. In particular, each
communication operation is transformed into a new communication
operation (with initial duration determined by $B$) and a computation
operation (the overhead, which depends on the amount of transmitted
data).
Overlaps between communication and computation are modeled by the
simulation algorithm using different resources for parameter server
uplink/downlink and for computation on each node.
The simulation algorithm can be extended to $M$~parameter servers by
introducing distinct resources $\cn{downlink}_i$, $\cn{uplink}_i$,
$\cn{ps}_i$ for each parameters server $i=1,\dots,M$ and using the
model of \cref{sec:multiple_ps} to share $\cn{downlink}_i$ and
$\cn{uplink}_i$ among the workers of the configuration.

We note that the queue $Q$ is sorted at every iteration only for ease
of presentation at Line~\ref{line:sort} of \textsc{GenerateTrace}; our
implementation uses priority queues. Simulation time is proportional
to the number of steps $N$ simulated for each worker; multiple runs
can be performed in parallel on separate cores.

% -*- ispell-local-dictionary: "american"; TeX-master: "../tensorpredict.tex"; -*-

\section{Results} \label{results}

\subsection{Experimental Setup}

All the experiments are performed using TensorFlow 1.13 and the
official TensorFlow
benchmarks\footnote{\url{https://github.com/tensorflow/benchmarks}}
from the v1.13 branch, with slight modifications (less than 5 lines of
code) to turn on trace recording (used to acquire profiling
information). The prediction algorithm is validated on the following
platforms:
(1)~\textit{private CPU cluster}: 8~nodes equipped with quad-core AMD
Opteron Processor 2376 and 16~GB of memory, and connected by Gigabit
Ethernet;
(2)~\textit{cloud CPU cluster}: AWS \texttt{c4.8xlarge} instances (36
vCPUs, 2.9~GHz, Intel Xeon E5-2666v3, 60~GB memory) connected by 10
Gbps networking;
(3)~\textit{cloud GPU cluster}: AWS \texttt{p3.2xlarge} instances (8
vCPUs, 2.7~GHz, Intel Xeon E5-2686v4, 61~GB memory, 1~NVIDIA Tesla
V100 GPU with 32~GB memory), connected by 10~Gbps networking.

To perform throughput prediction, we collect the following
information:
(1)~For each platform, we use \texttt{iperf} to measure
network bandwidth and estimate the parameters $\alpha$ and $\beta$
of our communication overhead model (\cref{sec:overhead}), using the
time difference between TCP packets captured by \texttt{tcpdump} and
the end of communication operations recorded by TensorFlow.
(2)~We profile each training job (which specifies a DNN model and
hyperparameters such as the batch size) for \emph{100 steps with one
  parameter server and one worker} to obtain a trace of operations
within an SGD step (and their dependencies).

For each target configuration of $W$~workers, we run our trace
simulation procedure to generate a synthetic trace and use it to
evaluate training throughput (total number of examples/s processed
by the workers).
In practice, we find that a trace of 1000 steps is sufficient to
obtain a consistent estimate.
Since it requires some time for asynchronous SGD workers to get out of
the initial synchronization (training starts at the same time for all
workers) and generate stable training throughput, we exclude the first
50 simulated steps and compute a time-average over the remaining
steps.
The predicted throughput is compared with the throughput measured in a
real cluster with $W$~workers and $M$~parameter servers, as the
time-average over the last 50 SGD steps out of 100 measured.

\subsection{Private CPU Cluster}
\label{sec:private-cluster}

\begin{figure}[tb]
  \centering
  \begin{subfigure}[b]{.33\linewidth}
    \centering
    \includegraphics[width=\linewidth]{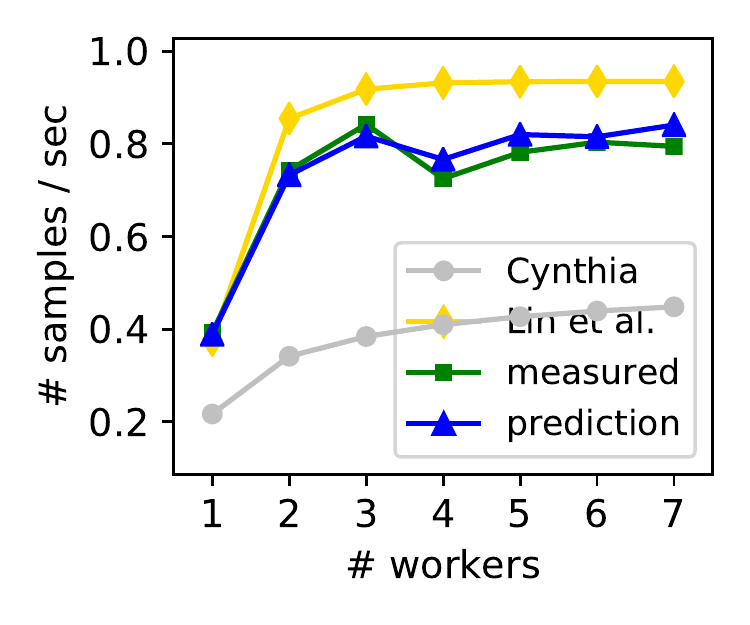}
    \vspace*{-6mm}
    \captionsetup{justification=centering,margin={-3mm,0mm}}
    \caption{AlexNet,\\batch size = 2}\label{fig:cluster_batch_alexnet_2}
  \end{subfigure}%
  \begin{subfigure}[b]{.33\linewidth}
    \centering
    \includegraphics[width=\linewidth]{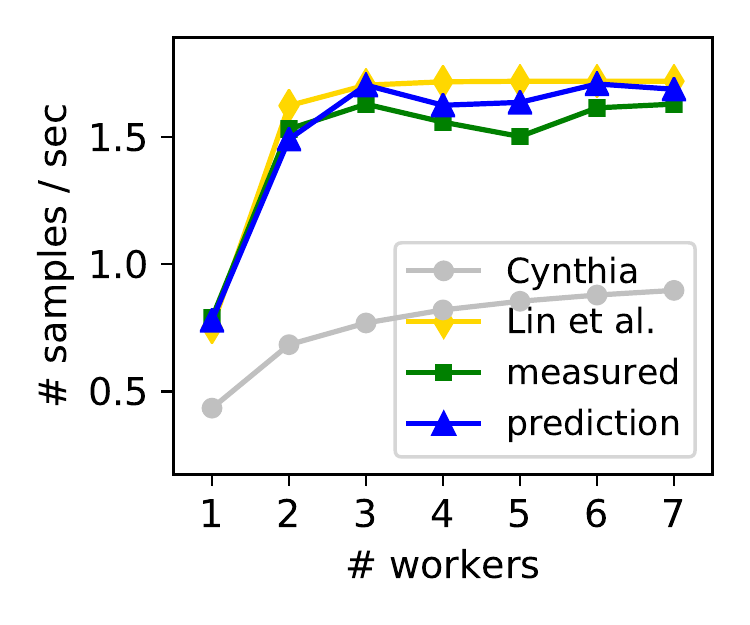}
    \vspace*{-6mm}
    \captionsetup{justification=centering,margin={-3mm,0mm}}
    \caption{AlexNet,\\batch size = 4}\label{fig:cluster_batch_alexnet_4}
  \end{subfigure}%
  \begin{subfigure}[b]{.33\linewidth}
    \centering
    \includegraphics[width=\linewidth]{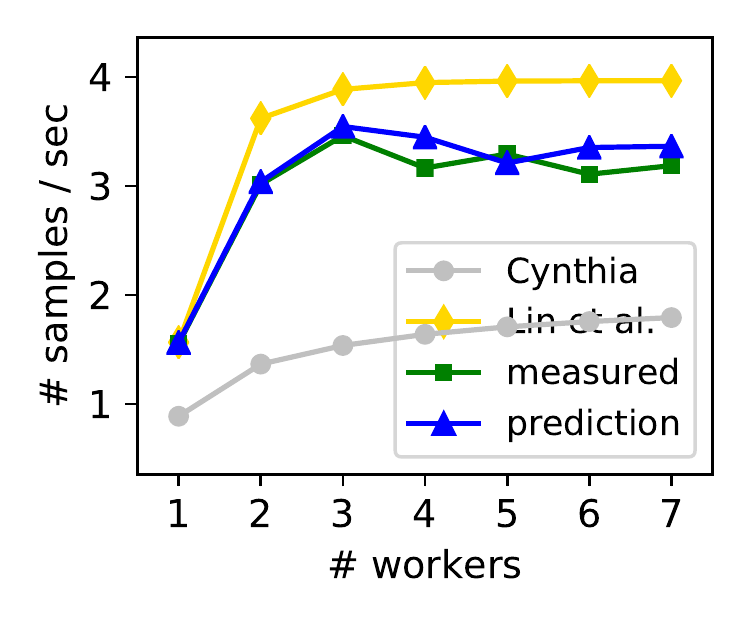}
    \vspace*{-6mm}
    \captionsetup{justification=centering,margin={-2mm,0mm}}
    \caption{AlexNet,\\batch size = 8}\label{fig:cluster_batch_alexnet_8}
  \end{subfigure}%

  \begin{subfigure}[b]{.33\linewidth}
    \centering
    \includegraphics[width=\linewidth]{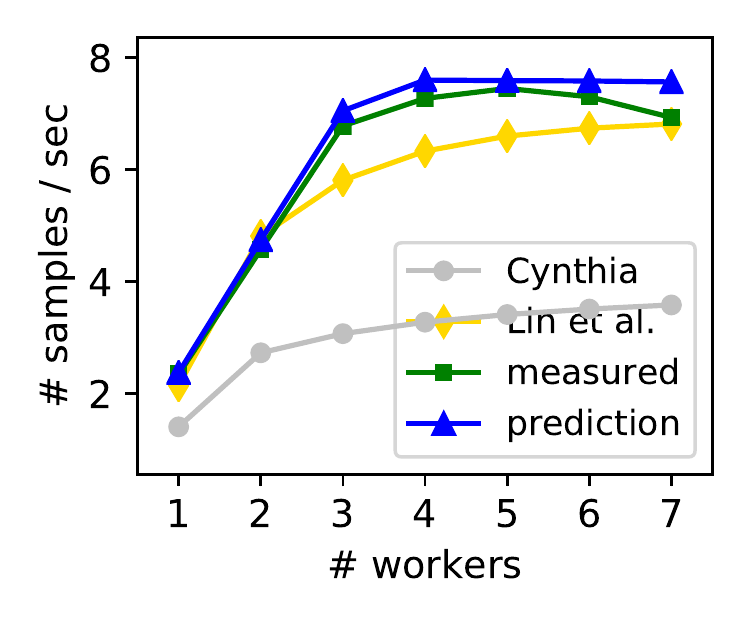}
    \vspace*{-6mm}
    \captionsetup{justification=centering,margin={-2mm,0mm}}
    \caption{AlexNet,\\batch size = 16}\label{fig:cluster_batch_alexnet_16}
  \end{subfigure}%
  \begin{subfigure}[b]{.33\linewidth}
    \centering
    \includegraphics[width=\linewidth]{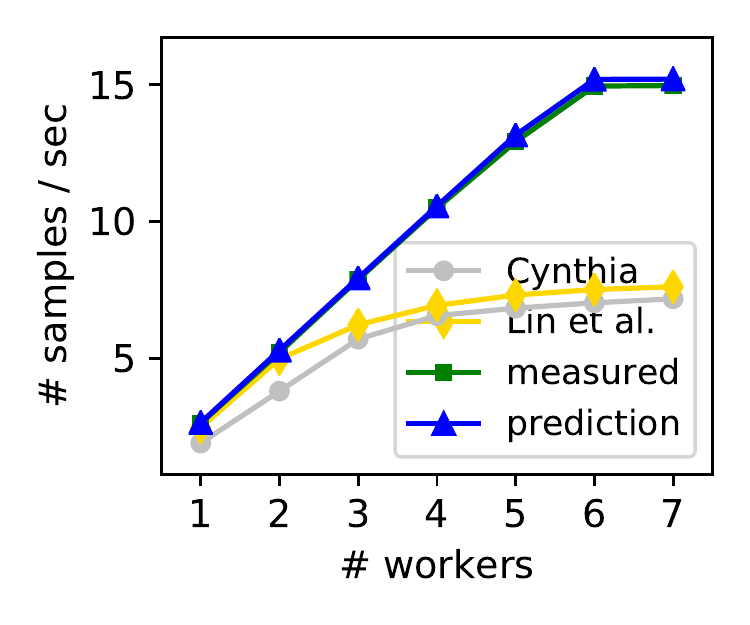}
    \vspace*{-6mm}
    \captionsetup{justification=centering,margin={-3mm,0mm}}
    \caption{AlexNet,\\batch size = 32}\label{fig:cluster_batch_alexnet_32}
  \end{subfigure}%
  \hspace{.33\linewidth}
  \caption{Results on private CPU cluster, varying batch size}\label{fig:cluster_batch}
\end{figure}

First, to illustrate the ability of our approach to accurately predict
throughput with different batch sizes on our local CPU cluster, we
consider a fixed DNN model (AlexNet~\cite{krizhevsky2012imagenet}) and
vary the batch size (batch sizes are small compared to GPU experiments
of \cref{sec:cloud} because of the limited processing power of
CPUs). \cref{fig:cluster_batch} presents the results, showing that
prediction error is within 10\% for all batch sizes (the figure also
includes quantitative comparisons to related work detailed in
\cref{sec:accuracy-comparison}).

\begin{figure}[tb]
  \centering
  \begin{subfigure}[b]{.33\columnwidth}
    \centering
    \includegraphics[width=\linewidth]{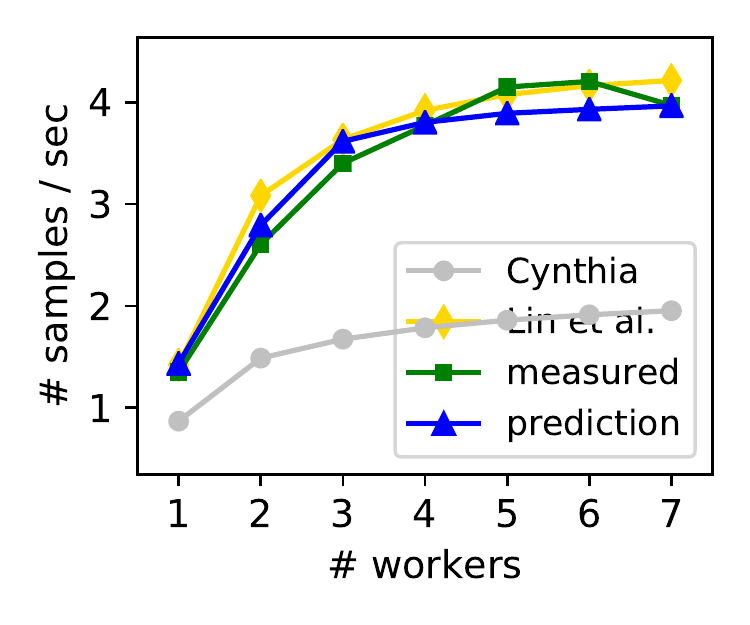}
    \vspace*{-6mm}
    \captionsetup{justification=centering,margin={-2mm,0mm}}
    \caption{GoogLeNet,\\batch size = 1}\label{fig:cluster_model_googlenet_1}
  \end{subfigure}%
  \begin{subfigure}[b]{.33\columnwidth}
    \centering
    \includegraphics[width=\linewidth]{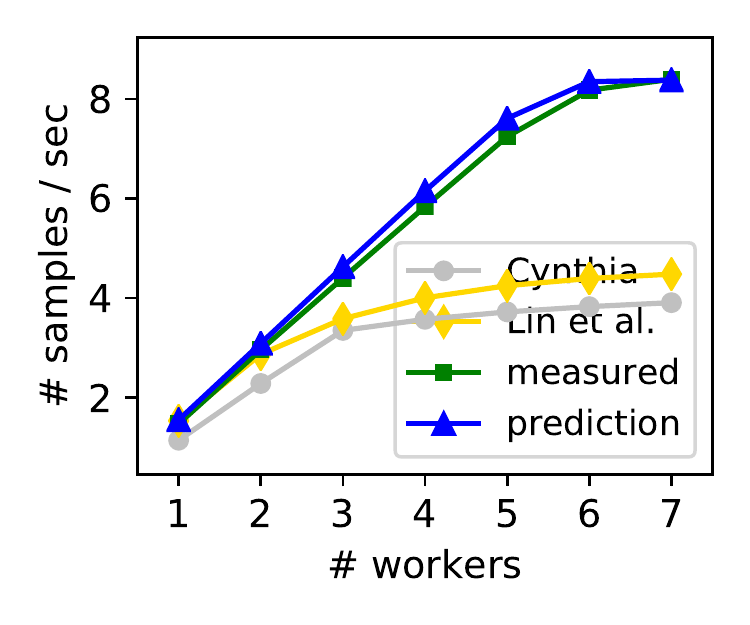}
    \vspace*{-6mm}
    \captionsetup{justification=centering,margin={-2mm,0mm}}
    \caption{GoogLeNet,\\batch size = 2}\label{fig:cluster_model_googlenet_2}
  \end{subfigure}%
  \begin{subfigure}[b]{.33\columnwidth}
    \centering
    \includegraphics[width=\linewidth]{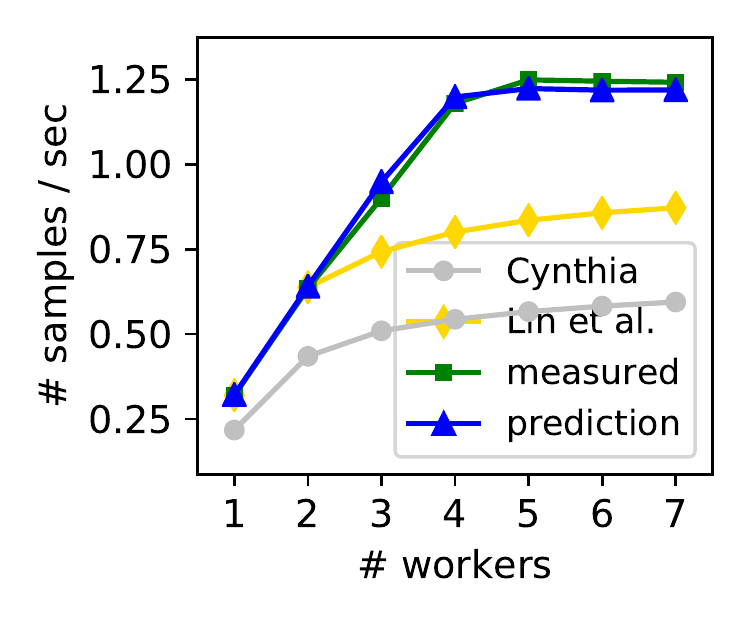}
    \vspace*{-6mm}
    \captionsetup{justification=centering,margin={-3mm,0mm}}
    \caption{Inception-v3,\\batch size = 1}\label{fig:cluster_model_inception3_1}
  \end{subfigure}%
  \hfill
  \begin{subfigure}[b]{.33\columnwidth}
    \centering
    \includegraphics[width=\linewidth]{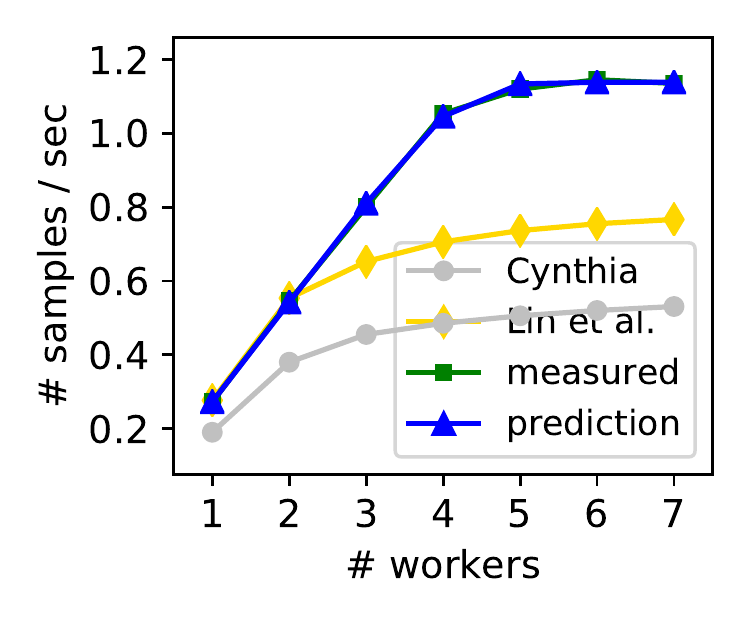}
    \vspace*{-6mm}
    \captionsetup{justification=centering,margin={-3mm,0mm}}
    \caption{ResNet-50,\\batch size = 1}\label{fig:cluster_model_resnet50_1}
  \end{subfigure}%
  % \iffalse
  % \begin{subfigure}[b]{.33\columnwidth}
  %   \centering
  %   \includegraphics[width=\linewidth]{figures/4-results/cluster/vgg11_2.pdf}
  %   \vspace*{-6mm}
  %   \captionsetup{justification=centering,margin={-3mm,0mm}}
  %   \caption{VGG-11,\\batch size = 2}\label{fig:cluster_model_vgg11_2}
  % \end{subfigure}%
  % \fi
  \begin{subfigure}[b]{.33\columnwidth}
    \centering
    \includegraphics[width=\linewidth]{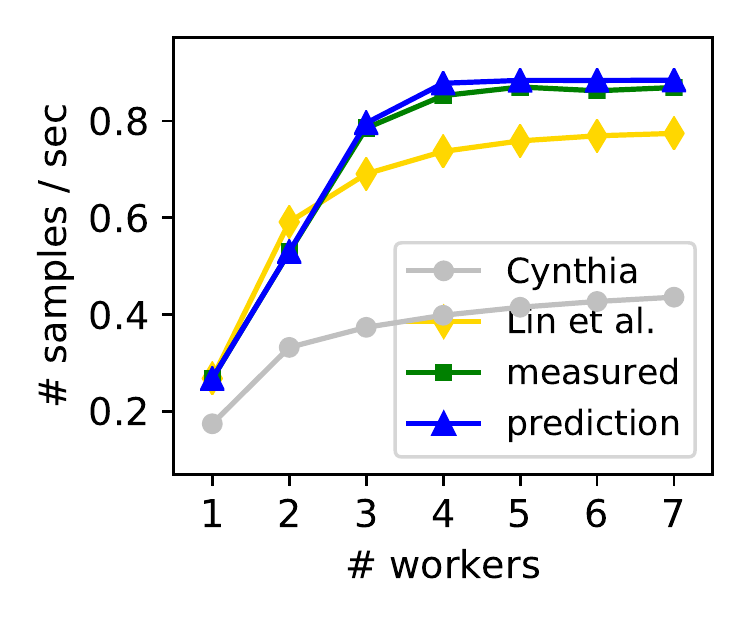}
    \vspace*{-6mm}
    \captionsetup{justification=centering,margin={-3mm,0mm}}
    \caption{VGG-11,\\batch size = 4}\label{fig:cluster_model_vgg11_4}
  \end{subfigure}%
  \begin{subfigure}[b]{.33\columnwidth}
    \centering
    \includegraphics[width=\linewidth]{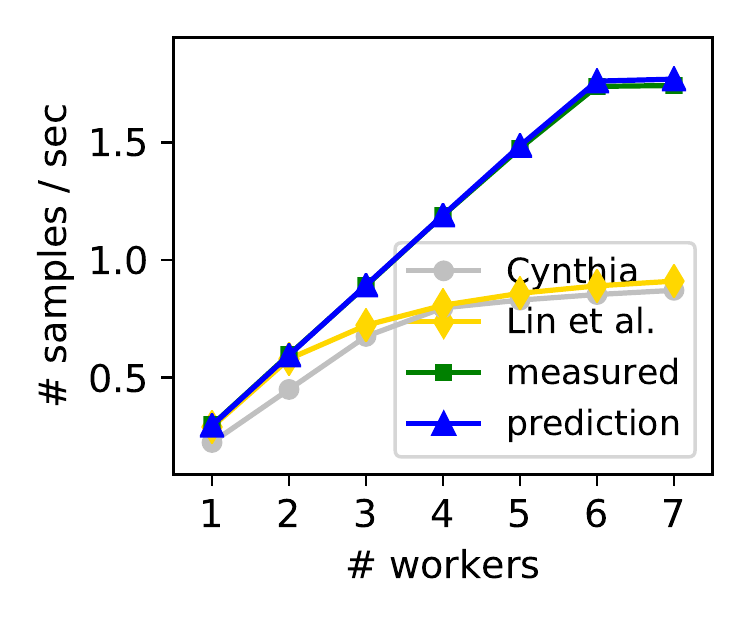}
    \vspace*{-6mm}
    \captionsetup{justification=centering,margin={-3mm,0mm}}
    \caption{VGG-11,\\batch size = 8}\label{fig:cluster_model_vgg11_8}
  \end{subfigure}%
  \caption{Results on private CPU cluster, different models}\label{fig:cluster_model}
\end{figure}

Next, we evaluate the accuracy of our throughput prediction method
across different DNN models, including GoogLeNet~\cite{googlenet},
Inception-v3~\cite{inceptionv3}, ResNet-50~\cite{resnet},
VGG-11~\cite{vgg}. \cref{fig:cluster_model} shows that, in this
case, prediction error is also within 10\%.

\begin{figure}[tb]
  \centering
  \begin{subfigure}[b]{.57\linewidth}
    \centering
    \includegraphics[height=1.25cm]{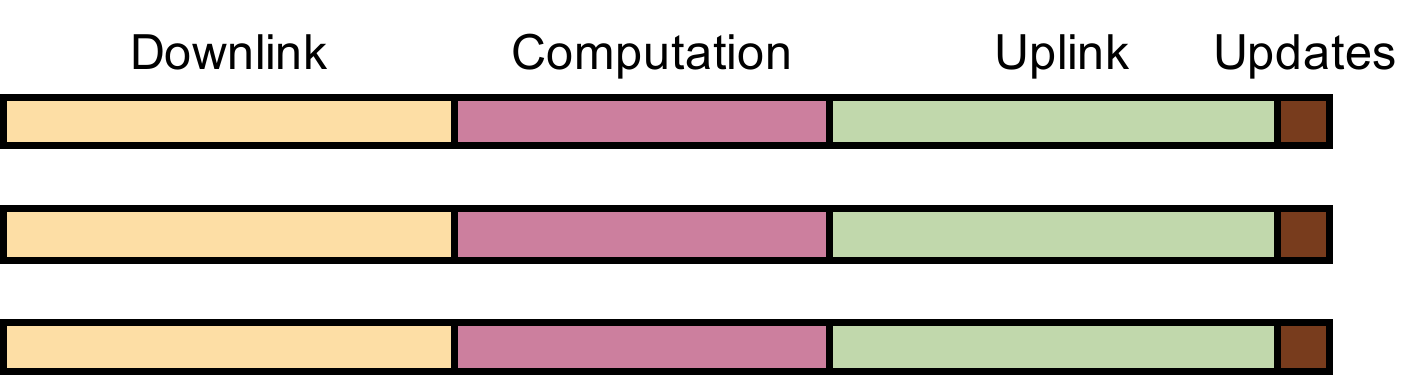}
    \captionsetup{justification=centering}
    \caption{Completely overlap}\label{fig:overlap}
  \end{subfigure}%
  \begin{subfigure}[b]{.43\linewidth}
    \centering
    \includegraphics[height=1.25cm]{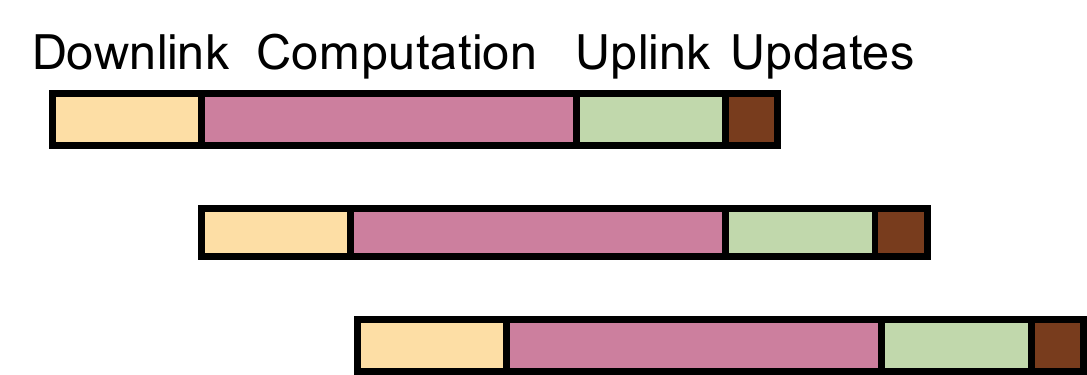}
    \captionsetup{justification=centering}
    \caption{Completely interleave}\label{fig:interleave}
  \end{subfigure}%
  \caption{Workers start at the same time and gradually get out of
    synchronization; training steps are fastest when workers alternate
    using the network.}
\end{figure}

\begin{figure}[tb]
  \centering
  \begin{subfigure}[b]{\linewidth}
    \centering
    \includegraphics[width=\linewidth]{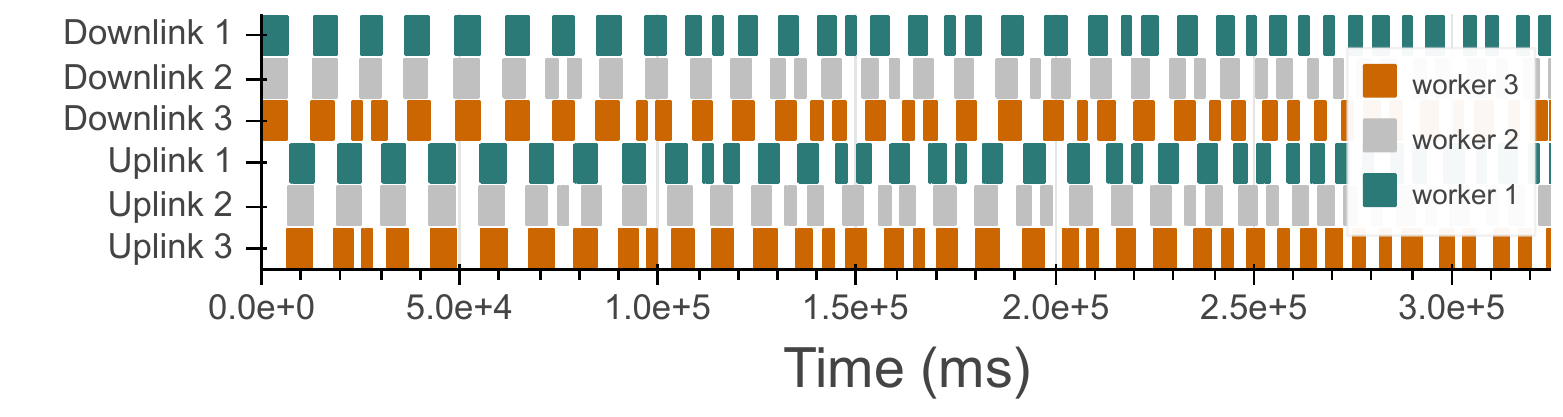}
    \captionsetup{justification=centering}
  \end{subfigure}%
  \hfill
  \begin{subfigure}[b]{\linewidth}
    \centering
    \includegraphics[width=\linewidth]{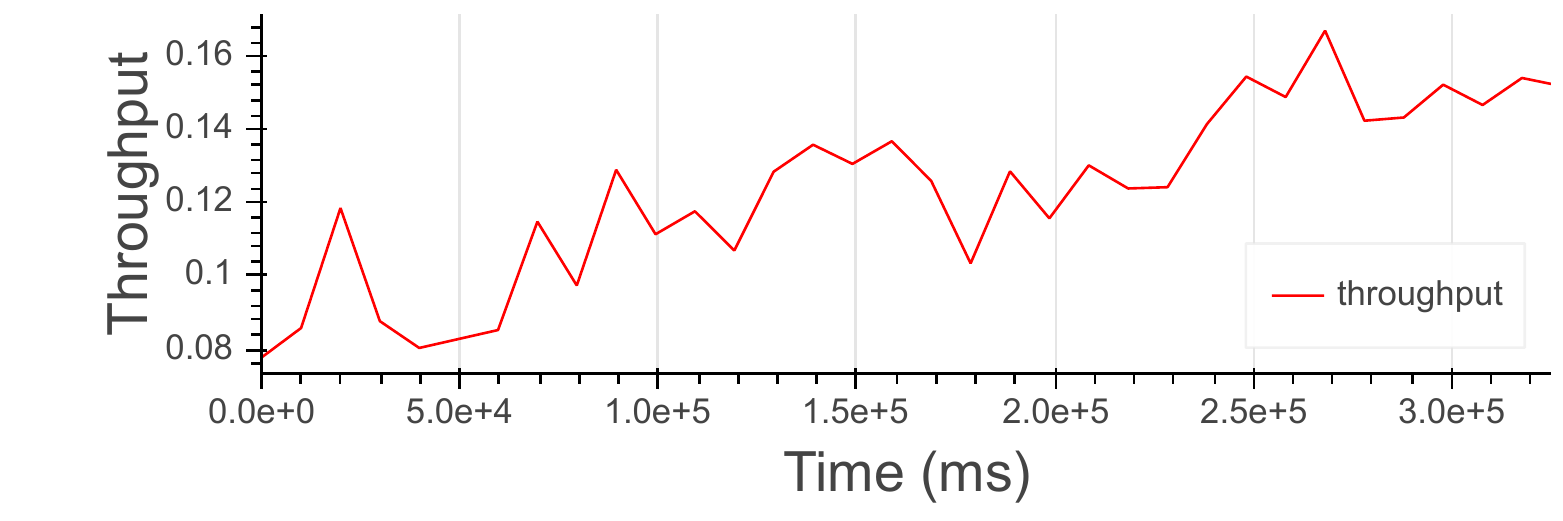}
    \captionsetup{justification=centering}
  \end{subfigure}%
  \caption{Distributed training of AlexNet with batch size of 8 and 3
    workers on private CPU cluster. As downlinks/uplinks start
    interleaving, throughput increases.}\label{fig:step_sync_real}
  \vspace{-1em}
\end{figure}

By inspecting the measured traces, we find that all workers start the
downlink phase of the first SGD step at the same time, as depicted in
\cref{fig:overlap}. As time advances, their training steps gradually
start at different times due to small variations in computation and
communication times.
Ideally, workers continue to interleave and eventually stabilize when
their downlink and uplink transmissions completely get out of
synchronization, leading to higher throughput
(\cref{fig:interleave}).
Such communication pattern could be enforced through \emph{network
  traffic control}, where the parameter server exchanges parameters
with the workers in a strictly sequential order
\cite{huang2019tensorlights}.
However, in TensorFlow different workers contend for bandwidth without
being regulated, potentially resulting only in partial interleaving of
communication operations (\cref{fig:step_sync_real}), leading to a
more challenging environment for predicting throughput, as explored
also in \cref{sec:cloud}.

\begin{figure}[tb]
  \centering
  \begin{subfigure}[b]{.33\columnwidth}
    \centering
    \includegraphics[width=\linewidth]{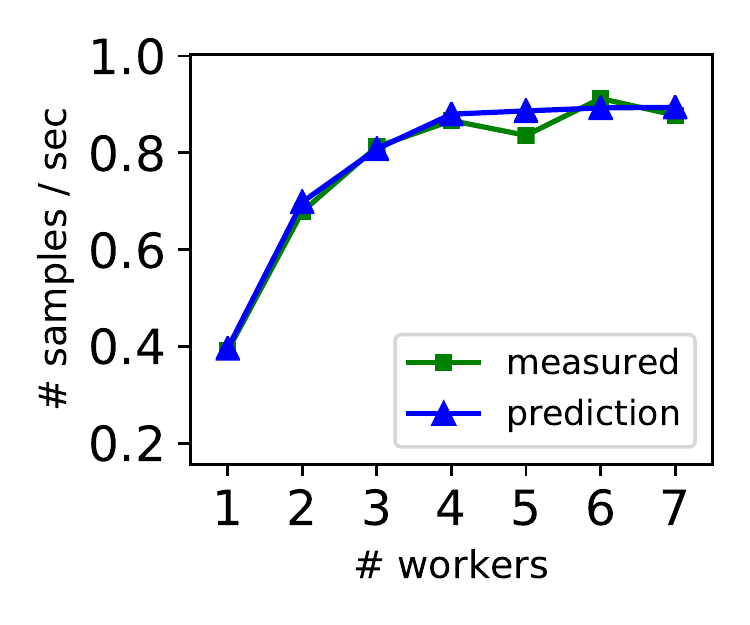}
    \vspace*{-6mm}
    \captionsetup{justification=centering,margin={3mm,0mm}}
    \caption{AlexNet,\\batch size = 2}\label{fig:alexnet_2_nofc}
  \end{subfigure}%
  \begin{subfigure}[b]{.33\columnwidth}
    \centering
    \includegraphics[width=\linewidth]{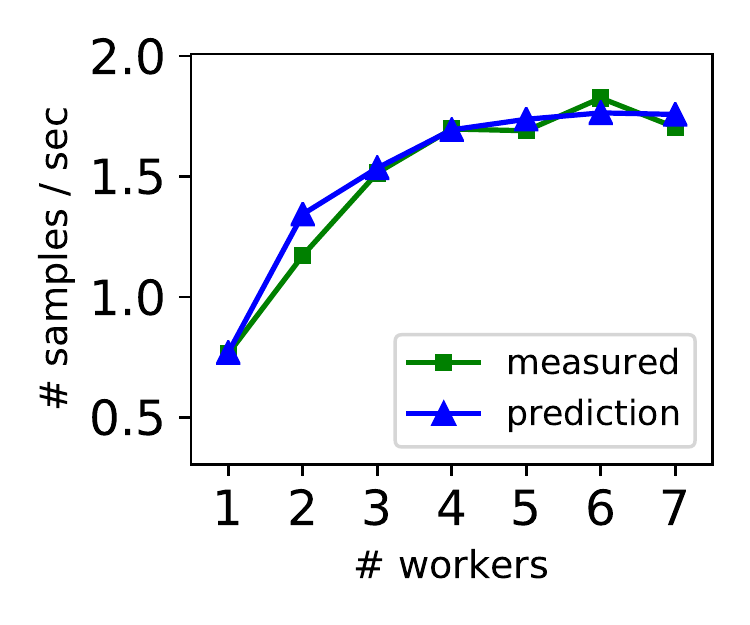}
    \vspace*{-6mm}
    \captionsetup{justification=centering,margin={3mm,0mm}}
    \caption{AlexNet,\\batch size = 4}\label{fig:alexnet_4_nofc}
  \end{subfigure}%
  \begin{subfigure}[b]{.33\columnwidth}
    \centering
    \includegraphics[width=\linewidth]{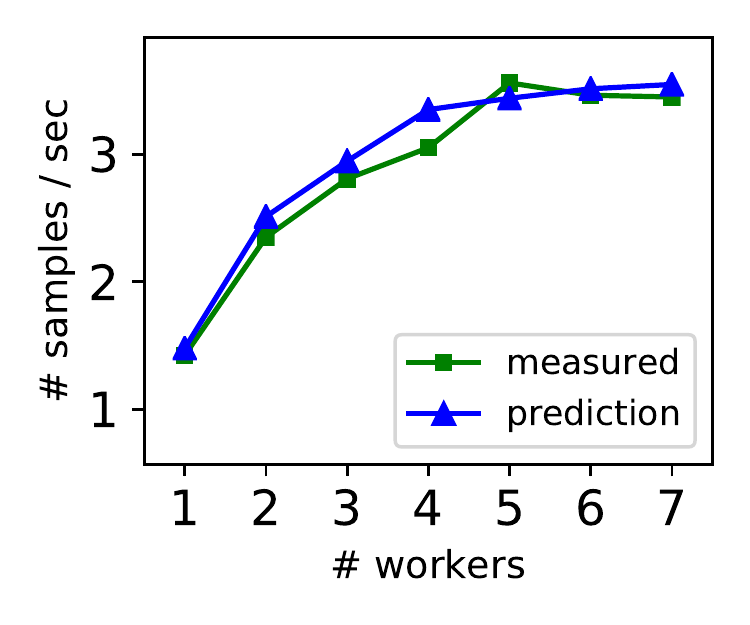}
    \vspace*{-6mm}
    \captionsetup{justification=centering,margin={2mm,0mm}}
    \caption{AlexNet,\\batch size = 8}\label{fig:alexnet_8_nofc}
  \end{subfigure}%
  \hfill
  \begin{subfigure}[b]{.33\columnwidth}
    \centering
    \includegraphics[width=\linewidth]{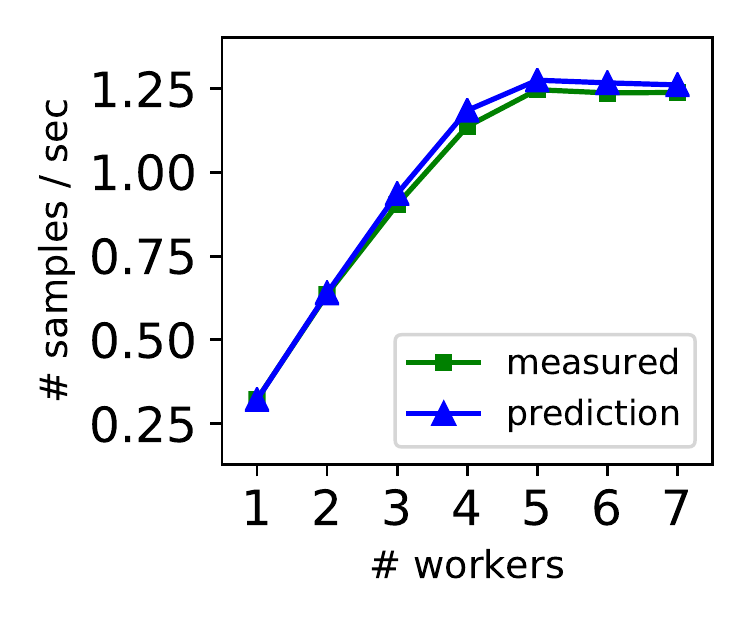}
    \vspace*{-6mm}
    \captionsetup{justification=centering,margin={4mm,0mm}}
    \caption{Inception-v3,\\batch size = 1}\label{fig:inception3_1_nofc} %
  \end{subfigure}%
  \begin{subfigure}[b]{.33\columnwidth}
    \centering
    \includegraphics[width=\linewidth]{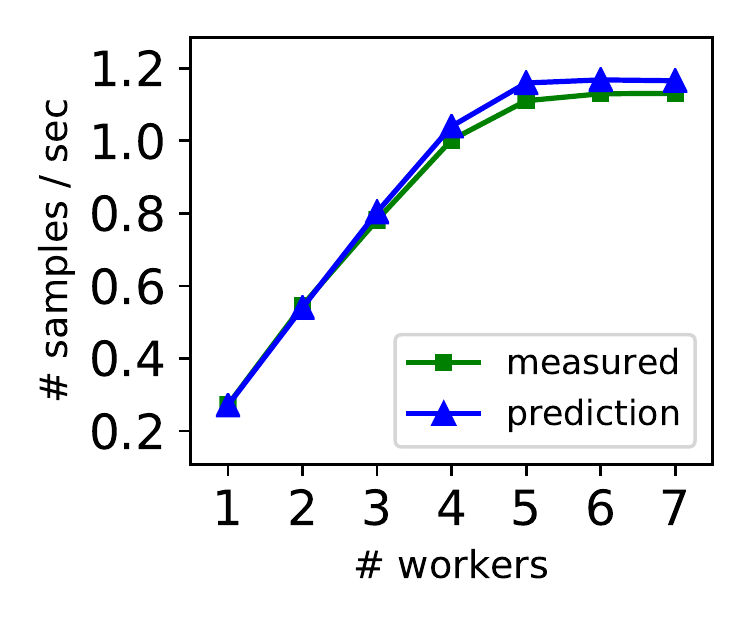}
    \vspace*{-6mm}
    \captionsetup{justification=centering,margin={3mm,0mm}}
    \caption{ResNet-50,\\batch size = 1}\label{fig:resnet50_1_nofc}
  \end{subfigure}%
  \begin{subfigure}[b]{.33\columnwidth}
    \centering
    \includegraphics[width=\linewidth]{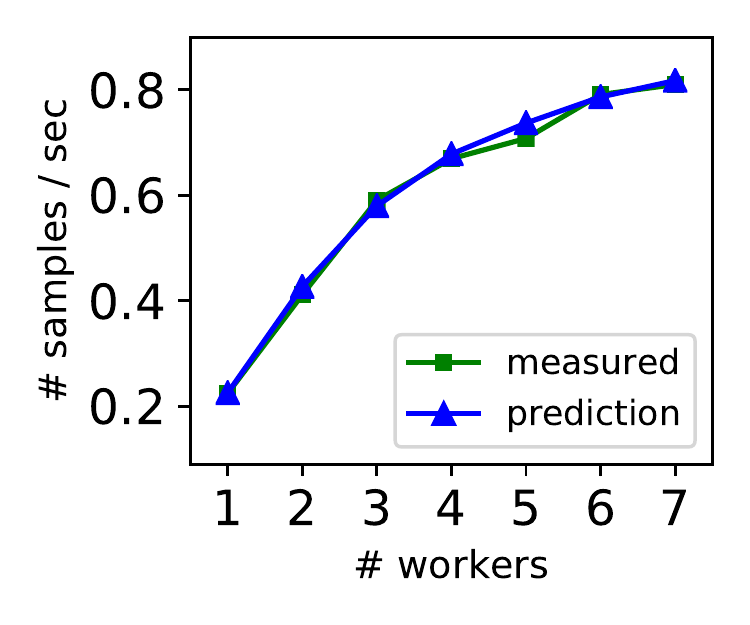}
    \vspace*{-6mm}
    \captionsetup{justification=centering,margin={3mm,0mm}}
    \caption{VGG-11,\\batch size = 4}\label{fig:vgg_4_nofc}
  \end{subfigure}%
  \caption{Prediction of different models on CPU cluster, with flow control disabled}\label{fig:cluster_no_flow_control}
  \vspace{-1em}
\end{figure}

Next, we explore the effects of HTTP/2 stream multiplexing. In
\cref{communication_order_optimization}, we motivated approaches to
disable HTTP/2 flow control and enforce a specific communication
ordering.
To evaluate the prediction accuracy of our method in these scenarios,
we disable flow control and repeat the experiments of
\cref{fig:cluster_batch_alexnet_2,fig:cluster_batch_alexnet_4,fig:cluster_batch_alexnet_8}
and
\cref{fig:cluster_model_inception3_1,fig:cluster_model_resnet50_1,fig:cluster_model_vgg11_4};
corresponding results, presented in
\cref{fig:alexnet_2_nofc,fig:alexnet_4_nofc,fig:alexnet_8_nofc} and
\cref{fig:inception3_1_nofc,fig:resnet50_1_nofc,fig:vgg_4_nofc},
highlight good prediction accuracy, with errors of at most 10\% in all
cases but \cref{fig:alexnet_4_nofc} for $W=2$ workers, where the error
is 20\%; we observed that this was due to lower than expected
interleaving in the measurements.

\begin{figure}[tb]
  \centering
  \begin{subfigure}[b]{.33\columnwidth}
    \centering
    \includegraphics[width=\linewidth]{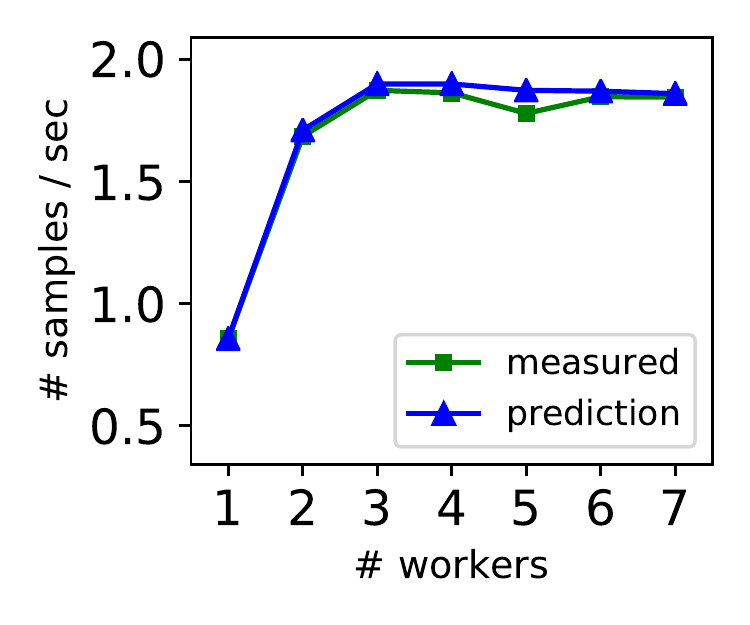}
    \vspace*{-6mm}
    \caption{TIC Order}\label{fig:alexnet_4_tic}
  \end{subfigure}%
  \begin{subfigure}[b]{.33\columnwidth}
    \centering
    \includegraphics[width=\linewidth]{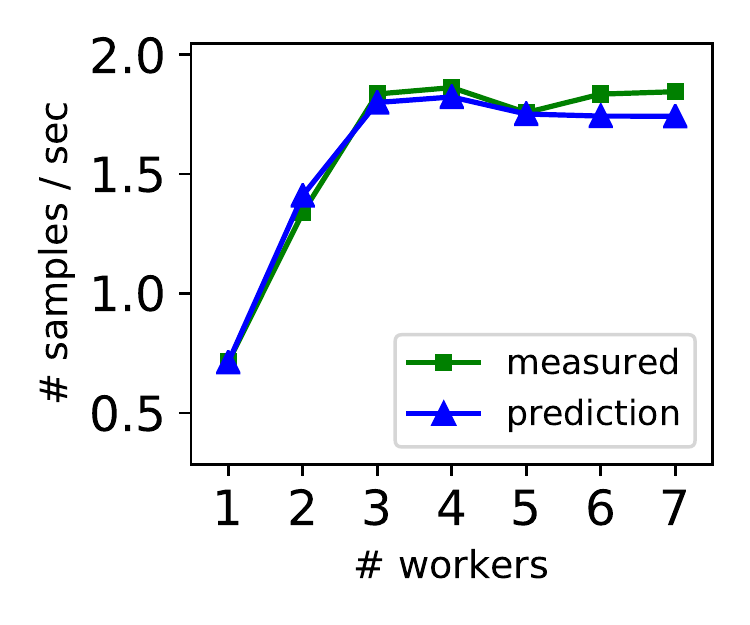}
    \vspace*{-6mm}
    \caption{TIC Reverse Order}\label{fig:alexnet_4_tic_reverse}
  \end{subfigure}%
  \begin{subfigure}[b]{.33\columnwidth}
    \centering
    \includegraphics[width=\linewidth]{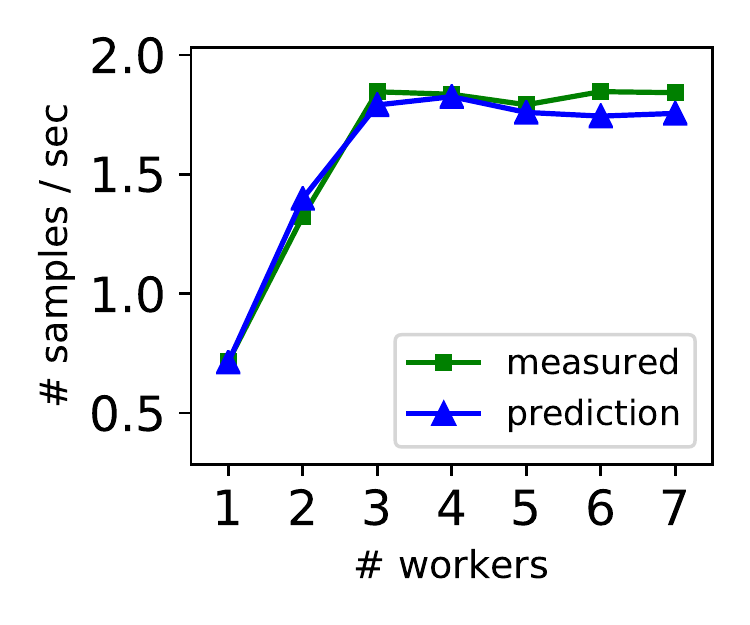}
    \vspace*{-6mm}
    \caption{Random Order}\label{fig:alexnet_4_random1}
  \end{subfigure}%
  % \iffalse
  % \begin{subfigure}[b]{.33\columnwidth}
  %   \centering
  %   \includegraphics[width=\linewidth]{figures/4-results/cluster_nofc/alexnet_4_random2.pdf}
  %   \vspace*{-6mm}
  %   \caption{Random Order 2}\label{fig:alexnet_4_random2}
  % \end{subfigure}%
  % \fi
  \caption{Prediction of AlexNet, batch size = 4, with flow control disabled, enforcing different orders on CPU cluster}\label{fig:cluster_alexnet_order}
  \vspace{-1em}
\end{figure}

\begin{figure}[tb]
  \centering
  \begin{subfigure}[b]{.33\columnwidth}
    \centering
    \includegraphics[width=\linewidth]{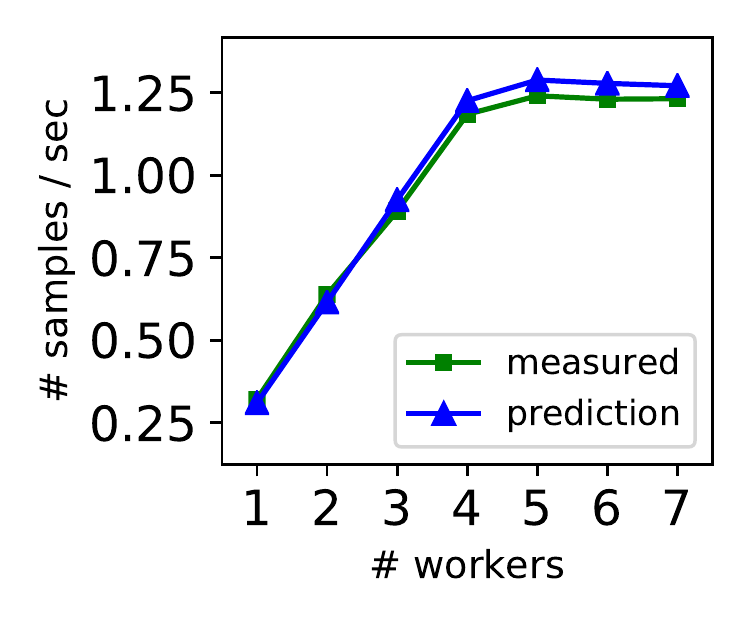}
    \vspace*{-6mm}
    \captionsetup{justification=centering}
    \caption{Inception-v3,\\batch size = 1}\label{fig:inception3_1_tic}
  \end{subfigure}%
  \begin{subfigure}[b]{.33\columnwidth}
    \centering
    \includegraphics[width=\linewidth]{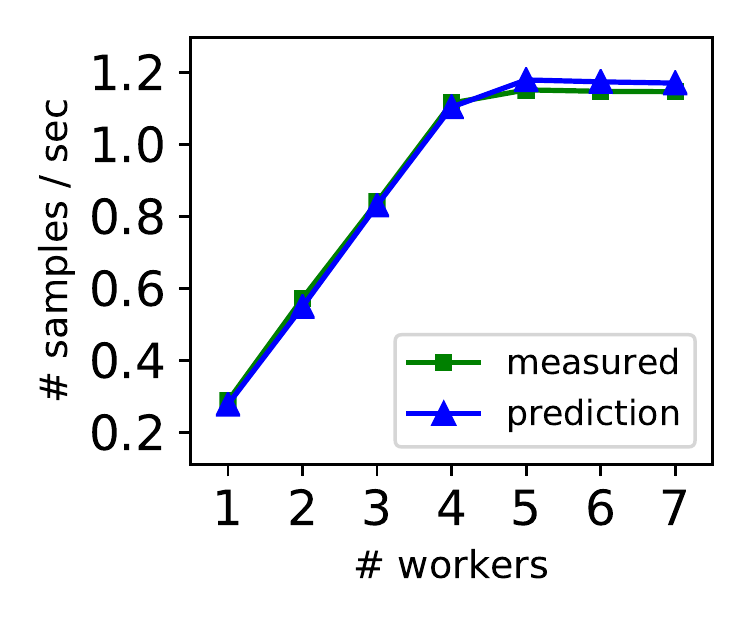}
    \vspace*{-6mm}
    \captionsetup{justification=centering}
    \caption{ResNet-50,\\batch size = 1}\label{fig:resnet50_1_tic}
  \end{subfigure}%
  \begin{subfigure}[b]{.33\columnwidth}
    \centering
    \includegraphics[width=\linewidth]{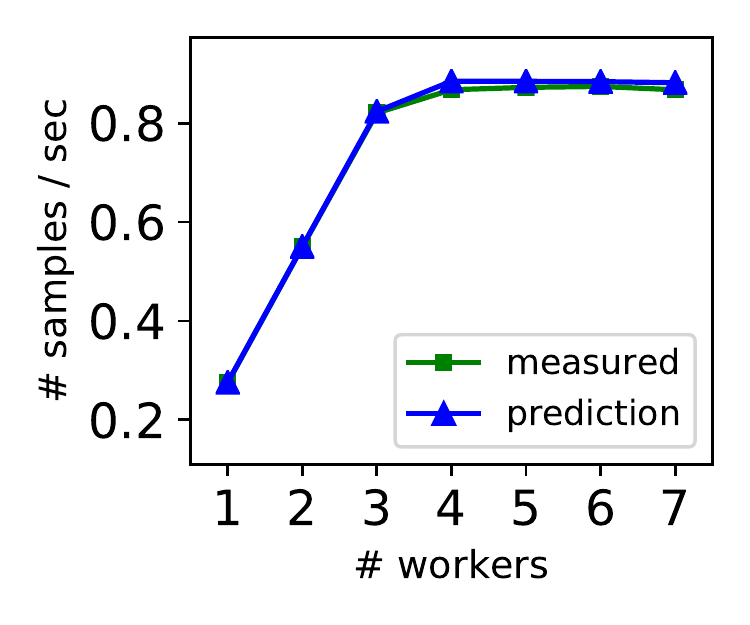}
    \vspace*{-6mm}
    \captionsetup{justification=centering}
    \caption{VGG-11,\\batch size = 4}\label{fig:vgg11_4_tic}
  \end{subfigure}%
  % \iffalse
  % \begin{subfigure}[b]{.33\columnwidth}
  %   \centering
  %   \includegraphics[width=\linewidth]{figures/4-results/cluster_nofc/googlenet_2_tic.pdf}
  %   \vspace*{-6mm}
  %   \caption{GoogLeNet,\\batch size = 2}\label{fig:googlenet_2_tic}
  % \end{subfigure}%
  % \fi
  \caption{Prediction of different models with flow control disabled, enforcing TIC order on CPU cluster}\label{fig:cluster_tic}
\end{figure}

We also enforce different stream transmission orderings on the model
in \cref{fig:cluster_batch_alexnet_4,fig:alexnet_4_nofc}: the TIC
ordering suggested by TicTac~\cite{hashemi2018tictac}, the reverse of
such order, and a random order. Results, presented in
\cref{fig:cluster_alexnet_order}, illustrate predictions within 10\%
error.
Finally, in \cref{fig:cluster_tic} we explore prediction accuracy for
the TIC ordering on the DNN models in
\cref{fig:cluster_model_inception3_1,fig:cluster_model_resnet50_1,fig:cluster_model_vgg11_4};
prediction error is less than 5\% for these experiments.

In conclusion, our approach can accurately predict throughput for
different stream selection orders, batch size, DNN model, and number
of workers. This indicates that the prediction algorithm can adapt to
different communication settings, and it has potential of accurately
predicting further optimizations and modifications of the current
TensorFlow implementation.

\subsection{Public Cloud}
\label{sec:cloud}

We evaluate our approach on the Amazon Web Services~(AWS) cloud
platform.
This environment is less stable than our private CPU cluster, as
networking performance may be affected by background traffic and by
the deployment of virtual machines to racks with different latency.

In addition, communication overhead due to parsing of received data
(\cref{sec:overhead}) plays a much more important role, since
networking is $10\times$ faster (10~Gbps).
For example, if overhead accounts for 10\% of the duration of
communication operations recorded in profiling traces on a 1~Gbps
network, it will account for 52.6\% of communication operations
recorded on a 10~Gbps network.

\subsubsection{CPU-Only Instances}

\begin{figure}[tb]
        \centering
        \begin{subfigure}[b]{.33\columnwidth}
                \centering
                \includegraphics[width=\linewidth]{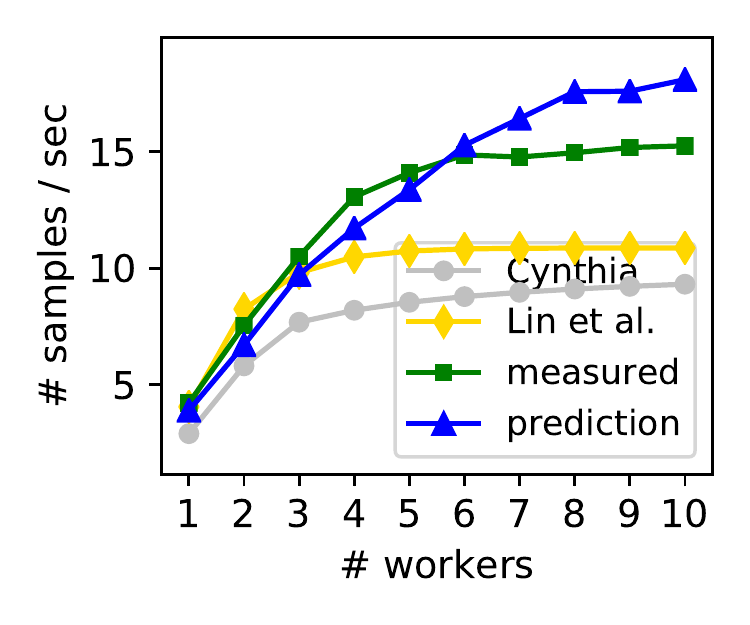}
                \vspace*{-6mm}
                \captionsetup{justification=centering,margin={-3mm,0mm}}
                \caption{AlexNet,\\batch size = 4}\label{fig:aws_cpu_alexnet_4}
        \end{subfigure}%
        \begin{subfigure}[b]{.33\columnwidth}
                \centering
                \includegraphics[width=\linewidth]{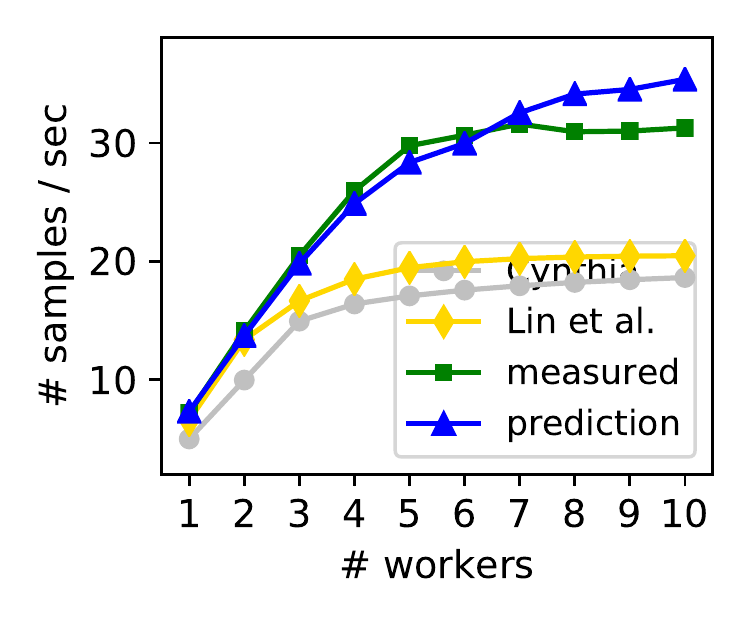}
                \vspace*{-6mm}
                \captionsetup{justification=centering,margin={-3mm,0mm}}
                \caption{AlexNet,\\batch size = 8}\label{fig:aws_cpu_alexnet_8}
        \end{subfigure}%
        \begin{subfigure}[b]{.33\columnwidth}
                \centering
                \includegraphics[width=\linewidth]{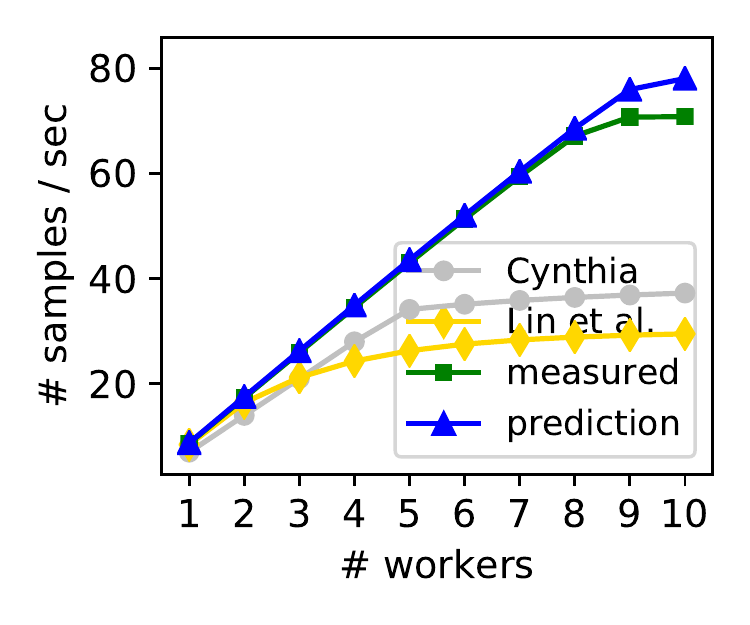}
                \vspace*{-6mm}
                \captionsetup{justification=centering,margin={-3mm,0mm}}
                \caption{AlexNet,\\batch size = 16}\label{fig:aws_cpu_alexnet_16}
        \end{subfigure}%
        \hfill
        \begin{subfigure}[b]{.33\columnwidth}
                \centering
                \includegraphics[width=\linewidth]{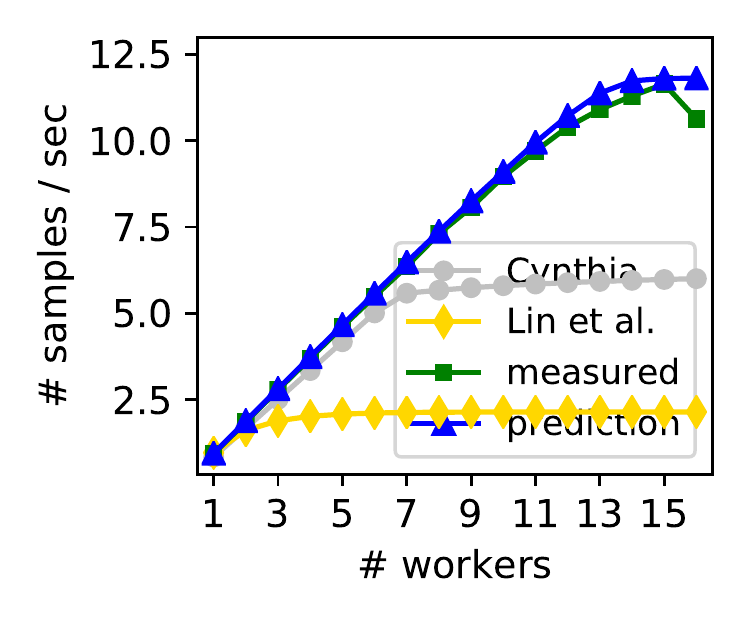}
                \vspace*{-6mm}
                \captionsetup{justification=centering,margin={-3mm,0mm}}
                \caption{ResNet-50,\\batch size = 1}\label{fig:aws_cpu_resnet50}
        \end{subfigure}%
        \begin{subfigure}[b]{.33\columnwidth}
                \centering
                \includegraphics[width=\linewidth]{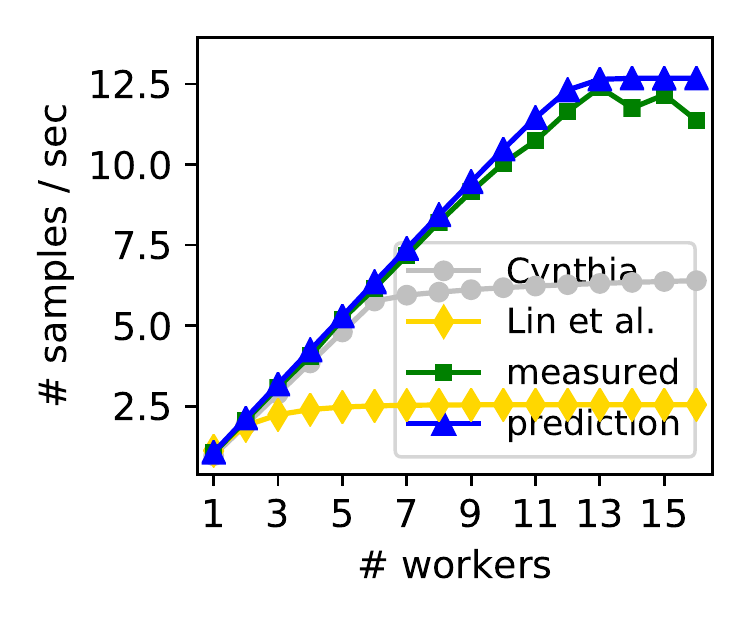}
                \vspace*{-6mm}
                \captionsetup{justification=centering,margin={-3mm,0mm}}
                \caption{Inception-v3,\\batch size = 1}\label{fig:aws_cpu_inception3_1}
        \end{subfigure}%
        \begin{subfigure}[b]{.33\columnwidth}
                \centering
                \includegraphics[width=\linewidth]{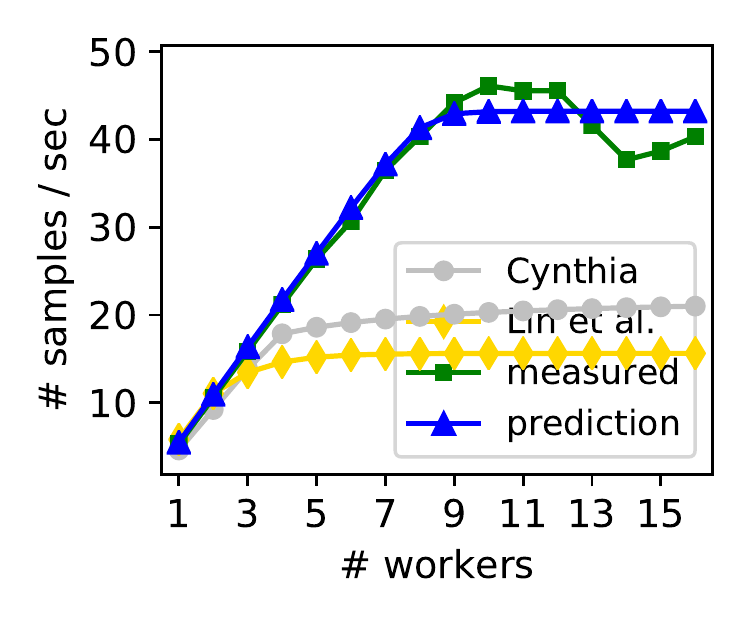}
                \vspace*{-6mm}
                \captionsetup{justification=centering,margin={-3mm,0mm}}
                \caption{GoogLeNet,\\batch size = 1}\label{fig:aws_cpu_googlenet_1}
        \end{subfigure}%
        \caption{Prediction of training on AWS cloud (CPU cluster)}\label{fig:aws_cpu}
        \vspace{-1em}
\end{figure}

As shown in \cref{fig:aws_cpu}, in most cases the error in throughput
predictions on the AWS CPU cluster is within 20\% for various DNN
models and batch sizes, except for \cref{fig:aws_cpu_googlenet_1} with
$W=14$ workers, where the error is 22.8\%.
The prediction error on AWS CPU cluster is larger than that on our
private CPU cluster, mainly because the network on the cloud is less
predictable. In fact, intermittent background traffic can cause the
HTTP/2 flow control window to change over time, leading to prediction
errors.

\subsubsection{GPU Instances}

We also validate our approach on AWS GPU training. \cref{fig:aws_gpu}
illustrates that prediction error is within 20\% across most DNN
models and configurations. Prediction error is within 30\% in
\cref{fig:aws_gpu_small_inception3_16} for $W=2,4$,
\cref{fig:aws_gpu_small_inception4_64} for $W=4$, and
\cref{fig:aws_gpu_small_resnet50_32,fig:aws_gpu_small_resnet101_32,fig:aws_gpu_small_resnet101_64}
for $W=3$; and within 40\% in
\cref{fig:aws_gpu_small_inception4_32,fig:aws_gpu_small_resnet152_32,fig:aws_gpu_small_resnet152_64}
for $W=3$.
We believe that larger errors occur in scenarios with few workers and
smaller computation times (relative to communication), where it is
more difficult to accurately predict the interleaving of data transfers
between the parameter server and the workers; critical scenarios have
2 to 4 workers and smaller batch sizes (e.g., the error is higher in
\cref{fig:aws_gpu_small_inception3_16} than
\cref{fig:aws_gpu_small_inception3_32}).

\subsection{Prediction Accuracy Comparison}
\label{sec:accuracy-comparison}

\begin{figure}[tb]
  \centering
  \begin{subfigure}[b]{.33\columnwidth}
    \centering
    \includegraphics[width=\linewidth]{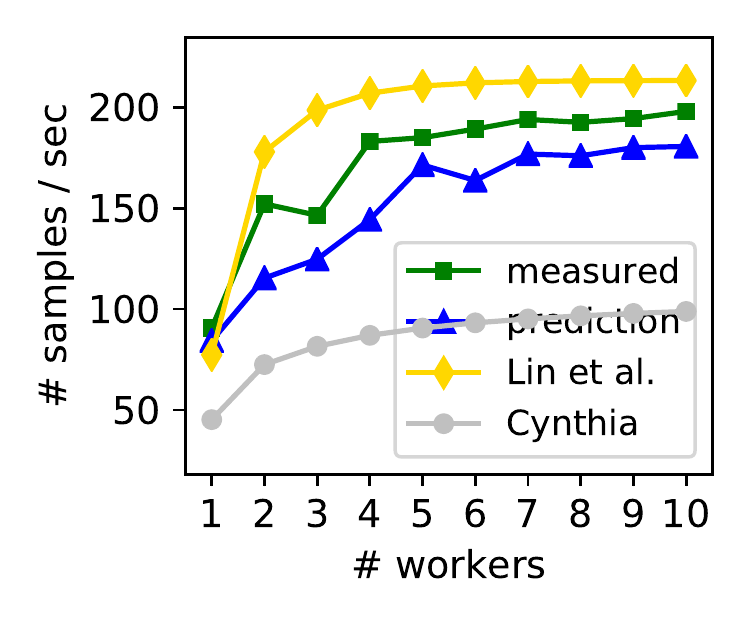}
    \vspace*{-6mm}
    \captionsetup{justification=centering}
    \caption{Inception-v3,\\batch size = 16}\label{fig:aws_gpu_small_inception3_16}
  \end{subfigure}%
  \begin{subfigure}[b]{.33\columnwidth}
    \centering
    \includegraphics[width=\linewidth]{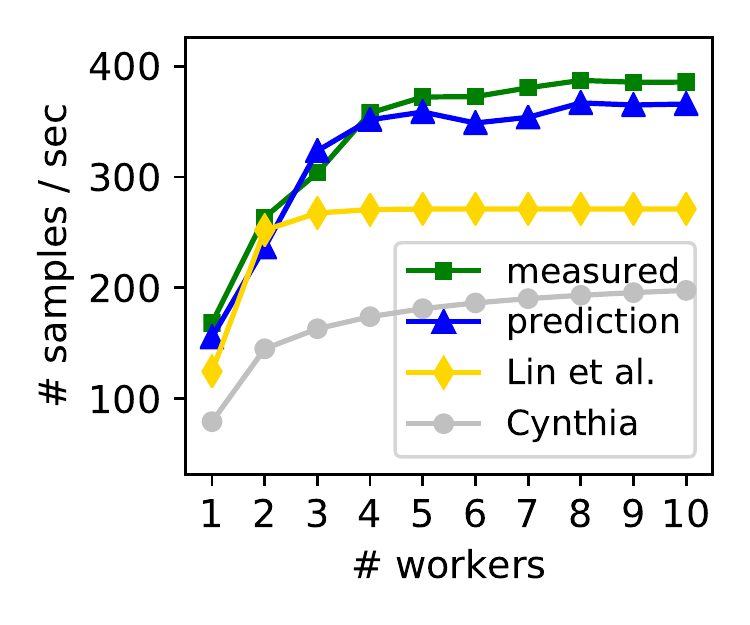}
    \vspace*{-6mm}
    \captionsetup{justification=centering}
    \caption{Inception-v3,\\batch size = 32}\label{fig:aws_gpu_small_inception3_32}
  \end{subfigure}%
  \begin{subfigure}[b]{.33\columnwidth}
    \centering
    \includegraphics[width=\linewidth]{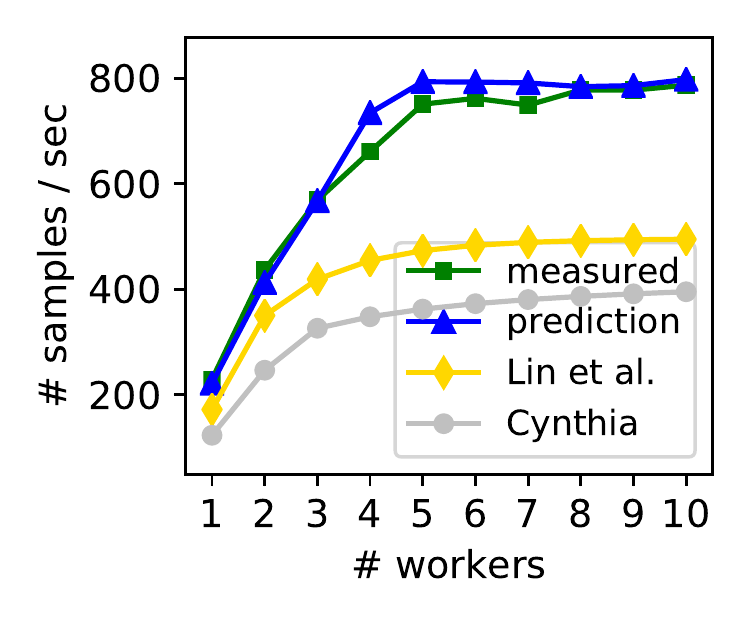}
    \vspace*{-6mm}
    \captionsetup{justification=centering}
    \caption{Inception-v3,\\batch size = 64}\label{fig:aws_gpu_small_inception3_64}
  \end{subfigure}%
  \hfill
  \begin{subfigure}[b]{.33\columnwidth}
    \centering
    \includegraphics[width=\linewidth]{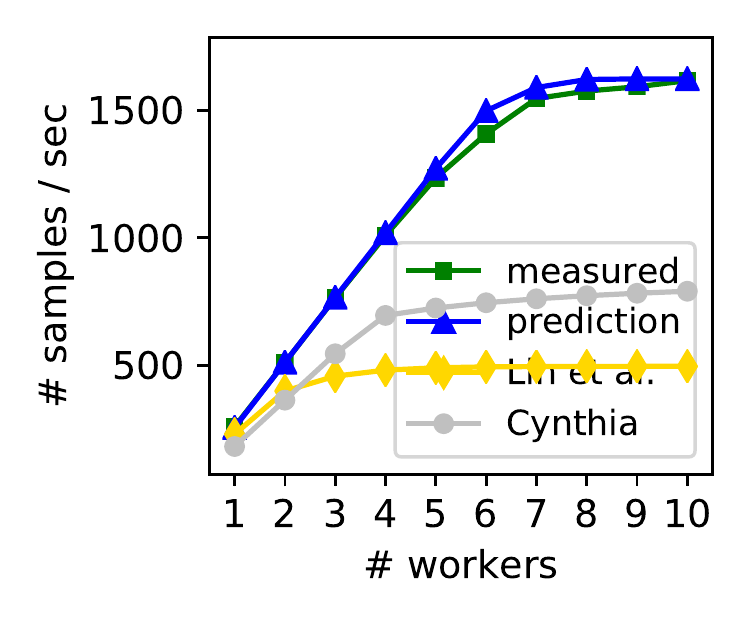}
    \vspace*{-6mm}
    \captionsetup{justification=centering}
    \caption{Inception-v3,\\batch size = 128}\label{fig:aws_gpu_small_inception3_128}
  \end{subfigure}%
  \begin{subfigure}[b]{.33\columnwidth}
    \centering
    \includegraphics[width=\linewidth]{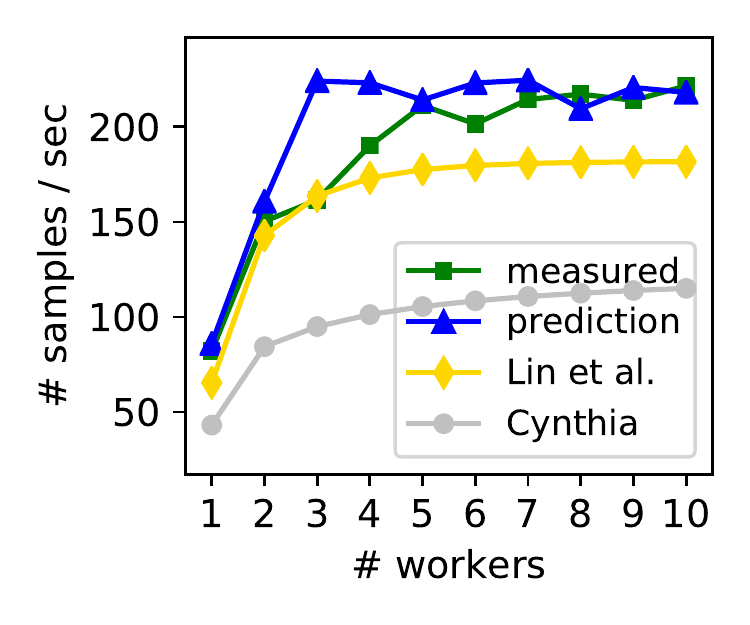}
    \vspace*{-6mm}
    \captionsetup{justification=centering}
    \caption{Inception-v4,\\batch size = 32}\label{fig:aws_gpu_small_inception4_32}
  \end{subfigure}%
  \begin{subfigure}[b]{.33\columnwidth}
    \centering
    \includegraphics[width=\linewidth]{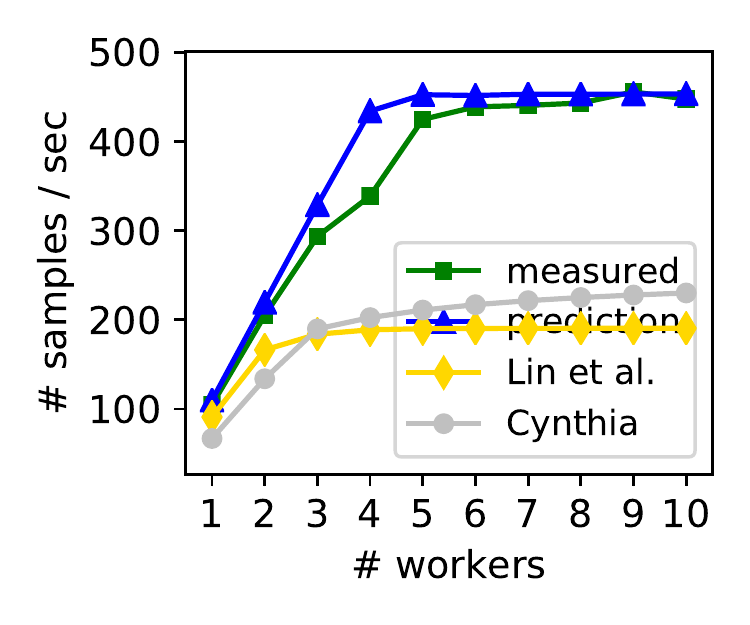}
    \vspace*{-6mm}
    \captionsetup{justification=centering}
    \caption{Inception-v4,\\batch size = 64}\label{fig:aws_gpu_small_inception4_64}
  \end{subfigure}%
  \hfill
  \begin{subfigure}[b]{.33\columnwidth}
    \centering
    \includegraphics[width=\linewidth]{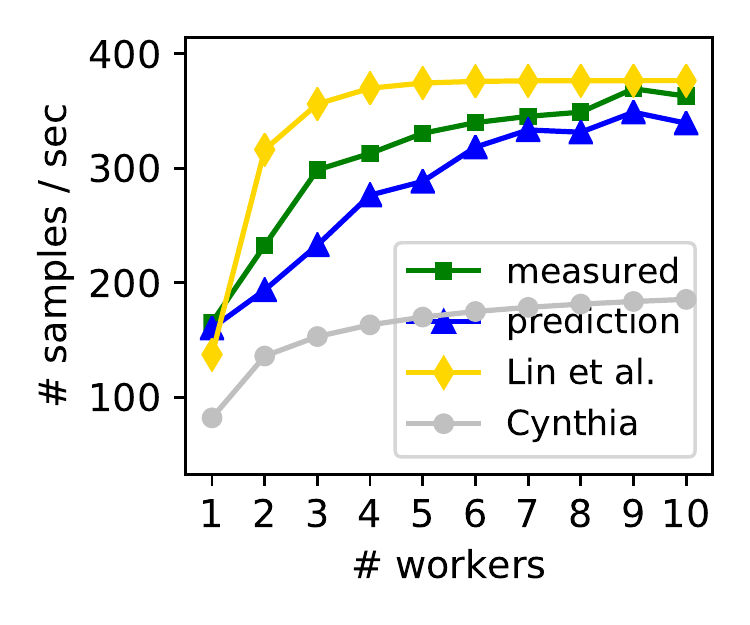}
    \vspace*{-6mm}
    \captionsetup{justification=centering}
    \caption{ResNet-50,\\batch size = 32}\label{fig:aws_gpu_small_resnet50_32}
  \end{subfigure}%
  \begin{subfigure}[b]{.33\columnwidth}
    \centering
    \includegraphics[width=\linewidth]{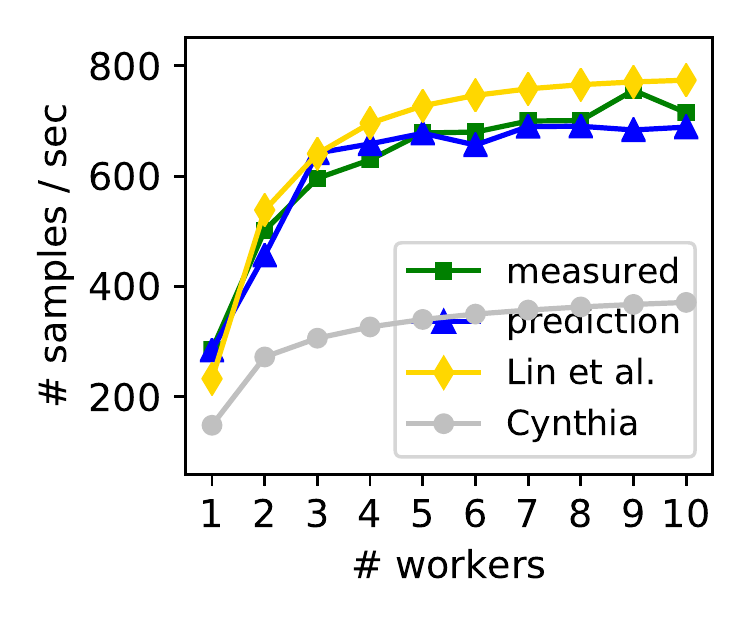}
    \vspace*{-6mm}
    \captionsetup{justification=centering}
    \caption{ResNet-50,\\batch size = 64}\label{fig:aws_gpu_small_resnet50_64}
  \end{subfigure}%
  \begin{subfigure}[b]{.33\columnwidth}
    \centering
    \includegraphics[width=\linewidth]{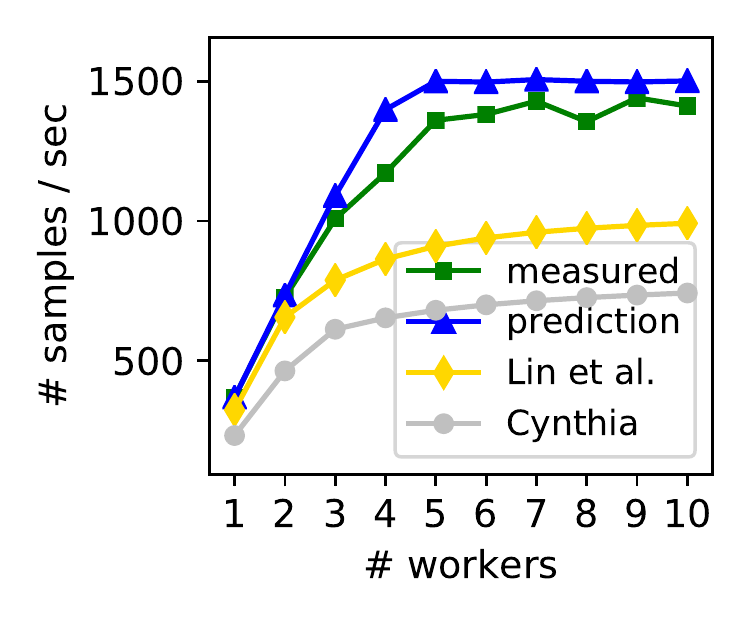}
    \vspace*{-6mm}
    \captionsetup{justification=centering}
    \caption{ResNet-50,\\batch size = 128}\label{fig:aws_gpu_small_resnet50_128}
  \end{subfigure}%
  \hfill
  \begin{subfigure}[b]{.33\columnwidth}
    \centering
    \includegraphics[width=\linewidth]{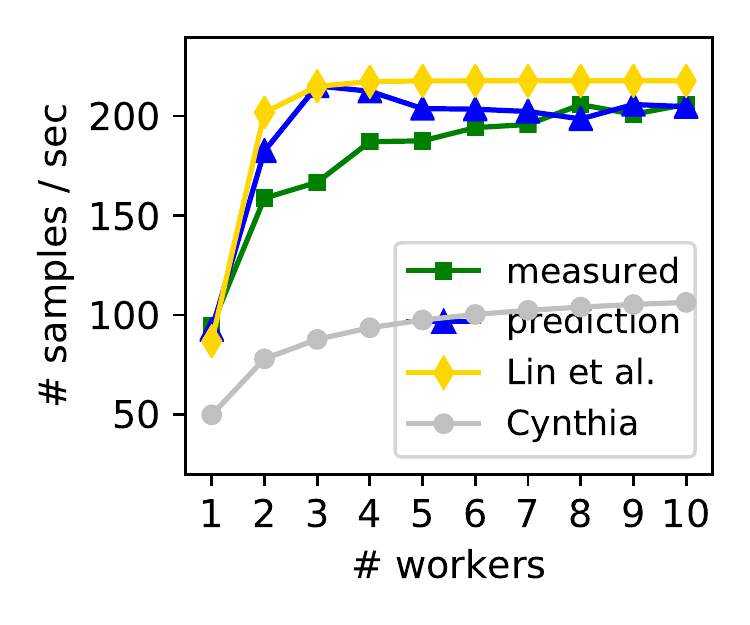}
    \vspace*{-6mm}
    \captionsetup{justification=centering}
    \caption{ResNet-101,\\batch size = 32}\label{fig:aws_gpu_small_resnet101_32}
  \end{subfigure}%
  \begin{subfigure}[b]{.33\columnwidth}
    \centering
    \includegraphics[width=\linewidth]{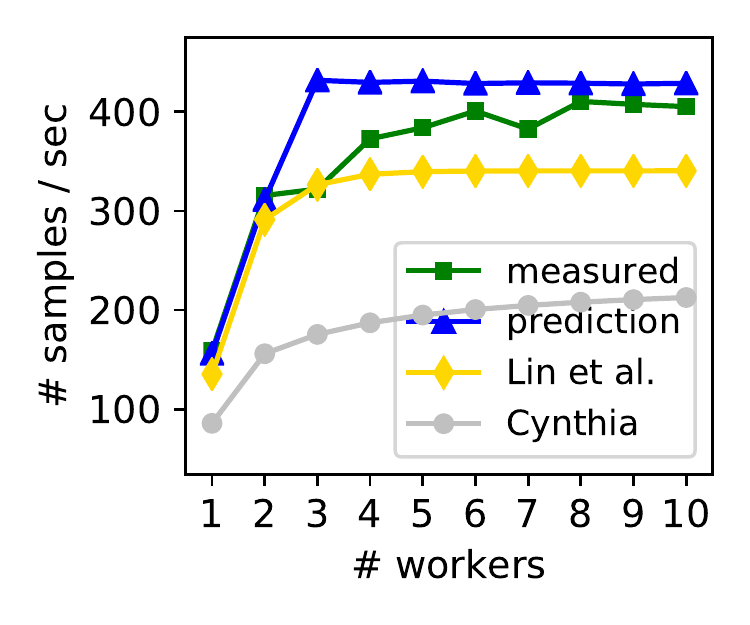}
    \vspace*{-6mm}
    \captionsetup{justification=centering}
    \caption{ResNet-101,\\batch size = 64}\label{fig:aws_gpu_small_resnet101_64}
  \end{subfigure}%
  \begin{subfigure}[b]{.33\columnwidth}
    \centering
    \includegraphics[width=\linewidth]{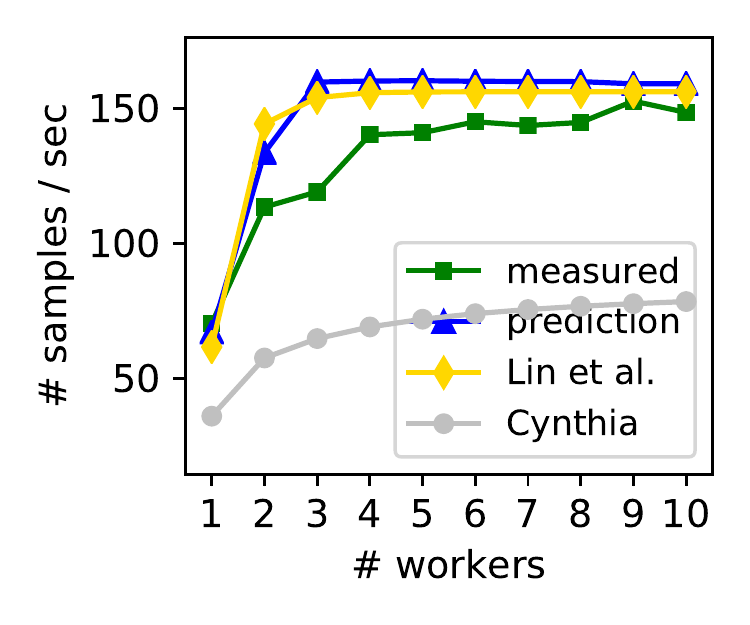}
    \vspace*{-6mm}
    \captionsetup{justification=centering}
    \caption{ResNet-152,\\batch size = 32}\label{fig:aws_gpu_small_resnet152_32}
  \end{subfigure}%
  \hfill
  \begin{subfigure}[b]{.33\columnwidth}
    \centering
    \includegraphics[width=\linewidth]{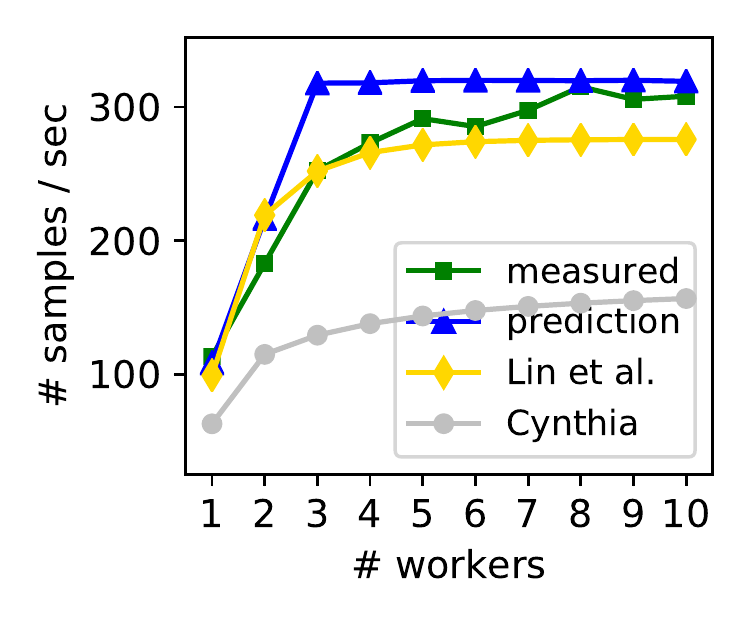}
    \vspace*{-6mm}
    \captionsetup{justification=centering}
    \caption{ResNet-152,\\batch size = 64}\label{fig:aws_gpu_small_resnet152_64}
  \end{subfigure}%
  \begin{subfigure}[b]{.33\columnwidth}
    \centering
    \includegraphics[width=\linewidth]{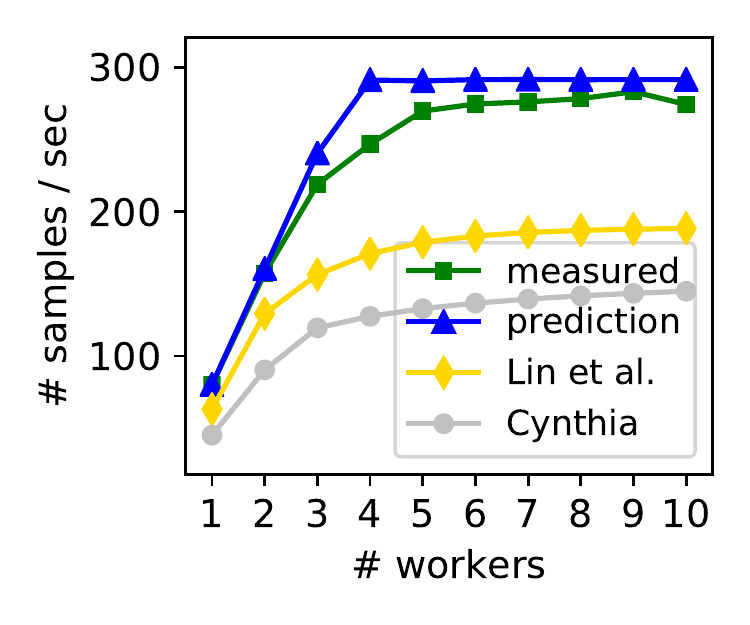}
    \vspace*{-6mm}
    \captionsetup{justification=centering}
    \caption{VGG-11,\\batch size = 128}\label{fig:aws_gpu_small_vgg11_128}
  \end{subfigure}%
  \begin{subfigure}[b]{.33\columnwidth}
    \centering
    \includegraphics[width=\linewidth]{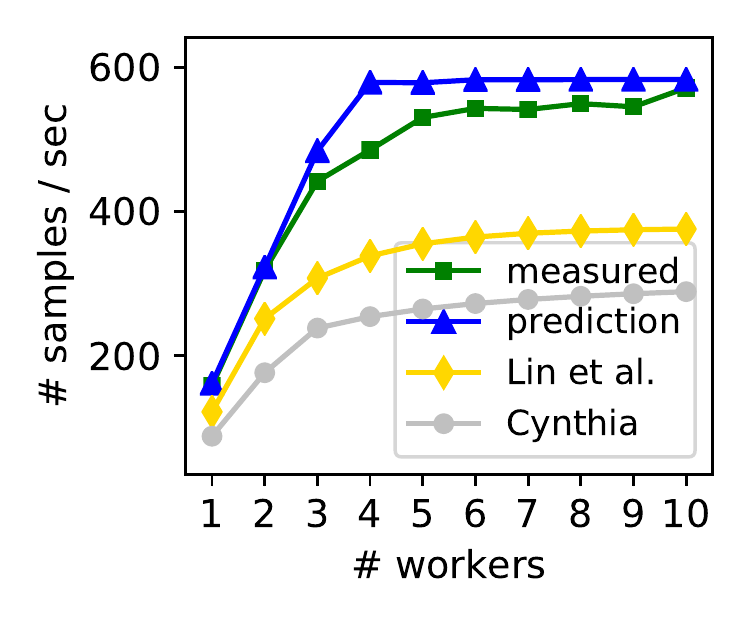}
    \vspace*{-6mm}
    \captionsetup{justification=centering}
    \caption{VGG-11,\\batch size = 256}\label{fig:aws_gpu_small_vgg11_256}
  \end{subfigure}%

  \caption{Prediction of training on AWS cloud (GPU cluster)}\label{fig:aws_gpu}\vspace{-1em}
\end{figure}

In \cref{fig:cluster_batch,fig:cluster_model,fig:aws_cpu,fig:aws_gpu},
we compared prediction accuracy of our method with Lin
et~al.~\cite{lin2018model} and Cynthia~\cite{zheng2019cynthia}. We
observe that the model of Lin et al. predicts throughput accurately in
DNN models with small batch sizes, while predicted throughput
saturates much earlier than real measurements for larger batch sizes,
where the overlap between communication and computation is
large. Throughput predicted by Cynthia is lower than measured. To
investigate whether a simple change could improve predictions, we
modified Cynthia's model by reducing communication times $T_C$~(which
affect utilization $U_1$ and throughput) in half, accounting for
separate uplink/downlink network resources; in a small set of test
cases, this modification seemed to improve Cynthia's prediction,
although still with large errors in some scenarios.

\subsection{Runtime Evaluation}

\begin{figure}[tb]
        \centering
        \begin{subfigure}[b]{\columnwidth}
                \centering
                \includegraphics[width=\linewidth]{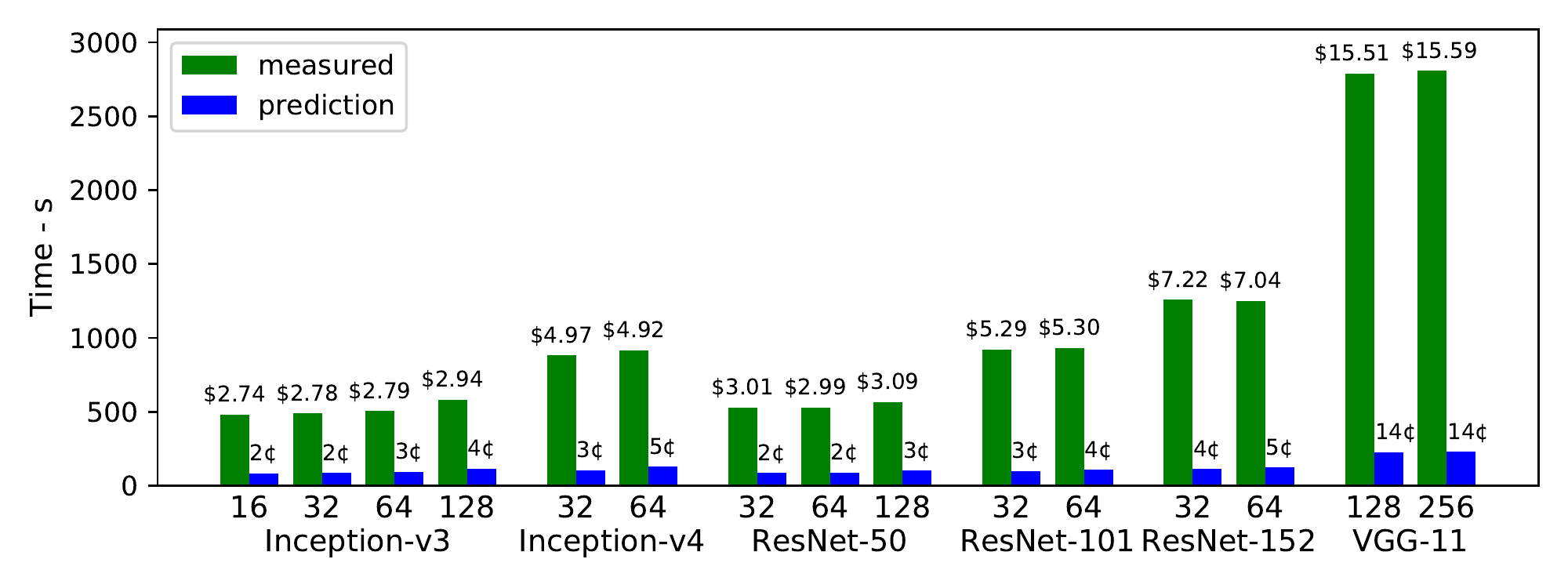}
        \end{subfigure}%
        \caption{Cost and time comparison between throughput measurement
          and profiling/simulation-based prediction}\label{fig:runtime}
        \vspace{-1em}
\end{figure}

\cref{fig:runtime} compares the execution time of 100 SGD steps on a
cloud GPU cluster with that of our prediction algorithm. The algorithm
execution time is evaluated using a single CPU core of an AWS
\texttt{c4.large} instance (2 vCPUs, 2.9~GHz, Intel Xeon E5-2666v3,
3.75~GB memory).
For example, directly measuring throughput of Inception-v3 with batch
size 128 using 1 to 10 workers (100 steps for each setting) took 581
seconds.  Our prediction method, including throughput measurement for
100 steps using 1 worker and prediction of 1000 steps for settings
with 2 to 10 workers took only 117s.  The prediction algorithm runs
faster than actual training, and it could be further optimized by
executing simulation runs in parallel over multiple cores or CPUs.

While the time required to obtain throughput measurements depends on
the batch size and network (training with many workers can reach
network bottlenecks), the execution time of the prediction algorithm
depends on the number of operations in an SGD step and on the number
of workers in the cluster.
Predicting throughput (instead of direct measurements) allows not only
considerable savings (each GPU instance is over $35\times$ more
expensive than the only CPU instance used for simulation) but also
shorter evaluation times.
Evaluation is particularly fast for DNNs with few operations (e.g.,
AlexNet, VGG), when network speed is slow (e.g., private CPU cluster
with 1~Gbps Ethernet), when computation is slow (e.g., slow CPU or
GPU), and when batch size is large.

% -*- ispell-local-dictionary: "american"; TeX-master: "../tensorpredict.tex"; -*-

\section{Multiple Parameter Servers}
\label{sec:multiple_ps}

\begin{figure}[tb!]
  \centering
  \begin{subfigure}[b]{.6\columnwidth}
    \centering
    \includegraphics[width=\linewidth]{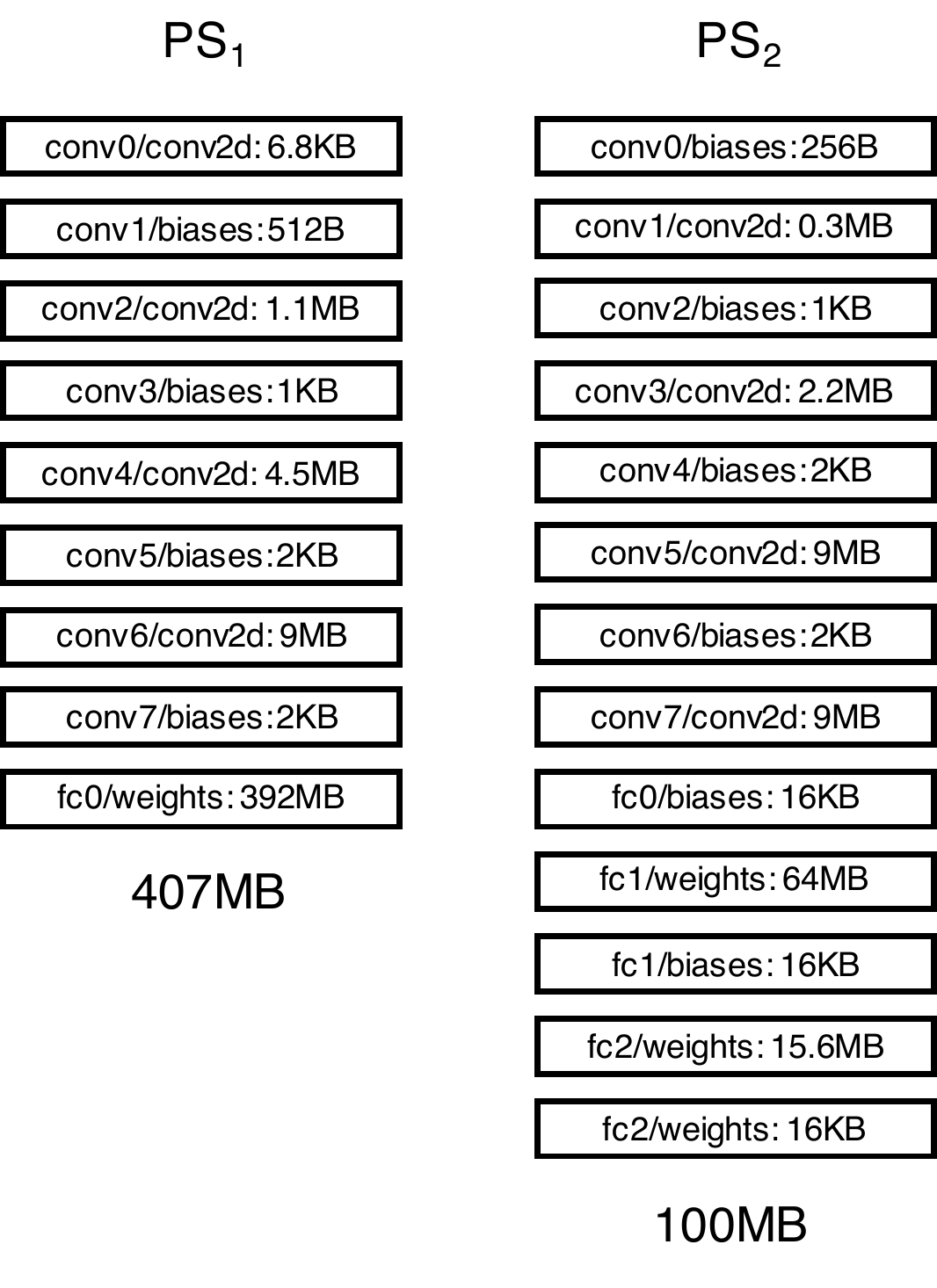}
  \end{subfigure}%
  \vspace{-1em}
  \caption{Partition of VGG-11 parameters among 2 PS}\label{fig:unbalanced_split}
  \vspace{-1em}
\end{figure}

\begin{figure}[t!]
  \begin{minipage}[b]{\linewidth}
    \begin{subfigure}[t]{.45\linewidth}
      \centering
      \includegraphics[width=\linewidth]{figures/3-prediction/downlink_1ps.pdf}
      \caption{One parameter server}\label{fig:downlink_1ps_states}
    \end{subfigure}%
    \hfill
    \begin{subfigure}[t]{.52\linewidth}
      \centering
      \includegraphics[width=\linewidth]{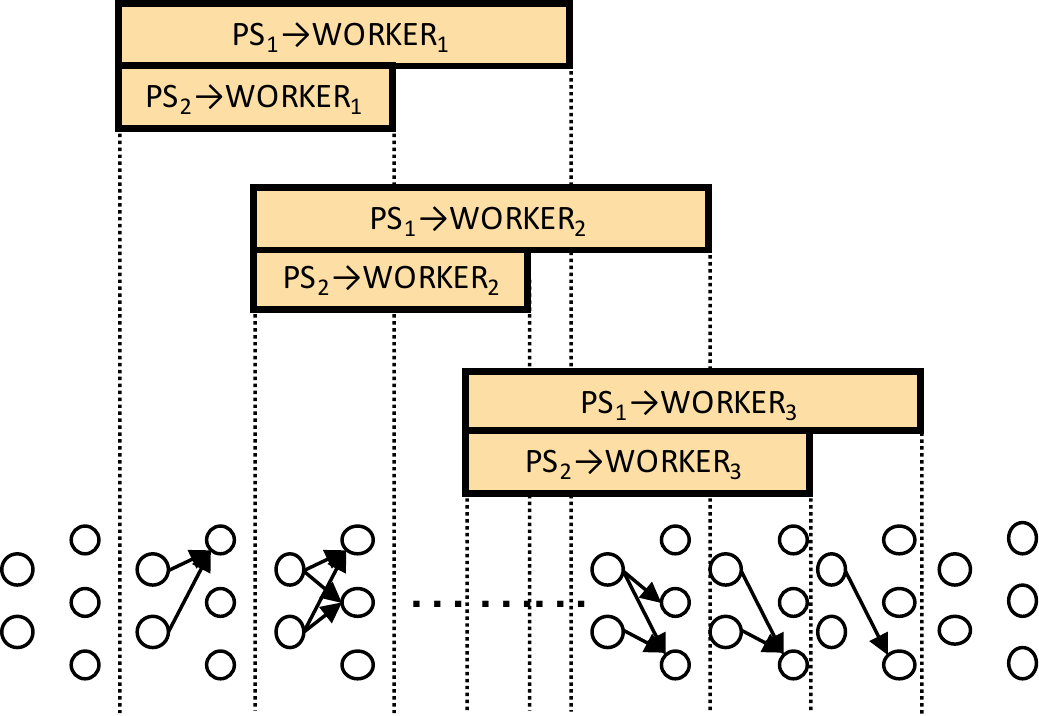}
      \caption{Two parameter servers}\label{fig:downlink_2ps_states}
    \end{subfigure}%
    \caption{Possible states (active/inactive) of download links}
    \label{fig:downlink_2ps_2ps_states}
  \end{minipage}%
\end{figure}

When a single parameter server becomes a
bottleneck, more parameter servers can be added to the cluster with
TensorFlow.
In this case, model parameters are partitioned among parameter
servers: for each part of the model, workers send updates and receive
new parameters from a specific parameter server.
We observe that the partition of model parameters among parameter
servers is often uneven: since a DNN layer is the minimum unit of
model parameters assigned in TensorFlow, parameters of entire DNN
layers are assigned to parameter servers so as to balance the amount
of data and networking load due to all workers.
Since the size of different layers can vary greatly, this split can be
uneven, as illustrated in \cref{fig:unbalanced_split} for VGG-11 when
model parameters are partitioned among 2~parameter servers $\cn{ps}_1$
and $\cn{ps}_2$.
Each layer is assigned to the parameter server that is currently
holding parameters with smallest total size (in bytes); therefore,
$\cn{ps}_1$ receives model parameters of a larger size
than~$\cn{ps}_2$ (407~MB instead of 100~MB): as a result, during
training each worker will exchange more data with~$\cn{ps}_1$ than
with~$\cn{ps}_2$.
This asymmetry complicates network communication patterns in the multiple
parameter server scenario.
Furthermore, \cref{fig:downlink_2ps_states} shows that for 2~parameter
servers, there are many more network states than for a single
parameter server: with $W$ workers and $M$ parameter servers, there
are $WM$ distinct links that can be used to download model parameters;
each can be active or inactive depending on whether the worker is
currently downloading parameters from the specific server, resulting
in $2^{\mathit{WM}}$ downlink states during simulation (and,
similarly, $2^{\mathit{WM}}$ uplink states).

Based on our observations from running \texttt{iperf} benchmarks, we
adopt a simple model for the case of 2 parameter servers: all active
connections with the same parameter server equally share its bandwidth
(for each uplink/downlink direction).
In addition, we account for configurations of active links where some
worker is the only worker exchanging data with $\cn{ps}_1$ but has to
contend with $n-1$ other workers to exchange data with $\cn{ps}_2$. In
this case, we assign bandwidth $\frac{1}{n}$ to the connections with
$\cn{ps}_2$ (equal sharing), but only up to $1-\frac{1}{n}$ to those
with $\cn{ps}_1$ (because the worker is already using $\frac{1}{n}$ of
its transmission bandwidth for $\cn{ps}_2$).

\begin{figure}[t!]
  \centering
  \begin{subfigure}[b]{.33\columnwidth}
    \centering
    \includegraphics[width=\linewidth]{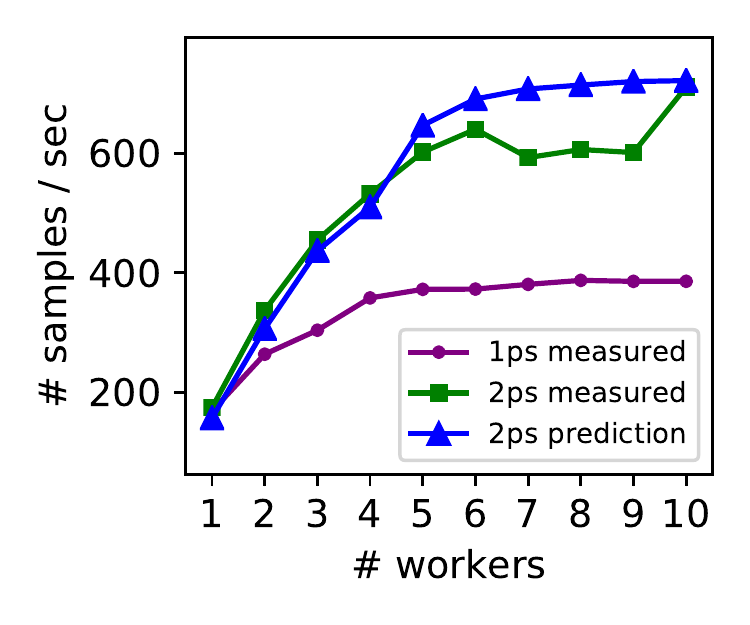}
    \vspace*{-6mm}
    \captionsetup{justification=centering}
    \caption{Inception-v3,\\batch size = 32}\label{fig:aws_gpu_small_inception3_32_2ps}
  \end{subfigure}%
  \begin{subfigure}[b]{.33\columnwidth}
    \centering
    \includegraphics[width=\linewidth]{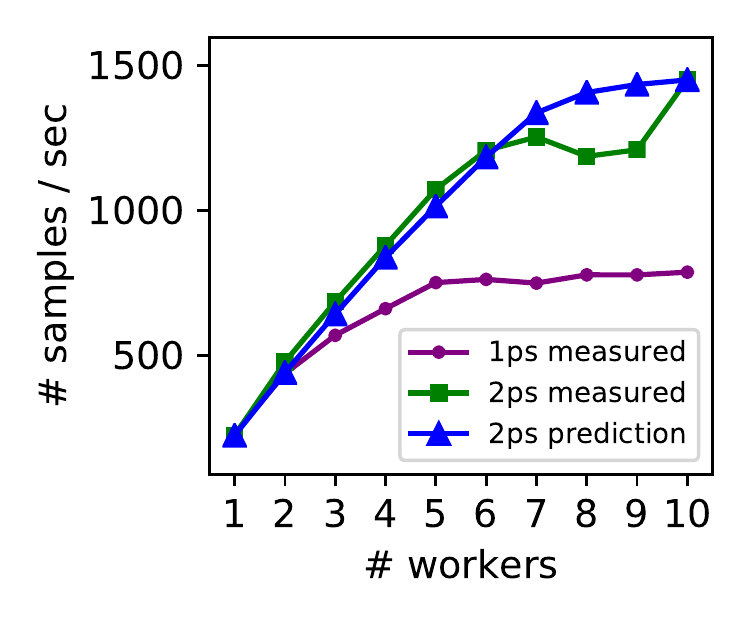}
    \vspace*{-6mm}
    \captionsetup{justification=centering}
    \caption{Inception-v3,\\batch size = 64}\label{fig:aws_gpu_small_inception3_64_2ps}
  \end{subfigure}%
  \begin{subfigure}[b]{.33\columnwidth}
    \centering
    \includegraphics[width=\linewidth]{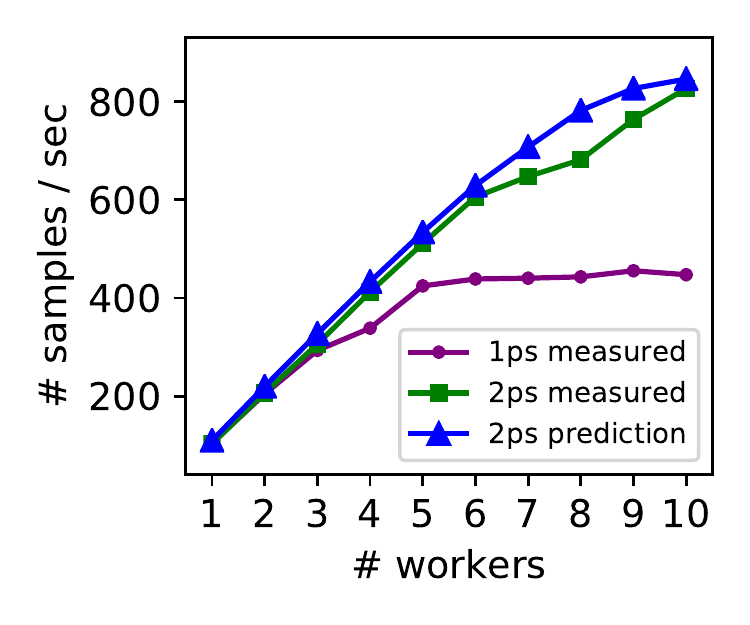}
    \vspace*{-6mm}
    \captionsetup{justification=centering}
    \caption{Inception-v4,\\batch size = 64}\label{fig:aws_gpu_small_inception4_64_2ps}
  \end{subfigure}%
  \hfill
  \begin{subfigure}[b]{.33\columnwidth}
    \centering
    \includegraphics[width=\linewidth]{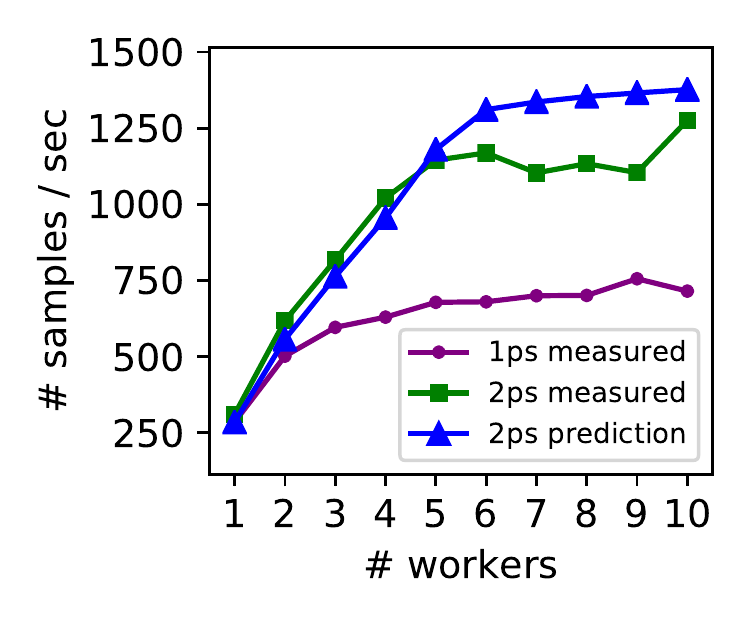}
    \vspace*{-6mm}
    \captionsetup{justification=centering}
    \caption{ResNet-50,\\batch size = 64}\label{fig:aws_gpu_small_resnet50_64_2ps}
  \end{subfigure}%
  \begin{subfigure}[b]{.33\columnwidth}
    \centering
    \includegraphics[width=\linewidth]{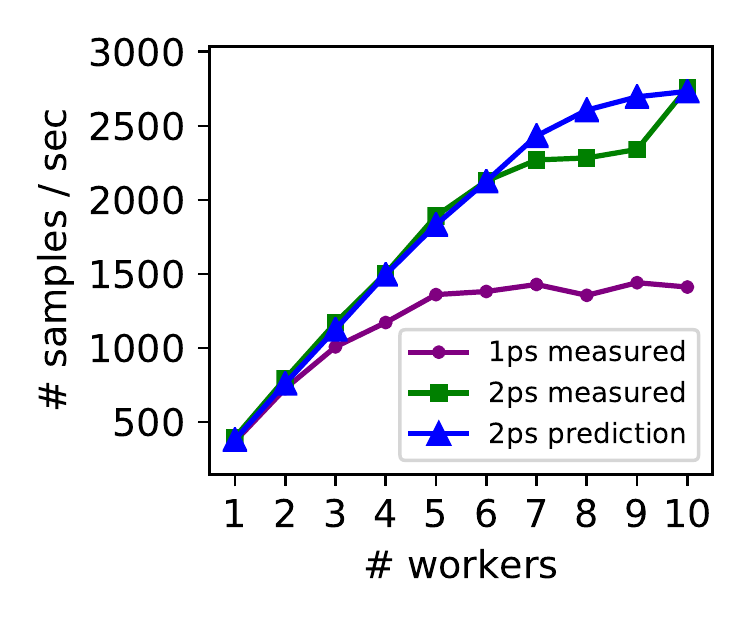}
    \vspace*{-6mm}
    \captionsetup{justification=centering}
    \caption{ResNet-50,\\batch size = 128}\label{fig:aws_gpu_small_resnet50_128_2ps}
  \end{subfigure}%
  \begin{subfigure}[b]{.33\columnwidth}
    \centering
    \includegraphics[width=\linewidth]{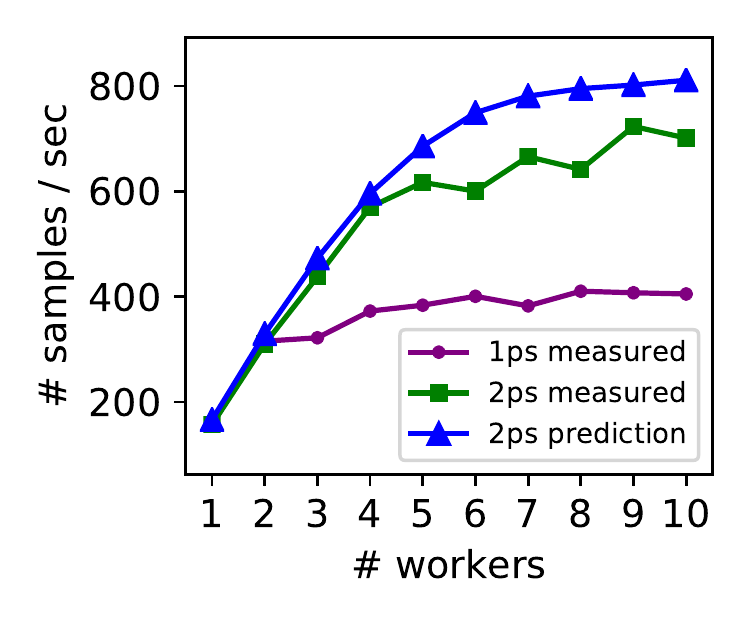}
    \vspace*{-6mm}
    \captionsetup{justification=centering}
    \caption{ResNet-101,\\batch size = 64}\label{fig:aws_gpu_small_resnet101_64_2ps}
  \end{subfigure}%
  \hfill
  \begin{subfigure}[b]{.33\columnwidth}
    \centering
    \includegraphics[width=\linewidth]{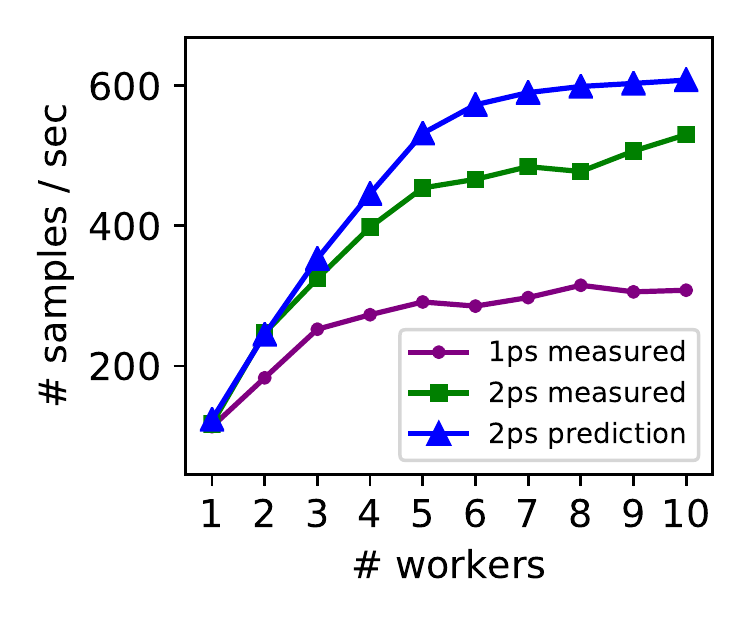}
    \vspace*{-6mm}
    \captionsetup{justification=centering}
    \caption{ResNet-152,\\batch size = 64}\label{fig:aws_gpu_small_resnet152_64_2ps}
  \end{subfigure}%
  \begin{subfigure}[b]{.33\columnwidth}
    \centering
    \includegraphics[width=\linewidth]{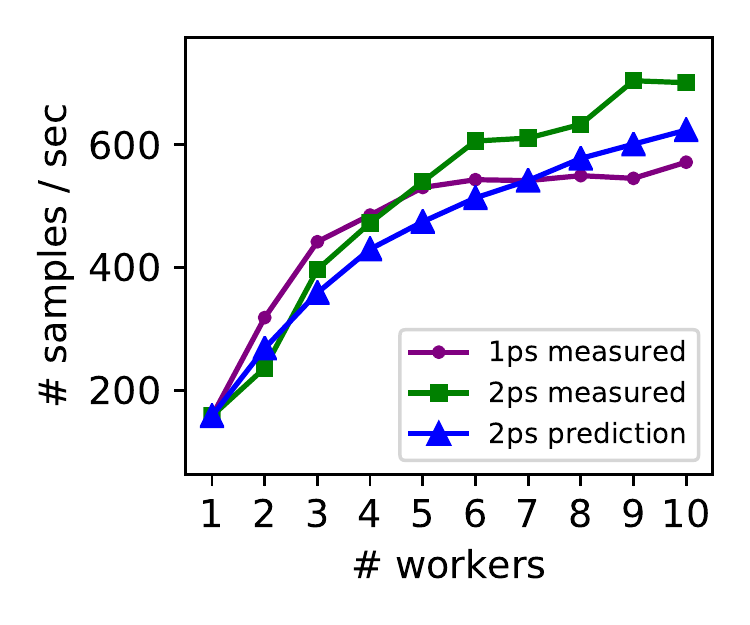}
    \vspace*{-6mm}
    \captionsetup{justification=centering}
    \caption{VGG-11,\\batch size = 256}\label{fig:aws_gpu_small_vgg11_256_2ps}
  \end{subfigure}%
  \hspace{.33\linewidth}
  \caption{Prediction with 2 PS on AWS GPU cloud}\label{fig:aws_gpu_multiple_ps}\vspace{-1em}
\end{figure}

Using this model of bandwidth sharing, we modify the trace generation
approach presented in \cref{simulation}: first, we collect profiling
traces with 2~parameter servers and 1~worker using \texttt{p3.2xlarge}
AWS instances (1~$\times$~NVIDIA V100 GPU); then, we run the
simulation algorithm (\cref{alg:simulation}) for $W$~workers and
resources $\cn{downlink}_i$, $\cn{uplink}_i$, $\cn{ps}_i$ for $i=1,2$.
We also run a real cluster with $W$ workers with GPUs and compare
throughput measured with 2~parameter servers (green curve) with our
predictions (blue curve), and with throughput measured with
1~parameter server (purple curve).
The results, presented in \cref{fig:aws_gpu_multiple_ps}, illustrate
that the error of throughput predictions on AWS with 2~parameter
servers is within 20\%, for most DNN models and batch sizes.
Prediction error is within 25\% in
\cref{fig:aws_gpu_small_resnet50_64_2ps} for $W = 7, 9$,
\cref{fig:aws_gpu_small_resnet101_64_2ps} for $W = 6, 8$,
and \cref{fig:aws_gpu_small_resnet152_64_2ps} for $W = 6, 7, 8$. 
Measurements with 1 and 2~parameter servers illustrate that throughput
is improved because of the additional parameter server.
Note that, due to the uneven split of parameters in VGG-11, adding a
second parameter server offers only a marginal improvement in
\cref{fig:aws_gpu_small_vgg11_256_2ps}; since our method is based on
real profiling traces collected in a configuration with 2~parameter
servers and 1~worker, we can account for the uneven split and
accurately predict throughput.

% -*- ispell-local-dictionary: "american"; TeX-master: "../tensorpredict.tex"; -*-

\section{Related Work}

\iffalse

Modeling of DNN workloads

High-level, curve fitting
- Jockey, recurring jobs
- Ernest, black box analytical model
- Optimus, ?

Difficult to predict accurately in different scenarios (unless lots of experiments are tried).
Our model is on SGD.

Fine-grained models, based on FLOPS:
- yan, ops profiled
- paleo, only data
- ulanov synchronous ..
Don't consider optimizations by the framework

Moreover, most model synchronous SGD => Cynthia and Li

\fi

Performance models of machine learning and data processing jobs have
been proposed in the literature with different levels of granularity
for computation and communication operations.
Ernest \cite{venkataraman2016ernest} and Optimus
\cite{peng2018optimus} use black-box analytical models of throughput
and perform, for each job, test runs with different numbers of workers
to estimate model parameters; in contrast, our work uses only one
profiling run, with a single worker.
Jockey \cite{ferguson2012jockey} predicts performance of data
processing jobs based on execution times and dependencies of each
phase, but operations in DNN training jobs are more fine-grained than
in data processing jobs.

At a lower level of granularity, works predating modern machine
learning frameworks \cite{yan2015performance} estimate speedups of
distributed DNN training jobs through detailed analysis of elementary
computations (such as matrix operations) which are profiled through
micro-benchmarks; Paleo \cite{qi2017paleo} models DNN computations
layer-wise, while \cite{shi2018performance} and
\cite{pei2019iteration} analyze individual computations executed on
a GPU.
A common characteristic of these works is that throughput predictions
are made as a function of computing speed (measured in FLOPS) and
complexity of each layer (modeled as number of operations). However,
additional factors including optimization strategies of the machine
learning framework, operation scheduling, transmission overheads, can
potentially affect training performance. Moreover, new optimization
strategies are being implemented in machine learning frameworks;
hence, such fine-grained models need to be adapted over time. In this
work, we base our model on the profiling of individual operations in a
computational graph, which is the intermediate representation in
TensorFlow: any optimization in the framework is reflected and
accounted for in the profiling information, so that our approach is
not limited to specific hardware platforms or framework
implementations.

While the majority of analytical approaches focus on synchronous SGD
\cite{ulanov2017modeling}, modeling asynchronous SGD is more difficult
because communication patterns between parameter servers and workers
are more complex and can change over time. Previous
work~\cite{lin2018model} builds a queueing model to estimate
throughput of asynchronous SGD, using this model for scheduling of
heterogeneous training jobs; coarse model parameters (the duration of
downlink/uplink and worker/server computation) are estimated from
single-worker scenario profiling.
Cynthia~\cite{zheng2019cynthia} predicts training time through an
analytical model based on network and CPU utilization, in order to
provision cloud instances.
These models are also at a high-level of granularity and model
communication and computation as \emph{sequential} phases, which is
contradictory to our findings: overlaps between communication and
computation are significant and play an important role in the
optimization of training performance, as illustrated by
\cref{fig:inceptionv3_bandwidth_trace}.
We compared our prediction results to those obtained with these
methods in \cref{sec:accuracy-comparison}
(\cref{fig:cluster_batch,fig:cluster_model,fig:aws_cpu,fig:aws_gpu}),
highlighting major accuracy improvements.

% -*- ispell-local-dictionary: "american"; TeX-master: "../tensorpredict.tex"; -*-

\section{Conclusions}

We proposed an approach to predicting training throughput of
asynchronous SGD in TensorFlow that extracts operation-level tracing
information from minimal single-worker profiling data and performs
discrete-event simulation to generate synthetic traces. Experimental
results show good prediction accuracy across DNN models, batch sizes,
and platforms as well as variants of TensorFlow, including
optimizations of the training process. Our experiments also indicate
that the more challenging cases are those with small number of workers
and smaller computation (relative to communication) times. Although
our simulation-based approach outperforms existing analytical modeling
approaches, it is still of interest to develop fine-grained analytical
models that address the shortcomings of previous work.

Future directions include heterogeneous hardware settings, multiple
($> 2$) parameter servers (with more complex bandwidth sharing
models), and other (than gRPC) communication mechanisms, including
TensorFlow with MPI and RDMA support.

\begin{acks}
This work was supported in part by the NSF CNS-1816887 and CCF-1763747 awards.
\end{acks}

\bibliographystyle{ACM-Reference-Format}
\bibliography{tensorpredict}
\end{document}